\documentclass[12pt]{JHEP3}

\usepackage{amsmath,epsfig}
\usepackage{amssymb,amsfonts}
\usepackage{latexsym}
\usepackage[latin1]{inputenc}
\usepackage{tocvsec2}
\usepackage{subeqnarray}
\usepackage{xcolor}

\usepackage{graphicx}
\usepackage{longtable}

\relax
\renewcommand{\theequation}{\arabic{section}.\arabic{equation}}
\def\be{\begin{equation}}
\def\ee{\end{equation}}
\def\bea{\begin{eqnarray}}
\def\eea{\end{eqnarray}}

\newcommand\fverb{\setbox\pippobox=\hbox\bgroup\verb}
\newcommand\fverbdo{\egroup\medskip\noindent%
                        \fbox{\unhbox\pippobox}\ }
\newcommand\fverbit{\egroup\item[\fbox{\unhbox\pippobox}]}

\newcommand{\bear}{\begin{eqnarray}}

\newcommand{\eear}{\end{eqnarray}}

\newcommand{\bsea}{\begin{subeqnarray}}
\newcommand{\esea}{\end{subeqnarray}}
\newbox\pippobox

\def\6{\partial}

\def\a{\alpha}

\def\m{\mu}
\def\n{\nu}

\def\sq
\def\a{\alpha}

\def\hri#1#2{\href{http://arxiv.org/abs/#1}{[arXiv:#1]#2}}
\def\hre#1#2{\href{http://arxiv.org/abs/#1/#2}{[arXiv:#1/#2]}}
\def\hspi#1#2{\href{http://www.slac.stanford.edu/spires/find/hep/www?irn=#1}{#2}}

\newcommand{\comments}[1]{}
\allowdisplaybreaks[3]

\title{Holographic Competition of Phases and Superconductivity }
\author{\large  Elias Kiritsis$^{a,b}$ and Li Li$^{a}$\\
~\\$^a$ \href{http://hep.physics.uoc.gr}{Crete Center for Theoretical Physics},
Department of Physics, University of Crete, 71003 Heraklion, Greece.\\
~\\
$^b$\href{http://www.apc.univ-paris7.fr}
{APC, Universit\'e Paris 7}, CNRS/IN2P3, CEA/IRFU, Obs. de Paris, Sorbonne Paris Cit\'e, B\^atiment Condorcet, F-75205, Paris Cedex 13, France (UMR du CNRS 7164).\\
~\\
\\\\
E-mail:\email{ lili@physics.uoc.gr}, \href{http://hep.physics.uoc.gr/~kiritsis/}{http://hep.physics.uoc.gr/~kiritsis/}
}

\preprint{CCTP-2015-19\\CCQCN-2015-102}

\abstract{We use a holographic theory to model and study the competition of four phases: an antiferromagnetic phase, a superconducting phase, a metallic phase and a striped phase, using as control parameters temperature and a doping-like parameter. We analyse the various instabilities and determine the possible phases. One class of phase diagrams, that we analyse in detail, is similar to that of high-temperature superconductors as well as other strange metal materials. }

\keywords{Holography, AdS/CMT, High-Tc superconductor, Pseudogap, Phase competition}

\begin{document}

\newpage
\section{Introduction and Results}
The understanding of the properties of high-temperature superconductivity, especially its microscopic origin, is significant in theory and application. Conventional superconductors are well described by BCS theory in which the condensate is a Cooper pair of electrons weakly bound together through lattice vibrations known as phonons.  However, many materials of significant theoretical and practical interests, such as high-$T_c$ cuprates and heavy fermion systems, are beyond BCS theory. The intrinsic interaction for those materials are believed to be in the strongly coupled regime, thus the conventional approaches, such as quasi-particle paradigm of Fermi liquid theory, seem at least in part of the phase diagram inapplicable. Some new concepts have been developed, such as the spin fluctuation superglue~\cite{Moriya2000}, the resonating valence bond  (RVB) gauge approach~\cite{anderson1987,weng:2007} and the SO(5) theory~\cite{demler2004}, which indeed provide useful insights into such  strongly-coupled materials and superconductors. Nevertheless, a deeper understanding of the unconventional superconductivity is still an open question~\cite{review,norman2011}. Any new idea would be helpful to shed light on the quest for understanding strongly correlated electron systems.

The holographic duality, also known as the gauge/gravity duality or AdS/CFT correspondence~\cite{Maldacena:1997re,Gubser:1998bc,Witten:1998qj}, originated  from string theory and has been used recently to understand theories that are at finite density and may be in the universality class of interesting, strongly-coupled condensed matter systems. The holographic duality relates the dynamics of a lower-dimensional quantum field theory to a dual gravitational/string theory living in higher dimensions. The key point is that, in the classical limit, the gravity side can be well described by generalisations of general relativity, while the dual field theory involves dynamics with strong interactions, thus providing an invaluable source of physical intuition as well as computational power to deal with strongly coupled problems. Progress has been made in the application of the holographic description towards phenomena at finite densities, relevant for condensed matter systems, in the last few years, such as strange metals~\cite{pol,pal,kkp}, (non-) Fermi liquid~\cite{Cubrovic, Liu1,Liu2}, superconductivity (superfluidity)~\cite{Gubser:2008px,Hartnoll:2008vx} and N\'eel phase transition~\cite{Iqbal:2010eh}, see~\cite{Hartnoll:2009sz,H,Sachdev:2011wg,Green:2013fqa} for reviews and \cite{ZLSS} for an excellent book on this emergent topic.

\begin{figure}[ht!]
\begin{center}
\includegraphics[width=.55\textwidth]{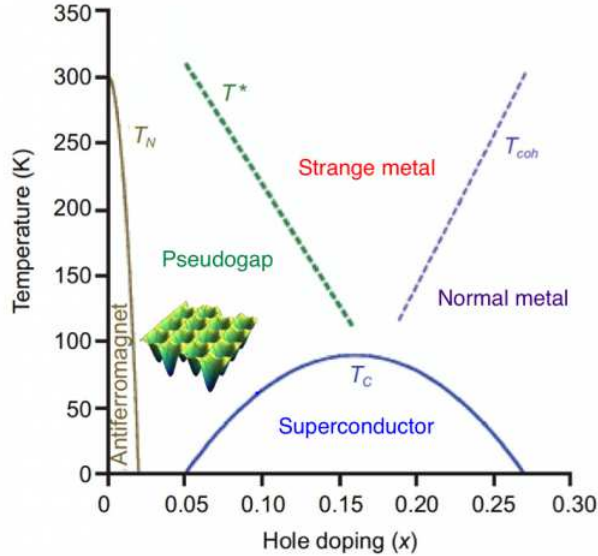}
\caption{ Schematic phase diagram of the high-$T_c$ cuprates. The subplot in the pseudogap region is used to stress the possible striped order and checkerboard order. This figure is adapted from~\cite{Cai:2015cya}.}
\label{fig:cuprates}
\end{center}
\end{figure}

In an attempt to build a consistent overall picture of those unconventional materials, mapping out and forming a rudimentary understanding of the temperature-doping phase diagram becomes a primary focus of research. The phase diagram is a landscape of exotic states of matter.
Typical phase diagrams of many unconventional superconductors, such as cuprates and iron-based pnictides, have some similarities among them. At low temperature $T$ and doping (or in general any other control parameter) $\textbf{x}$ the phase diagram is dominated by an antiferromagnetic (AF) phase. As the doping is increased to some intermediate region, a superconducting (SC) order appears below a critical temperature and will finally take over the phase diagram. Nevertheless, for sufficiently large $\textbf{x}$ the superconductivity can be destroyed. Therefore, there is typically a superconducting dome in the global phase diagram.  Besides antiferromagnetism and superconductivity as two fundamental and common states of matter, in other regions of $(T, \textbf{x})$ plane there are different kinds of phases that could emerge, depending on the concrete materials one is considering. One particularly interesting regime, typically in cuprates, is the underdoped pseudogap phase. Although a true understanding of this mysterious state is an open question, it is plausible that competing orders are at work and the evidence points to  a whole collection of exotic phenomena. The exotic orders that have been identified in the pseudogap regime include the stripes, the quantum liquid crystal orders, the spontaneous diamagnetic currents and so on~\cite{review,zaanen1,Berg:2009dga,vojta2009}. To illustrate this, in figure~\ref{fig:cuprates} we provide a schematic sketch of the phase diagram for high-$T_c$ cuprates. Although many efforts have been made, (see, for example,~\cite{weng:2007,demler2004}), so far the phase diagram in the temperature-doping plane,  putting all ingredients together, has not been assembled.

The purpose of the present work is to  study a similar class of problems: namely the competition of various phases that are similar to those we discussed above, namely the antiferromagnetic order, the striped order and the superconducting order, using the techniques of holography. We will use the techniques of effective holographic theories~\cite{Charmousis:2010zz} in order to analyse the competition of phases and the related phase diagrams.

We construct a holographic theory  by picking the relevant bulk fields corresponding to the most important operators of the dual field theory and then writing down a natural bulk action.\footnote{There is a complementary approach called top-down in which a theory is uniquely determined by a consistent truncation from string theory or supergravity. } This effective holographic  description has the advantage  of being general, i.e., the results might be valid for many different dual field theories upon varying parameters in the bulk Lagrangian. The bulk gravitational theory captures the macroscopic dynamics of the order parameters and symmetry breaking pattern, but can not explain their microscopic origin. Nevertheless, we hope that the holographic approach would uncover some universal aspects of the strongly-coupled superconductivity.

There are two key steps in this direction. The theory, should be able to accommodate the phases that we are interested in. Depending on the values of the temperature and the doping,  the boundary system can be in antiferromagnetic phase, superconducting phase, normal metallic phase\,\footnote{ Notice that in figure~\ref{fig:cuprates} there is also a state above the SC regime known as strange metal characterised by a resistivity linear in temperature. We do not include the strange metal phase in the present study, but we shall do it in the future. There are several holographic strange metal candidates,~\cite{pol,pal,kkp}, that can be included.} or some exotic phase in pseudogap region.  Next and more importantly, the theory should have a degree of freedom to realise the desired patten of each phase in the $(T, \textbf{x})$ plane. Comparing with figure~\ref{fig:cuprates}, we should have a superconducting dome, a corner region of antiferromagnetism at low doping level, a normal metallic phase at high doping and a pseudogap region between antiferromagnetic and superconducting phases. As a consequence of holography, our dual field theory is scale invariant which typically appears at some quantum critical point of strongly correlated systems. The quantum phase transition is one major theme in the context of theories of unconventional superconductivity.

The $(3+1)$ dimensional (gravitational) ``toy" theory we shall consider contains the following bulk fields.
 \begin{itemize}

\item A metric $g_{\m\n}$  dual to the stress tensor of the boundary theory.

 \item A triplet real scalar $\Phi^a$ which is dual to the order parameter for antiferromagnetism.

 \item An SU(2) gauge field $g_{\m}^a$ dual to the conserved spin (antiferromagnetic) current. The scalar field $\Phi^a$ transforms in the adjoint of the SU(2) gauge group.

 \item Two U(1) gauge fields $A_{\m},B_{\m}$ that are dual to two conserved U(1) currents in the theory.  One corresponds to the total charge of the dual system and the other one is associated with the doping $\textbf{x}$. The association to the doping parameter is motivated as follows. In standard cuprates there are two charge densities: one is intrinsic to the basic material and the other is related to added charges. In such materials such an addition is associated to the doping parameter. This resembles our case where there are two charge densities and their relative charge density defines our ``doping" variable $\textbf{x}$. This is our definition of the doping variable and in the rest of the paper we will use this name without any additional explanation

 \item A scalar $\chi$ which is in general charged under two U(1) gauge fields and in particular under electromagnetism. It will be the order parameter for superconductivity.

 \item A neutral (real) pseudo-scalar $\alpha$ which is dual to antisymmetric tensor that couples to string defects. It will play a pivotal role in the realisation of a striped phase.

\end{itemize}

The two control parameters in our holographic setup are temperature $T$ and a doping-like parameter $\textbf{x}$. By tuning $T$ and $\textbf{x}$ several phases can appear:
\begin{itemize}

\item  A normal metallic phase without any symmetry breaking.
\item  An antiferromagnetic phase.
\item  A superconducting phase.
\item  A spatially modulated phase with charge density wave (CDW) order.

\end{itemize}
In our setup the latter three kinds of phases appear spontaneously without any deformation from the the dual field theory point of view. The spatially modulated phase can have striped or checkerboard structure. The AF and SC condensates can also be spatially modulated  and there are in principle coexisting phases in which more than one orders can exist simultaneously, resulting in, such as AF+SC, SC+CDW and AF+CDW states.

\begin{figure}[ht!]
\begin{center}
\includegraphics[width=.55\textwidth]{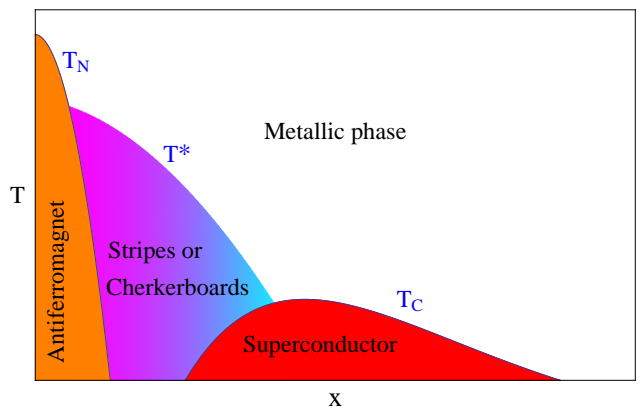}
\caption{ A schematic figure of a plausible phase diagram in the temperature-doping plane that can be proposed from our holographic setup.}
\label{fig:holoydiagram}
\end{center}
\end{figure}

We find that this simple bulk theory can exhibit many interesting features. A schematic picture of a likely phase diagram incorporating our study is given in  figure~\ref{fig:holoydiagram}.\footnote{To obtain figure~\ref{fig:holoydiagram} some assumptions are made which will be discussed in detail in the text. Moreover, a much more complex phase diagram will also be discussed.} At low doping on the far left, the phase diagram is dominated by the antiferromagnetic phase. As one increases the doping parameter, the antiferromagnetic phase will be destroyed and be replaced by a striped or checkerboard phase which would be reminiscent of  the ``pseudogap" phase.  More interestingly, there is a dome-shaped superconducting region in the $(T, \textbf{x})$ phase diagram, separated from the spatially modulated state at smaller values of the doping parameter and a metallic normal phase at larger values of the doping parameter. In particular, the characteristic temperature for each broken phase is comparable with others. Moreover, the temperature at which the phase transition occurs can be tuned to zero by the doping parameter $\textbf{x}$. Therefore, the dual system would contain quantum phase transitions.

It should be stressed that if one scans all possible parameters of the effective holographic theory, then many possible phase diagrams can appear. These are classified and discussed in section~\ref{sec:diagram}.
We argue however that simple assumptions about the stability properties of the undoped (AF ground state and no superconductivity) and extrermely doped theory (no AF ground state, no superconductivity) give, for natural values of parameters, a phase diagram that is qualitatively similar to the one in figure~\ref{fig:holoydiagram}.

The rest of the paper is arranged in the following way:
 \begin{itemize}
\item Section~\ref{sec:hmodel} motivates our choice of holographic theory and fields and introduces the holographic theory that will be investigated in detail in the following sections.

\item The instabilities towards developing antiferromagnetic phase and superconducting phase are discussed in section~\ref{sec:normal} by studying the modes that violate the Breitenlohner-Freedman (BF)  bound of the $AdS_2$ IR geometry of the normal phase at zero temperature.

\item In section~\ref{sec:scphase} and section~\ref{sec:afphase}  we construct the superconducting phase and the antiferromagnetic phase at finite temperatures, respectively. We explicitly show that there can   be a superconducting dome as well as a corner region of antiferromagnetic phase in the $(T, \textbf{x})$ plane.

\item Section~\ref{sec:spatial} is devoted to studying the striped instabilities which result in charge density wave seen from the dual field theory perspective. We also briefly discuss the checkerboard structure.

\item In section~\ref{sec:diagram} we classify possible phase diagrams for each broken phase that could be realised in our theory. Then we combine the basic necessary ingredients together to construct the full phase diagram.

\item We conclude with a discussion of some possible future research directions in section~\ref{sec:outlook}.

\item In appendix~\ref{app:stuck} we briefly introduce the theory which is a natural generalisation of holographic superconductors.

\item The most general equations of motion for homogeneous and isotropic background are presented in appendix~\ref{app:eoms}.

\item In appendix~\ref{app:afspin} we demonstrate that the linearised fluctuations of SU(2) gauge field and triplet scalar in symmetry breaking directions form a closed system, from which we show that the holographic description displays the associated spin wave with linear dispersion.

\item The method we use to search for the static zero modes of the striped instabilities is described in appendix~\ref{app:doushot}.

\item We give much more details of checkerboard instabilities in appendix~\ref{app:checkerboard}.

\item Finally in appendix~\ref{app:stripedonaf}, the striped zero mode is analysed around the condensed phase of antiferromagnetism.
\end{itemize}


\section{Construction of the Holographic Theory}
\label{sec:hmodel}
\subsection{Condensed matter related motivations}
During the past three decades, intense experimental and theoretical research have discovered many common features in the properties of materials that are on the border with magnetism. Such materials show a strong interplay and interaction between spin and charge (electrons), lead generically to strong electron correlations and in some cases show ``high-temperature" superconductivity. Their phase diagrams have similar topologies. This definition encompasses a wider variety of unconventional superconductors, such as some organic superconductors and heavy fermion compounds despite the critical temperature can be lower than many conventional superconductors. To develop a rigorous theory for such strongly correlated materials and high-temperature superconductivity remains one of the major problems in physics and has proven to be a difficult challenge.

High-temperature superconducting materials are characterised by strong electronic correlations which are responsible for a number of anomalous properties and competing orders. Those low temperature orders interact with superconductivity in an intricate way.  Besides the antiferromagnetic (AF) and the superconducting (SC) states,  one of the prominent orders is known as the stripe order and has been identified, for example, in the pseudogap region of high-$T_c$ cuprates~\cite{review,zaanen1,Berg:2009dga,vojta2009}. The phase diagram of the high-temperature superconductors is in that sense a landscape of exotic states of matter. Although different classes of high-temperature superconductors have their unique characters, their phase diagrams share many similarities. A representative phase diagram is the one of cuprates as shown in figure~\ref{fig:cuprates}. To establish a consistent overall picture of those unconventional superconductors, mapping out and forming a rudimentary understanding of the temperature-doping phase diagram can be a primary focus of research.

Holography provides a novel approach for studying strongly correlated systems. It has been widely used to describe dual systems in various phases and to discuss phase transitions at finite temperature and densities. To reproduce the phase diagram of the high-temperature superconductors by a dual theory is an idea worth trying. Since a number of important unconventional superconductors, such as the cuprates, have layered structure and much of the physics is $(2+1)$ dimensional, we will consider a dual gravitational  theory in  $(3+1)$ dimensions.

In order to construct a holographic dual to a  doped superconductor, the first challenge is how one can introduce the effect of doping into the holographic theory.
One crucial point in our theory is to introduce two bulk U(1) gauge fields, which in turn correspond to two currents in the dual boundary theory. The two independently conserved currents can be related to different kinds of charge in the theory. One may be reflecting ``bound" charge while the other itinerant charge but this is not necessarily the only interpretation.

The ratio of these two charges plays the role of the additional control parameter that we will call  ``doping" $\textbf{x}$ in the holographic setup.
Theories with two tuneable chemical potentials or charge densities are often called imbalanced mixtures, because the two kinds of charges are present in imbalanced numbers. Typical examples include two component Fermi gas~\cite{Shin2008} and QCD at finite baryon and isospin chemical potential~\cite{He:2005nk}.\footnote{Holographic studies with such imbalanced mixtures can be found, for example, in~\cite{Erdmenger:2011hp,Bigazzi:2011ak,Musso:2013rnr,Amoretti:2013oia}.} It is the competition between them that determines the macroscopic  behaviours of the real system, such as the whole phase diagram and transport coefficients.
In the rest of the papers we will drop the quote marks from the ``doping" parameter $\textbf{x}$.

We will also not consider transitions that covert one kind of U(1) charge into the other. It is however an interesting project to investigate the effect of such transitions to the phase diagrams we are studying.

\subsection{The holographic theory and its possible phases}

The normal phase without any symmetry breaking is described by the AdS Reissner-Nordstri\"{o}m (AdS-RN) black brane with two charges corresponding to two U(1) gauge fields. The zero temperature solution is a domain wall interpolating between $AdS_2\times \mathbb{R}^2$ in the far infrared (IR) and $AdS_4$ in the ultraviolet (UV). The near horizon $AdS_2$ geometry with two charges will be used to analyse the global $(T, \textbf{x})$ phase diagram.

Besides the normal phase, there are three kinds of phases, i.e., superconducting phase, antiferromagnetic phase as well as spatially modulated phase, such as striped phase and checkerboard phase. Those phases are associated with a particular type of symmetry breaking. We call them broken phases. In the following, we will give the gravity duals  for the three kinds of broken phases in our setup and then write down a general effective action.

\subsubsection{Gravity dual of broken phases}

The superconducting phase transition is associated with spontaneously breaking of U(1) gauge symmetry. The first holographic superconductor is known as the Abelian-Higgs model~\cite{Hartnoll:2008vx}, in which a charged scalar, say $\psi$, is minimally coupled with a bulk U(1) gauge field. When the temperature is below a critical one, the scalar obtains a nontrivial bulk configuration, which means that the dual operator acquires a non-vanishing vacuum expectation value (VEV) breaking the U(1) symmetry spontaneously. This s-wave model was soon generalised to p-wave models~\cite{Gubser:2008wv,Donos:2011ff,Aprile:2010ge,Cai:2013pda} as well as d-wave models~\cite{Chen:2010mk,Benini:2010pr}. Reviews can be found in~\cite{Cai:2015cya,Herzog:2009xv,Horowitz:2010gk,Musso:2014efa}. In this paper we adopt a generalised version of previous holographic superconductors, i.e., in the generalised symmetry breaking theories, with the most general allowed interactions at the two-derivative level~\cite{gk2}. In particular we allow  kinetic terms to depend in general on the condensates.

The complex scalar $\psi$ (dual to the superconducting order parameter)  is replaced by two real scalars $\chi$ and $\theta$ (see appendix~\ref{app:stuck} for a short introduction). Spontaneous breaking of the boundary U(1) symmetry at a particular temperature is dual to the condensation of the scalar $\chi$ in the bulk. A nontrivial solution of $\chi$ corresponds to the superconducting phase and the scalar $\theta$ is the source of the Goldstone excitation of the broken U(1) symmetry as the breaking is spontaneous. Holographic superconductors in similar models have been considered in a number of works~\cite{Franco:2009yz,Aprile:2009ai,Peng:2011gh,Cai:2012es}.

In the low energy limit of an electron system, spin rotation is a symmetry even though the rotational symmetry could be broken by a presence of lattice or other effects. The low-energy theory is then characterised by an SU(2) global symmetry which describes the spin rotation. From the symmetry breaking point of view, antiferromagnetic ordering is described by a spontaneous breaking from SU(2) to U(1) which corresponds to rotations about the remaining axis. The gravity dual of AF phase can be constructed by adopting a real scalar $\Phi^a$ which transforms as a triplet under SU(2) in the bulk~\cite{Iqbal:2010eh}. Spontaneous breaking of the boundary SU(2) symmetry at some critical temperature translates in the holographic dual into the condensation of the scalar $\Phi^a$.
A bulk background with vanishing SU(2) gauge field but with a normalisable $\Phi^a$ in one direction is then holographically dual to an AF phase or spin density wave phase in the boundary theory.~\footnote{An alternative approach for antiferromagnetism in holography was realised by introducing two real antisymmetric tensor fields on the gravity side~\cite{Cai:2014jta,Cai:2015mja}.}

Spatially modulated phases such as charge density wave (CDW) phase and checkerboard structure in high-temperature superconductors are associated with spontaneously breaking spatially translational invariance. There are some studies of the spontaneous breaking of translational invariance in variety of holographic theories~\cite{Domokos:2007kt, Nakamura:2009tf,Ooguri:2010xs,Takeuchi:2011uk,Donos:2012gg,Donos:2011bh,Donos:2011qt,Bergman:2011rf,Iizuka:2013ag,Alsup:2012kr,Donos:2013gda,Krikun:2013iha}. The corresponding broken solutions could be related to the spatially modulated phases of condensed matter systems in which, for example, superconducting order can compete or coexist with CDWs or checkerboard orders in the pseudogap region. In practice, in order to break the translational invariance spontaneously, one needs to introduce particular static mixing modes, such that the stable branch of the solution to the holographic theory shifts from the homogeneous state to the state with nonzero momentum, resulting in spontaneous translational symmetry breaking. In this paper we introduce striped instabilities while keeping the charged AdS-RN as the unbroken phase by adopting a pseudo-scalar $\alpha$ (dual to a CP-odd operator) which couples with Chern-Simons terms of U(1) gauge fields.~\footnote{This is the simplest way to generate such an order. There are more involved ways that can use CP-even operators~\cite{Donos:2013gda}, but we would need to introduce extra interactions from the most general scalar operator in our theory. This will be included in future work. }

\subsubsection{The holographic theory}
Now we have gravity duals for each broken phase from a symmetry breaking point of view. We can then write down a general effective gravitational theory,
\begin{eqnarray}\label{action1}
&&S=\frac{1}{2\kappa_N^2}\int d^{4}x \sqrt{-g} \left[\mathcal{R}-2\Lambda+\mathcal{L}_{m}+\mathcal{L}_{cs}\right],\\
\mathcal{L}_{m}=&&-\frac{Z_G}{4}G^a_{\mu\nu}G^{a\mu\nu}-\frac{1}{2}D_\mu \Phi^{a} D^\mu \Phi^a-\frac{Z_A}{4}A_{\mu\nu}A^{\mu\nu}-\frac{Z_B}{4}B_{\mu\nu}B^{\mu\nu}-\frac{Z_{AB}}{2}A_{\mu\nu}B^{\mu\nu}\nonumber\\ &&-\frac{1}{2}\nabla_{\mu}\chi \nabla^{\mu}\chi-\mathcal{F}(\chi)(\nabla_\mu\theta-q_A A_\mu-q_B B_\mu)^2-\frac{1}{2}\nabla_{\mu}\alpha \nabla^{\mu}\alpha-V_{int},\\ \nonumber
\mathcal{L}_{cs}=&&-\vartheta_1(\alpha) \epsilon^{\mu\nu\lambda\sigma}A_{\mu\nu}A_{\lambda\sigma}-\vartheta_2(\alpha) \epsilon^{\mu\nu\lambda\sigma}A_{\mu\nu}B_{\lambda\sigma}.
\label{Smatter}
\end{eqnarray}
Further explanation of the action is given as follows.
\begin{itemize}

\item   $g^a_\mu$ is an SU(2) gauge field, dual  to the global SU(2) spin currents of the boundary theory. The field strength is
\begin{equation}\nonumber
G^a_{\mu\nu}=\partial_\mu g^a_\nu-\partial_\nu g^a_\mu+\epsilon^{abc}g^b_\mu g^c_\nu,
\end{equation}
where $a, b, c=(1,2,3)$ are the index of the SU(2) algebra and $\epsilon^{abc}$ is antisymmetric with $\epsilon^{123}=1$.

\item $\Phi^a$ is a triplet scalar charged under the bulk SU(2) gauge field. It corresponds to the AF order parameter. The covariant derivative of $\Phi^a$ is defined by
\begin{equation}\nonumber
D_\mu\Phi^a=\partial_\mu\Phi^a+\epsilon^{abc}g_\mu^b\Phi^c.
\end{equation}

\item $A_\mu$ with its field strength $A_{\mu\nu}=\partial_\mu A_\nu-\partial_\nu A_\mu$ is one of the two U(1) gauge fields dual to boundary conserved U(1) currents.

\item $B_\mu$ with its field strength $B_{\mu\nu}=\partial_\mu B_\nu-\partial_\nu B_\mu$ is the second U(1) bulk gauge field dual to a boundary conserved current.

\item Both $\chi$ and $\theta$ are real scalar fields and are related to the superconducting complex condensate. $\theta$ is the phase of the condensate while  $\chi$ is related non-linearly to the amplitude of the condensate. $\chi$ is charged under the two U(1) gauge fields with charges $q_A$ or $q_B$.~\footnote{We will show later that, in general, one can set either $q_A$ or $q_B$ to be zero without changing the global phase diagram.}

\item The field $\alpha$ is a CP-odd pseudo-scalar field (also known as an axion if it has no potential). The last two terms in $\mathcal{L}_{cs}$ are Chern-Simons like terms in which $\epsilon_{\mu\nu\lambda\sigma}=\sqrt{-g}\varepsilon_{\mu\nu\lambda\sigma}$ with $\varepsilon_{txyr}=1$. The pseudo-scalar $\alpha$ couples with Chern-Simons terms.~\footnote{In a general effective theory, one can also add a term $\vartheta_3(\alpha) \epsilon^{\mu\nu\lambda\sigma}B_{\mu\nu}B_{\lambda\sigma}$ into $\mathcal{L}_{cs}$. Nevertheless, this term will not affect our results, and therefore we do not include it in the action for the sake of simplicity.}

\item  The cosmological constant is $\Lambda=-\frac{3}{L^2}$ with $L$ the AdS radius.

\item $V_{int}$ is the gauge invariant interaction potential for the scalars and is further discussed below.

\end{itemize}
This action will be supplemented later by the Gibbons-Hawking term and boundary counter terms for renormalisation.

The solution with nontrivial $\chi$ and $\mathcal{F}(\chi)\neq 0$ describes holographic physics in the U(1) symmetry broken phase.  Note that the scalar $\theta$ is the source of the Goldstone boson of the broken U(1) symmetry if the breaking is spontaneous.  The symmetry broken solution with a nonzero order parameter $\Phi^a$ but zero background gauge field $g_\mu^a$ corresponds to the antiferromagnetic phase. We always turn on the third component of $\Phi$, i.e., $\Phi=(0, 0, \phi)$ and set $g_\mu^a=0$. The bulk solution with nontrivial $\alpha$ is dual to a phase breaking both time-reversal and parity symmetry. We will show in below that the axion-like terms can induce striped instabilities.

As a consequence of SU(2) symmetry, $\Phi^a$ should appear in any function in terms of the combination $\Phi^a\Phi^a$. We take the coupling $Z_G$ to be a constant while other couplings are analytic functions of $\Phi^a\Phi^a$, $\chi$ and $\alpha$. We choose to work with functions $Z_A$, $Z_B$, $Z_{AB}$ and $V_{int}$ that admit the electrically charged AdS Reissner-Nordstri\"{o}m (AdS-RN) black brane when $\Phi=\chi=\alpha=0$.

Without loss of generality, we will parametrise the functions in the following expansion\footnote{A way to think about this basis of currents is as follows. Consider a gauge field $A_1$ dual to the charge of electrons in material without doping and $A_2$ dual to the charge associated to doping. There will be terms, in the effective action proportional to $F_1F_2$ induced by higher operators that are charged under both kinds of U(1) symmetries, and we neglect a mass term that is due to transitions that convert one kind of charge into another. The superconducting condensate $\chi$  will be expected to involve only the second kind of charge and therefore $q_1=0$, $q_2\not=0$. By an appropriate rotation we can set the constant part of $Z_{12}$ to zero and we rotate $F_1, F_2$ to the basis above, i.e., $A_{\mu\nu}, B_{\mu\nu}$. Then $q_A\not=0, q_B\not=0$ in general.}
as $\Phi, \chi, \alpha\rightarrow 0$.
\begin{equation}\label{vpossi}
\begin{split}
&Z_A(\Phi^a\Phi^a, \chi,\alpha)=1+\frac{a_1}{2}\Phi^a\Phi^a+\frac{a_2}{2}\chi^2+\frac{a_3}{2}\alpha^2+\cdots,\\
&Z_B(\Phi^a\Phi^a, \chi,\alpha)=1+\frac{b_1}{2}\Phi^a\Phi^a+\frac{b_2}{2}\chi^2+\frac{b_3}{2}\alpha^2+\cdots,\\
&Z_{AB}(\Phi^a\Phi^a,\chi,\alpha)=\frac{c_1}{2} \Phi^a\Phi^a+\frac{c_2}{2}\chi^2+\frac{c_3}{2}\alpha^2+\cdots,\\
&V_{int}(\Phi^a\Phi^a, \chi,\alpha)=\frac{1}{2}m^2\chi^2+\frac{1}{2}M^2 \Phi^a\Phi^a+\frac{1}{2}\widetilde{m}^2\alpha^2\cdots,\\
&\vartheta_1(\alpha)=\frac{n_1}{2}\alpha+\cdots,\quad \vartheta_2(\alpha)=\frac{n_2}{2}\alpha+\cdots,\quad \mathcal{F}(\chi)=\frac{1}{2}\chi^2+\cdots,
\end{split}
\end{equation}
where the dots denote the higher order corrections, like $\chi^4$, $\Phi^a\Phi^a \chi^2$ and $\chi^2\alpha^4$.  As we will show below, these leading quadratic terms are enough to compute the temperature at which the AdS-RN black brane becomes perturbatively unstable with respect to an exponentially growing mode of the dual operator.
Note also that as the scalar fields are dimensionless the parameters $a_i,b_i,c_i$ are also dimensionless.

Note that in the effective holographic theory, coupling parameter can be taken to have  general values. However, such constants are constrained by unitarity constraints. We may obtain some of them from the requirement that the diagonal coupling constants of the two gauge fields should not diverge (should be regular) at finite values of the scalars. This requires that
\begin{equation}
Z_{\pm}={Z_A+Z_B\pm \sqrt{(Z_A-Z_B)^2+4Z_{AB}^2}\over 2},
\end{equation}
when expanded in terms of the scalars should have positive coefficients, indicating that the coefficients $c_i$ cannot be large\footnote{This argument can be eventually avoided, however it suggests what are natural values for the coefficients.} compared to $|a_i-b_i|$.

The coefficients $a_i,b_i,c_i$ etc control the overall size of the four-point correlators involving two scalars (condensates) and two currents.
In holographic theories the size of such correlators is naturally small controlled by a small string coupling constant in the dual theory.

From the action we can derive the equations of motion for the matter fields
\begin{equation}
\nabla_\mu[\mathcal{F}(\chi)(\nabla^\mu\theta-q_A A^\mu-q_B B^\mu)]= 0,
\end{equation}
\begin{equation}\label{eomgg}
Z_G \nabla_\nu G^{\nu\mu a}+Z_G\epsilon^{abc}g_\nu^b G^{\nu\mu c}= \epsilon^{abc} \Phi^b D^\mu \Phi^c,
\end{equation}
\begin{equation}
\begin{split}
 \nabla_\mu\nabla^\mu\alpha-\left(\frac{\partial_\alpha Z_A}{4}A^2+\frac{\partial_\alpha Z_B}{4}B^2+\frac{\partial_\alpha Z_{AB}}{2}A B\right)-\partial_\alpha V_{int}=\\
\partial_\alpha\vartheta_1\epsilon^{\mu\nu\lambda\sigma}A_{\mu\nu}A_{\lambda\sigma}+\partial_\alpha\vartheta_2\epsilon^{\mu\nu\lambda\sigma}A_{\mu\nu}B_{\lambda\sigma},
 \end{split}
\end{equation}
\begin{equation}\label{eomphi}
\nabla_\mu D^\mu \Phi^a+\epsilon^{abc}g_\mu^b D^\mu\Phi^c-\left(2\partial_\Phi V_{int}+\frac{\partial_\Phi Z_A}{2} A^2+\frac{\partial_\Phi Z_B}{2} B^2+\partial_\Phi Z_{AB}A B\right)\Phi^a=0,
\end{equation}
\begin{equation}
\begin{split}
 \nabla_\mu\nabla^\mu\chi-\partial_\chi \mathcal{F}(\partial_\mu\theta-q_A A_\mu-q_B B_\mu)^2-\left(\frac{\partial_\chi Z_A}{4}A^2+\frac{\partial_\chi Z_B}{4}B^2+\frac{\partial_\chi Z_{AB}}{2}A B\right) \\-\partial_\chi V_{int}=0,
 \end{split}
\end{equation}
\begin{equation}
\begin{split}
\nabla_\nu(Z_A A^{\nu\mu}+Z_{AB}B^{\nu\mu})+2\mathcal{F}q_A(\nabla^\mu\theta-q_A A^\mu-q_B B^\mu)=4\partial_\alpha\vartheta_1 \epsilon^{\mu\nu\lambda\sigma}A_{\lambda\sigma}\nabla_\nu\alpha\\+2\partial_\alpha\vartheta_2 \epsilon^{\mu\nu\lambda\sigma}B_{\lambda\sigma}\nabla_\nu\alpha,
\end{split}
\end{equation}
\begin{equation}
\nabla_\nu(Z_B B^{\nu\mu}+Z_{AB} A^{\nu\mu})+2\mathcal{F}q_B(\nabla^\mu\theta-q_A A^\mu-q_B B^\mu)=2\partial_\alpha\vartheta_2 \epsilon^{\mu\nu\lambda\sigma}A_{\lambda\sigma}\nabla_\nu\alpha,
\end{equation}
where we have used $A^2=A_{\mu\nu}A^{\mu\nu}$, $B^2=B_{\mu\nu}B^{\mu\nu}$ and $A B=A_{\mu\nu}B^{\mu\nu}$ for short, and $\partial_\Phi$, $\partial_\chi$ and $\partial_\alpha$ denote the derivative with respect to $\Phi^a\Phi^a$, $\chi$ and $\alpha$, respectively. The equations of motion for the metric $g_{\mu\nu}$ read
\begin{equation}
\mathcal{R}_{\mu\nu} -\frac{1}{2}\mathcal{R}g_{\mu\nu}-\frac{3}{L^2} g_{\mu\nu} = \mathcal{T}_{\mu\nu},
\end{equation}
where the energy-momentum tensor is given by
\begin{eqnarray}
\mathcal{T}_{\mu\nu} =&& \frac{Z_G}{2}G^a_{\mu\rho}{G^a_\nu}^\rho+\frac{1}{2}D_\mu\Phi^a D_\nu\Phi^a+\frac{Z_A}{2} A_{\mu\rho}{A_\nu}^\rho+\frac{Z_{AB}}{2}(A_{\mu\rho}{B_\nu}^\rho+A_{\nu\rho}{B_\mu}^\rho)\nonumber\\&&+\frac{Z_B}{2} B_{\mu\rho}{B_\nu}^\rho+\mathcal{F}(\nabla_\mu\theta-q_A A_\mu-q_B B_\mu)(\nabla_\nu\theta-q_A A_\nu-q_B B_\nu)\\ \nonumber&&
+\frac{1}{2}\partial_\mu\chi\partial_\nu\chi+\frac{1}{2}\partial_\mu\alpha\partial_\nu\alpha+\frac{1}{2}\mathcal{L}_{m}g_{\mu\nu}.
\end{eqnarray}
Note that the Chern-Simons terms in $\mathcal{L}_{cs}$ do not contribute to the energy-momentum tensor.

Since the dual system lives in a spatial plane, we choose Poincar\'{e} coordinates with $r$ the radial direction in the bulk. Near the AdS boundary, say $r\rightarrow\infty$, one obtains the following asymptotic form,
\begin{eqnarray}\label{uvexpansions}
\phi & = & \frac{\phi_s}{r^{\Delta^\phi_-}}+\cdots+\frac{\phi_v}{r^{\Delta^\phi_+}}+\cdots, \quad A_t=\mu+\cdots-\frac{\rho}{r}+\cdots,\nonumber\\
\chi & = &\frac{\chi_s}{r^{\Delta^\chi_-}}+\cdots+\frac{\chi_v}{r^{\Delta^\chi_+}}+\cdots, \quad  B_t=\mu_B+\cdots-\frac{\rho_B}{r}+\cdots,\\
\alpha&=&\frac{\alpha_s}{r^{\Delta^\alpha_-}}+\cdots+\frac{\alpha_v}{r^{\Delta^\alpha_+}}+\cdots,\nonumber
\label{expansion}\end{eqnarray}
with
\be
\Delta^\phi_\pm=\frac{3\pm\sqrt{9+4M^2L^2}}{2},\quad \Delta^\chi_\pm=\frac{3\pm\sqrt{9+4m^2L^2}}{2},\quad
\Delta^\alpha_\pm=\frac{3\pm\sqrt{9+4\widetilde{m}^2L^2}}{2}\;.
\ee
  The dots denote higher order corrections which can be extracted by analysing the equations of motion near $r\rightarrow\infty$ order by order.

According to the standard AdS/CFT dictionary, $\mu$ is regarded as the chemical potential in the dual boundary theory, while $\rho$ is the charge density and similarly for $\rho_B$ and $\mu_B$. $\phi$ is interpreted as corresponding to the staggered magnetisation, in which the source $\phi_s$ is considered as the staggered magnetic field while $\phi_v$ gives the response of the antiferromagnetic order parameter in the presence of source $\phi_s$. $\chi$ is dual to the superconducting order parameter. The leading term $\chi_s$ has the interpretation of the source, while $\chi_v$ is the expectation value of the dual operator. $\alpha_s$ is the source for the dual pseudo-scalar operator. Since we expect all the broken phases to arise spontaneously, we will turn off all source terms, i.e., $\phi_s=\chi_s=\alpha_s=0$.\footnote{In this paper we only consider the standard quantisation, i.e., the leading term is regarded as source while the subleading term as response.} We shall work in the canonical ensemble with the charge densities $\rho$ and $\rho_B$ fixed.

There are a number of parameters in the effective holographic  theory. We classify  all parameters into two types. One type is called ``theory parameters" which correspond to the parameters appearing in the bulk gravitational action~\eqref{action1}, including $Z_G$, $q_A$, $q_B$ as well as all the parameters that appear in~\eqref{vpossi}.

The other class are ``state parameters", such as temperature, charge density, and VEVs (condensates). They appear in the asymptotic expansions~\eqref{uvexpansions} as well as in the expansion of the metric. Such parameters are directly related to the solution or state of an action such as~\eqref{action1},  specified by theory parameters.

Since we want to give a complete phase diagram of different phases in a unified theory, we should fix all theory parameters to specify the action and the control parameters of our phase diagram should belong to the second type, i.e., state parameters.

We start from the  homogeneous and isotropic case at finite temperature and charge density. The bulk metric as well as matter part takes the generic form,
\begin{eqnarray}\label{fullansatz}
\begin{split}
ds^2=g_{\mu\nu} dx^\mu dx^\nu=E(r)dr^2-D(r)dt^2+C(r) (dx^2+dy^2),\\
g_\mu^a=0,\quad\quad \Phi=(0, 0, \phi(r)),\quad \alpha=\alpha(r),\\
A=A_t(r)dt,\quad B=B_t(r)dt,\quad \chi=\chi(r),\quad \theta=\theta(r).
\end{split}
\end{eqnarray}
Substituting the ansatz into the equations of motion, one obtains the concrete equations of motion for each field, that are presented in appendix~\ref{app:eoms}. Note that the equations of motion demand $\theta(r)$ to be a constant. We set it to zero without loss of generality.~\footnote{We show in appendix~\ref{app:afspin} that the present holographic description in terms of the spontaneous breaking of SU(2) symmetry can display the associated spin wave with linear dispersion, which is reminiscent of the N\'eel phase in a spin system.}


\section{The $T=0$ Normal Phase and Its Instabilities}
\label{sec:normal}
The development of new branches of black brane solutions at finite temperature is associated with the instability around the symmetric, normal phase black brane solutions.  A simple diagnostic can often be obtained by investigating the zero temperature limit of the symmetric (normal) solutions. In this section we will focus on the instability conditions for the normal phase towards the development of a nontrivial profile of each scalar. Our starting point is the extremal limit of the normal solution, with the semilocal geometry  $AdS_2\times \mathbb{R}^2$ in the far IR. The instability would appear if the IR dimension of the dual operator for each scalar can violate the $AdS_2$ Breitenlohner-Freedman (BF) bound. Then, by continuity, the finite temperature normal background would become unstable at a sufficiently low temperature.

\subsection{The normal phase and its extremal limit}
\label{sub:ads2a}

The normal phase corresponds to the solutions in which $\phi$, $\chi$ and $\alpha$ are vanishing. By considering the ansatz~\eqref{fullansatz} as well as the condition~\eqref{vpossi},
the normal phase is described by the AdS Reissner-Nordstri\"{o}m (AdS-RN) black brane,
\begin{equation}\label{RNads}
\begin{split}
ds^2=&\frac{1}{g(r)}dr^2- g(r)dt^2+r^2(dx^2+dy^2),\\
&g(r)=\frac{r^2}{L^2}(1-\frac{r_h^3}{r^3})+\frac{\rho^2+\rho_B^2}{4r^2}(1-\frac{r}{r_h}),\\
A_t=&\rho(\frac{1}{r_h}-\frac{1}{r})=\mu-\frac{\rho}{r},\\
B_t=&\rho_B(\frac{1}{r_h}-\frac{1}{r})=\mu_B-\frac{\rho_B}{r},
\end{split}
\end{equation}
where the boundary of the spacetime is at $r \rightarrow\infty$ and the planar outer horizon is at $r = r_h$. The black brane temperature reads
\begin{equation}
T=\frac{r_h}{4\pi}\left[\frac{3}{L^2}-\frac{\rho^2+\rho_B^2}{4r_h^4}\right].
\end{equation}

We are interested in the extremal limit $T=0$. In  this limit the horizon radius has a fixed value
\begin{equation}
r_h^4=\frac{L^2(\rho^2+\rho_B^2)}{12},
\end{equation}
and the near horizon limit of the blackening factor behaves
\begin{equation}
g(r)=\frac{1}{2}g''(r_h)\tilde{r}^2+\cdots, \quad \tilde{r}=r-r_h,\quad g''(r_h)=\frac{12}{L^2}.
\end{equation}
The near horizon metric then becomes
\begin{equation}\label{ads2bk}
ds^2=\frac{L_{(2)}^2}{\tilde{r}^2 }d\tilde{r}^2-\frac{\tilde{r}^2}{L_{(2)}^2}dt^2+r_h^2(dx^2+dy^2),
\end{equation}
which is  the $AdS_2\times \mathbb{R}^2$ metric with the $AdS_2$ radius
\begin{equation}\label{ads2radiu}
L_{(2)}=L/\sqrt{6}.
\end{equation}
The two U(1) gauge fields in this coordinate system can be rewritten as
\begin{equation}\label{ads2charge}
A_t=\frac{2\sqrt{3}}{L}\frac{\rho}{\sqrt{\rho^2+\rho_B^2}}\tilde{r},\quad
B_t=\frac{2\sqrt{3}}{L}\frac{\rho_B}{\sqrt{\rho^2+\rho_B^2}}\tilde{r}.
\end{equation}
The IR background~\eqref{ads2bk} and~\eqref{ads2charge} is our starting point to study instabilities towards broken phases.

\subsection{Instabilities in the $AdS_2$ picture}
\label{sub:ads2}

According to the AdS/CFT correspondence, the dual field theory in the IR contains a one dimensional CFT which depends on a single dimensionless  parameter $\rho_B/\rho$. We expect potential phase transitions towards a superconducting phase with nonzero $\chi$ and an antiferromagnetic phase with nonzero $\phi$ as we change $\rho_B/\rho=\textbf{x}$. Two kinds of phase transitions should be triggered by particular instabilities in the normal state. In order to obtain these instabilities we consider fluctuations of the charged scalar $\chi$  and the neutral scalar $\phi$ above the extremal background~\eqref{ads2bk} and~\eqref{ads2charge}. We also give the instability conditions for the pseudo-scalar $\alpha$ in the homogeneous case. In this paper we restrict ourselves to cases with $\textbf{x}\geqslant 0$.

\subsubsection{The neutral scalar $\phi$}

We first consider the neutral scalar which is used to describe the antiferromagnetism in the dual field theory. The linearised equation of motion for its fluctuation, $\delta\phi$, is given by
\begin{equation}
\delta\phi''+\frac{2}{\tilde{r}}\delta\phi'-\frac{M_{(2)}^2L_{(2)}^2}{\tilde{r}^2}\delta\phi=0,
\end{equation}
with
\begin{eqnarray}\label{ads2M}
\nonumber
M_{(2)}^2L_{(2)}^2& = & M^2L_{(2)}^2-\frac{a_1\rho^2+b_1\rho_B^2+2c_1\rho\rho_B}{\rho^2+\rho_B^2},  \\
& = & \frac{1}{6}\left[M^2L^2-\frac{6(b_1 \textbf{x}^2+2c_1\textbf{x}+a_1)}{(1+\textbf{x}^2)}\right],
\end{eqnarray}
where the prime denotes a derivative with respect to $\tilde{r}$. We have used $L_{(2)}=L/ \sqrt{6}$ as given in~\eqref{ads2radiu} and $\textbf{x}=\rho_B/\rho$. The theory parameters $(M^2, a_1, b_1, c_1)$ are defined in~\eqref{vpossi} before. This equation is the equation of motion for a scalar in $AdS_2$ with effective mass squared $M_{(2)}^2L_{(2)}^2$. It is clear that the IR dimension of the dual operator can be tuned by changing $\textbf{x}$. The fluctuation will become unstable if $M_{(2)}^2L_{(2)}^2$ is lower than the $AdS_2$ BF bound, i.e.,  $M_{(2)}^2L_{(2)}^2<-1/4$.

\begin{figure}[ht!]
\begin{center}
\includegraphics[width=.48\textwidth]{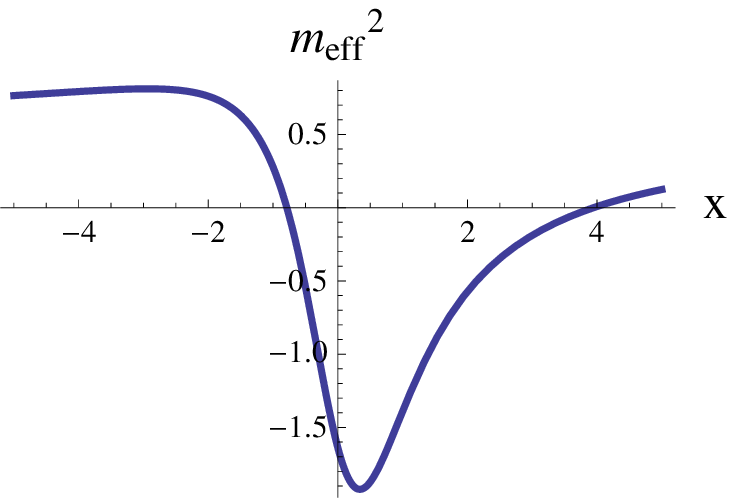}\quad\includegraphics[width=.48\textwidth]{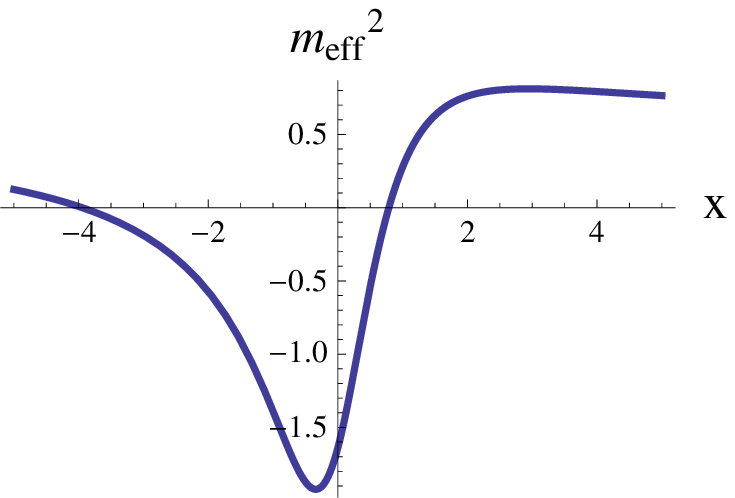}
\caption{The curve $m^2_{eff}(\textbf{x})$ for two distinct cases. (a) Left  $c_1>0$. (b) Right  $c_1<0$. For both cases $\Delta m_{eff}^2$ is negative. When $\Delta m_{eff}^2$ has other values the shape is qualitatively similar.}
\label{figm}
\end{center}
\end{figure}

The quantity that determines in which case we have a nontrivial condensation is
\begin{equation}
m^2_{eff}(\textbf{x})=M_{(2)}^2L_{(2)}^2+{1\over 4}=\frac{1}{6}\left[ M^2L^2+{3\over 2}-\frac{6(b_1 \textbf{x}^2+2c_1\textbf{x}+a_1)}{(1+\textbf{x}^2)}\right].
\end{equation}
The two important parameters that determine the shape of the curve $m^2_{eff}(\textbf{x})$ are
\begin{equation}
\Delta m_{eff}^2\equiv m^2_{eff}(\textbf{0})-m^2_{eff}(\pm\infty),
\end{equation}
with
\begin{equation}
m^2_{eff}(\textbf{0})=\frac{1}{6}\left[ M^2L^2+{3\over 2}-6b_1\right],\quad
m^2_{eff}(\pm\infty)=\frac{1}{6}\left[ M^2L^2+{3\over 2}-6a_1\right],
\end{equation}
as well as $c_1$.
In particular the sign of $c_1$ determines the shape of the curve as seen in figure~\ref{figm}.

The unbroken and the broken phases are determined by the part of the curve above or below the $m^2_{eff}=0$ axis.

We want to  determine the $(T,\textbf{x})$ phase diagram and in particular there is a corner region of the antiferromagnetism as $\textbf{x}$ is small. Therefore, we will assume that the AF phase appears at zero $\textbf{x}$.
We will also assume that at $\textbf{x}\to\infty$ there is no antiferromagnetism.

According to our discussion, in the effective holographic theory these two conditions translate to
demanding that $M_{(2)}^2L_{(2)}^2$ should below the $AdS_2$ BF bound $-1/4$ at $\textbf{x}=0$   and above for $\textbf{x}\to \infty$.
This requirement leads to
\begin{equation}\label{afcondition1}
 M^2L^2-6 a_1<-3/2,\quad M^2L^2-6 b_1>-3/2.
\end{equation}
Then there is always a unique critical point $\textbf{x}_0>0$ fixed by the condition $m_{eff}^2=0$.
There is also a unique critical point for negative $\bar {\bf x}_0<0$.
This implies that if we reverse the sign of the charge of the doping charge density $\rho_B\to -\rho_B$ there is still a quantum critical point for the AF transition.

The role of the parameter $c_1$ is subtler. When
\begin{equation}\label{afcondition}
c_1=0, \quad \text{or} \quad c_1<0,
\end{equation}
the mass squared~\eqref{ads2M} and therefore the critical temperature $T_N$ monotonically decreases until the critical point. On the other hand, if
\begin{equation}\label{ads2latter}
c_1>0,
\end{equation}
the critical temperature $T_N$ for the AF phase transition increases first and then decreases until it vanishes at the quantum critical point.
Interestingly, in hole-doped strange metals the shape of the N\'eel curve is similar to figure~\ref{figm}\,(a), while in electron-doped systems is still similar to the same figure in the sense that the condensed region is sensibly larger than the one in the hole-doped case.

\subsubsection{The charged scalar $\chi$}
We then consider the charged scalar $\chi$ dual to the superconducting condensate. The equation of motion for this fluctuation $\delta\chi$ is given by
\begin{equation}
\delta\chi''+\frac{2}{\tilde{r}}\delta\chi'-\frac{m_{(2)}^2L_{(2)}^2}{\tilde{r}^2}\delta\chi=0,
\end{equation}
where
\begin{eqnarray}\label{ads2m}
\nonumber
m_{(2)}^2L_{(2)}^2& = & m^2L_{(2)}^2-\frac{2(q_A \rho+q_B\rho_B)^2}{\rho^2+\rho_B^2}L_{(2)}^2-\frac{a_2\rho^2+b_2\rho_B^2+2c_2 \rho\rho_B}{\rho^2+\rho_B^2},  \\
& = & \frac{L^2}{6}\left[m^2-\frac{2(Q_A+Q_B \textbf{x})^2+6 a_2+12c_2 \textbf{x}+6 b_2 \textbf{x}^2}{(1+\textbf{x}^2)L^2}\right],\\ \nonumber
&=&\frac{1}{6}\left[m^2L^2-\frac{(6b_2+2Q_B^2)\textbf{x}^2+(12 c_2+4 Q_A Q_B)\textbf{x}+(6a_2+2 Q_A^2)}{(1+\textbf{x}^2)}\right],
\end{eqnarray}
with the prime denoting a derivative with respect to $\tilde{r}$.
We have also defined the dimensionless charges
\begin{equation}
Q_{A,B}=q_{A,B}\,L.
\end{equation}
The theory parameters have been given in the expansion~\eqref{vpossi}. This equation is the equation of motion for a scalar in $AdS_2$ with effective mass squared $m_{(2)}^2L_{(2)}^2$. It is clear that the IR dimension of the dual operator depends on the value of $\textbf{x}$. The fluctuation is stable until $m_{(2)}^2L^2$ is below $-3/2$, i.e., the $AdS_2$ BF bound.
For this we define again
\begin{equation}
m^2_{eff}(\textbf{x})=M_{(2)}^2L_{(2)}^2+{1\over 4}.
\end{equation}
When $m^2_{eff}<0$ we expect condensation.

We will now investigate  the $(T, \textbf{x})$ phase diagram. Although the general case is described in a later section, we will examine here the case of most interest. This involves the assumption that in the $\textbf{x}=0$ and $\textbf{x}\to\infty$ limits there is no superconducting instability.

Absence of a superconducting instability at $\textbf{x}=0$ implies the inequality
\begin{equation}
m^2L^2-2(Q_A^2+3a_2)>-3/2.
\end{equation}
Absence of a superconducting instability at $\textbf{x}\to\infty$ implies
\begin{equation}
 \quad m^2L^2-2(Q_B^2+3b_2)>-3/2.
 \end{equation}
 \begin{figure}[ht!]
\begin{center}
\includegraphics[width=.48\textwidth]{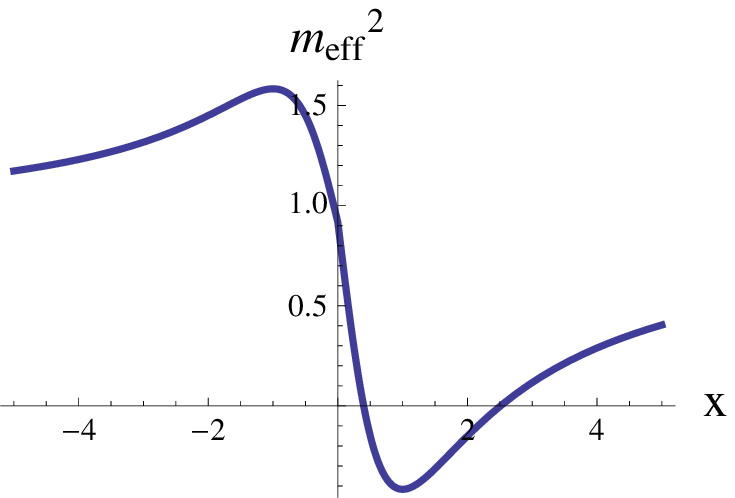}\quad\includegraphics[width=.48\textwidth]{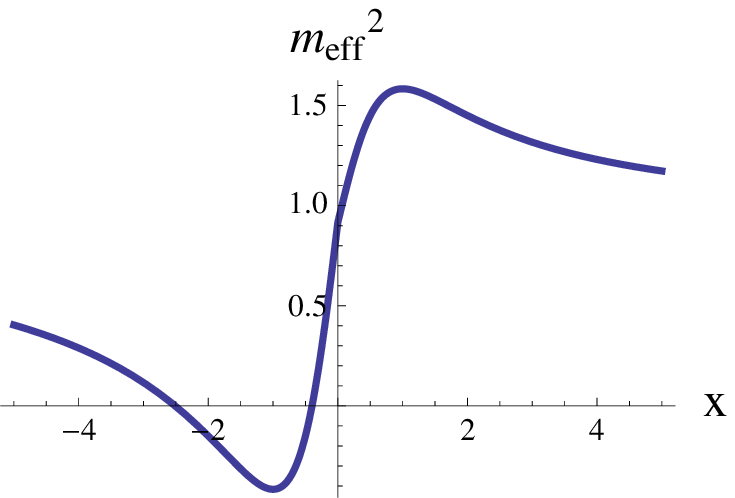}
\caption{The curve $m^2_{eff}(\textbf{x})$ for two distinct cases. (a) Left  $c_2>0$. (b) Right  $c_2<0$.  For both cases $\Delta m_{eff}^2=0$. When $\Delta m_{eff}^2$ has other values the shape is qualitatively similar. The right-hand diagram shows condensation for charges, while the left one shows condensation for holes.}
\label{figs}
\end{center}
\end{figure}
When the sign of  $Q_A Q_B+3c_2>0$, there might be a superconducting phase in an intermediate range
\begin{equation}
\textbf{x}_a<\textbf{x}<\textbf{x}_b.
\end{equation}
That follows if the $AdS_2$ BF bound is violated as $\textbf{x}_a<\textbf{x}<\textbf{x}_b$.
This will happen if the value at the minimum  violates the BF bound
\begin{equation}\label{sccondition}
 Q_A Q_B+3c_2>0,\quad \quad\quad m_{(2)}^2L_{(2)}^2(\textbf{x}=\textbf{x}_{min})<-1/4,
\end{equation}
where
\begin{equation}
\textbf{x}_{min}=\frac{Q_B^2-Q_A^2+3b_2-3a_2+\sqrt{(Q_B^2-Q_A^2+3b_2-3a_2)^2+4(Q_AQ_B+3c_2)^2}}{2Q_AQ_B+6c_2},
\end{equation}
is the minimum of the function $m_{(2)}^2L_{(2)}^2(\textbf{x})$. Two critical points $\textbf{x}_a$ and $\textbf{x}_b$ can be uniquely determined by the equation $m_{(2)}^2L_{(2)}^2(\textbf{x})=-1/4$.

According to our discussion, since the condensate is expected to form from the same charges as the ones in the charge density, we obtain
\begin{equation}
Q_AQ_B\rho_A\rho_B>0.
\end{equation}
Therefore, if the absolute value $|Q_AQ_B|$ is substantially larger then $c_2$, then a superconducting dome is expected. This is shown in figure \ref{figs}.

\subsubsection{The pseudo-scalar $\alpha$}

We now consider the pseudo-scalar, which is very similar with the neutral scalar case. The equation of motion for the fluctuation $\delta\alpha$ is given by
\begin{equation}\label{scpertubation}
\delta\alpha''+\frac{2}{\tilde{r}}\delta\alpha'-\frac{\widetilde{m}_{(2)}^2L_{(2)}^2}{\tilde{r}^2}\delta\alpha=0,
\end{equation}
with
\begin{eqnarray}\label{ads2axi}
\nonumber
\widetilde{m}_{(2)}^2L_{(2)}^2& = &\widetilde{m}^2L_{(2)}^2 -\frac{a_3\rho^2+b_3\rho_B^2+2c_3\rho\rho_B}{\rho^2+\rho_B^2},  \\
& = & \frac{1}{6}\left[\widetilde{m}^2L^2-\frac{6(b_3 \textbf{x}^2+2c_3\textbf{x}+a_3)}{(1+\textbf{x}^2)}\right].
\end{eqnarray}
Note that $L_{(2)}^2=L^2/6$ and $(\widetilde{m}^2, a_3, b_3, c_3)$ are given in~\eqref{vpossi}. This equation is just the equation of motion for a scalar in $AdS_2$ with effective mass squared $\widetilde{m}_{(2)}^2$. It is clear that the IR dimension of the dual operator is determined by $\textbf{x}$. The fluctuation will become unstable if $\widetilde{m}_{(2)}^2$ is lower than the $AdS_2$ BF bound, i.e., $\widetilde{m}_{(2)}^2L^2<-3/2$.  For the region $\textbf{x}\geqslant 0$, the function $\widetilde{m}_{(2)}^2(\textbf{x})$ can be monotonic or non-monotonic and  violate the BF bound or not, depending on the three theory parameters $(a_3, b_3, c_3)$.

One should keep in mind that our analysis on $\alpha$ is valid only for the homogeneous case where Chern-Simons couplings have no effect on the instability. Nevertheless, the presence of Chern-Simons terms in $\mathcal{L}_{cs}$ will lead to a new kind of instability towards developing spatially modulated solutions which we  discuss in section~\ref{sec:spatial}. We will show later that the parameter space of the instability region for $\alpha$ can be significantly enlarged  in general.

In principle, we can construct various kinds of $(T, \textbf{x})$ phase diagrams by tuning theory parameters in~\eqref{vpossi}. Those parameters, for example, determine the critical values of $\textbf{x}$, such as $\textbf{x}_0, \textbf{x}_a, \textbf{x}_b$, at which a broken phase may occur. In subsection~\ref{alldiagrams} we will classify different kinds of instability regions for each scalar by analysing the effective mass squared on the $AdS_2$ background and propose possible phase diagrams in the $(T,\textbf{x})$ plane. Two of them will be explicitly checked by numerical calculation in the next two sections. We believe that other types of phase diagrams can also be constructed by choosing appropriate couplings. Nevertheless, the exact shape of a phase diagram should depend on the  details of the theory itself.

In the rest of the paper, without loss of generality, we shall work  with $L=1$ for simplicity by fixing in such a way the length unit.


\section{The Superconducting Phase}
\label{sec:scphase}

In this section, we will illustrate how to realise the superconducting dome in the $(T, \textbf{x})$ plane in our theory. More precisely, we expect that the superconducting phase would appear at a nonzero $\textbf{x}=\textbf{x}_a$ and then vanish above a certain larger value $\textbf{x}=\textbf{x}_b$. We will focus on the superconducting order $\chi$ and work out the behaviour of the critical temperature $T_{c}$ by changing $\textbf{x}$.

\subsection{Linear analysis}
We have derived the instability conditions in the normal state at zero temperature in the previous section. We continue to study the case at a finite temperature. The superconducting transition corresponds to the process in the bulk as follows. As the black brane temperature is below a critical value $T=T_{c}$, the normal solution becomes unstable to developing nontrivial scalar hair of $\chi$. This is, in most cases, the point of a superconducting phase transition. The critical temperature $T_{c}$ can be determined by numerically solving the equations of motion in appendix~\ref{app:eoms} with $\phi(r)=\alpha(r)=0$. This approach is involved since one has to solve five coupled nonlinear ordinary differential equations (ODEs).
Nevertheless, one can, in some sense, complete this task by just turning the problem around. The phase transition is associated with the formation of scalar hair around the normal solution. In the vicinity of the temperature where the scalar hair begins to develop, the value of $\chi$ should be very small. Therefore it can be treated as a perturbation on the AdS-RN background,
\begin{equation}\label{scpertub}
\delta \chi''+\left(\frac{2}{r}+\frac{g'}{g}\right)\delta \chi'-\frac{1}{g}\left[m^2-\frac{b_2\textbf{x}^2+2 c_2 \textbf{x}+a_2}{2r^4}-\frac{(q_A+q_B\textbf{x})^2}{g}\left(\frac{1}{r_h}-\frac{1}{r}\right)^2\right]\delta \chi=0,
\end{equation}
where we have used the expansion in~\eqref{vpossi} and set $\rho=1$. Note that our perturbative analysis does not depend on the exact form of each coupling but only on the quadratic coefficients appearing in the leading expansion~\eqref{vpossi}.

An instability of the normal solutions towards formation of the scalar hair is associated with the existence of a static zero mode. The zero mode is a kind of marginally-unstable mode corresponding to a bulk solution that is regular at the horizon and source free at the boundary, i.e., a regular normalisable solution of~\eqref{scpertub}. After fixing seven theory parameters, i.e., $(m^2, q_A, q_B, a_2, b_2, c_2)$, we are left with two state parameters, namely, doping $\textbf{x}$ and temperature $T$ which is equivalent to $r_h$. Therefore, for a given $\textbf{x}$, one expects a normalisable solution to exist, if at all, for a certain critical value of temperature.\footnote{In practice, one will obtain a set of discrete values of temperature. It is the highest temperature that corresponds to the zero mode we are looking for.} This is precisely the static zero mode at the onset of the superconducting instability.

\begin{figure}[ht!]
\begin{center}
\includegraphics[width=.6\textwidth]{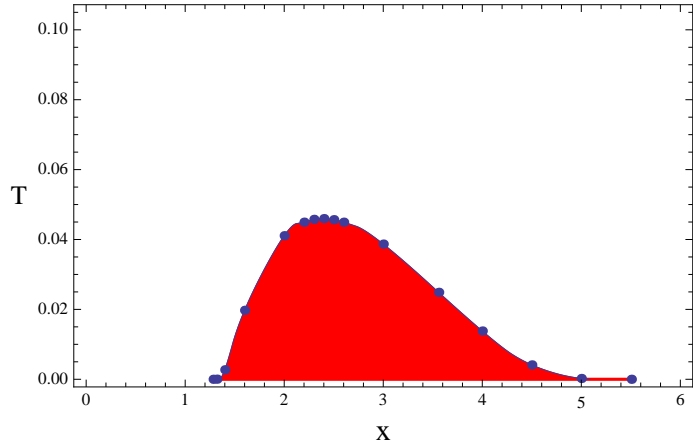}
\caption{ The critical temperature $T$ at the onset of the superconducting instability as a function of $\textbf{x}$. The red region in the $(T, \textbf{x})$ plane corresponds to the superconducting phase. Blue dots are directly from our numerical calculation. Two endpoints of the superconducting region are at $\textbf{x}_a\approx 1.28$ and $\textbf{x}_b\approx 5.51$. We have fixed the theory parameters as $m^2=-4/5, Q_A=1, Q_B=0, a_2=-10, b_2=-4/3, c_2=14/3$ and worked in the unites with $\rho=L=1$.}
\label{fig:scphase}
\end{center}
\end{figure}

According to the  analysis at zero temperature, to realise the superconducting dome one can choose theory parameters satisfying the condition~\eqref{sccondition}. A typical phase diagram is presented in figure~\ref{fig:scphase}, from which one can see a nice superconducting dome (red region). As a concrete example, to produce figure~\ref{fig:scphase} we have chosen a specific set of theory parameters. In particular, $q_B$ is chosen to be zero, which means that $\chi$ in only charged under the gauge field $A_\mu$. One may worry that this case with $q_B=0$ would be too special. Nevertheless, we also checked many other cases in which the effective mass squared~\eqref{ads2m} is the same as the one adopted in figure~\ref{fig:scphase} but with non-vanishing $q_A$ and $q_B$. The resulted critical temperature at each $\textbf{x}$ depends on the theory parameters we are taking, but the non-monotonic behaviour is similar to figure~\ref{fig:scphase}.

One should keep in mind that the linear approach only determines the value of $T(\textbf{x})$ at which the system becomes unstable but not the real phase boundary in general. For a continuous phase transition the two values coincide, while for a first order phase transition the transition always occurs before the instability is reached. Furthermore, to obtain a physical superconducting dome, one must make sure that the free energy of the broken phase should be lower than the unbroken phase. Nevertheless, the above calculation via marginally-unstable modes is helpful to obtain a general outline of the phase diagram. Our next task is to determine the order of the phase transition as well as the thermodynamically preferred phase by comparing free energies of the normal phase and broken phase.

\subsection{Superconducting transition in the canonical ensemble  \label{scandfree}}
Using perturbative analysis near the critical point, we have caught a glimpse of the existence of a superconducting dome. We will confirm the existence of the dome by solving the coupled equations for the fully back-reacted geometry. Although the leading expansion of each coupling in~\eqref{vpossi} is enough to determine the zero mode at the critical temperature, as we will show, the nonlinear details of the couplings are important away from the critical temperature.

This subsection aims at providing more numerical details we will adopt. We shall work in the canonical ensemble with two U(1) charges $\rho$ and $\rho_B$ (or equivalently $\textbf{x}=\rho_B/\rho$) fixed.

After fixing the gauge degrees of freedom, one can adopt a much simpler form of metric as well as matter fields,
\begin{equation}\label{scansatz}
\begin{split}
ds^2=-g(r) e^{-\xi(r)}d t^2+\frac{1}{g(r)} d r^2+r^2(dx^2+dy^2),\\
\chi=\chi(r),\quad A_t=A_t(r), \quad B_t=B_t(r),\quad \phi(r)=\alpha(r)=\theta(r)=0.
\end{split}
\end{equation}
The equations of motion can be obtained from general equations in appendix~\ref{app:eoms} by letting $E(r)= g(r) e^{-\xi(r)}, D(r)= 1/g(r), C(r)=r^2$ and $\phi(r)=\alpha(r)=\theta(r)=0$.
More precisely, $g$ and $\xi$ satisfy first order differential equations and other functions satisfy second order equations.

To solve the coupled ODEs, we need to specify suitable boundary conditions at the horizon $r_h$ as well as at the conformal boundary $r\rightarrow \infty$. We are looking for finite temperature black brane solutions that are regular at the horizon, which demands the following analytic expansion in terms of $(r-r_h)$, i.e.,
\begin{equation}
\begin{split}
&g=g_h\, (r-r_h)+\cdots, \quad A_t=a_h\, (r-r_h)+\cdots,  B_t=b_h\, (r-r_h)+\cdots,\\
&\xi=\xi^0_h+\xi^1_h\, (r-r_h)+\cdots, \quad \chi=\chi^0_h+\chi^1_h\, (r-r_h)+\cdots.
\end{split}
\end{equation}
By substituting into the equations of motion, one finds that there are only five independent coefficients $(r_h, a_h, b_h, \chi^0_h, \xi^0_h)$. Near the UV as $r\rightarrow \infty$,
we demand asymptotically AdS geometry with the falloff~\footnote{In general, $\xi$ approaches to a constant $\xi_s$ as $r\rightarrow \infty$. However, in order for the Hawking temperature of the black brane to be the temperature of the boundary field theory, the constant $\xi_s$ should vanish.}
\begin{equation}\label{scuvexpan}
\begin{split}
&g=r^2\left(1+\frac{g_v}{r^3}+\cdots\right), \quad \xi=\frac{\xi_v}{r^3}+\cdots, \quad \chi=\frac{\chi_v}{r^{\Delta^\phi_+}}+\cdots,\\
&A_t=\mu-\frac{\rho}{r}+\cdots,  B_t=\mu_B-\frac{\rho_B}{r}+\cdots,
\end{split}
\end{equation}
which is specified by seven parameters $(g_v, \xi_v, \chi_v, \mu, \rho, \mu_B, \rho_B)$. One will see later that as a consequence of the conformal invariant $\xi_v=0$. So one is left with six constants. Note that the expansion of $\chi$ is chosen such that the dual scalar operator has no source, therefore can spontaneously acquire an expectation value proportional to $\chi_v$. We have a useful scaling symmetry,
\begin{equation}
r\rightarrow c\, r,\quad (t, x, y)\rightarrow (t, x, y)/c,\quad g\rightarrow c^2 g,\quad (A_t, B_t)\rightarrow c\, (A_t, B_t),
\end{equation}
where $c$ is a constant. This scaling symmetry can be used to set $\rho=1$. To solve the coupled equations, we need to specify eight integration constants. We now have ten parameters on two boundaries at hand after fixing $\rho$. Thus, we expect to leave with a two parameter family of black brane solutions, which can be chosen as temperature $T$ and doping $\textbf{x}=\rho_B/\rho$.

To determine which phase is thermodynamically favoured, we should compare the free energy. Remind that we work in the canonical ensemble where $\rho$  and $\rho_B$ (and equivalently $\textbf{x}$) are fixed. The holographic dictionary tells us that Helmholtz free energy $F$ of the boundary thermal state is identified with temperature $T$ times the on-shell bulk action in Euclidean signature. Given the fact that the system is stationary, the Euclidean action is related to the Minkowski case by a total minus.
\begin{equation}
\begin{split}
-2\kappa^2_N S_{Euclidean}=\int dx^4\sqrt{-g} (\mathcal{R}+6+\mathcal{L}_m+\mathcal{L}_{cs})+\\
\int_{r\rightarrow\infty} dx^3 \sqrt{-\gamma}(A_\nu n_\mu Z_A A^{\mu\nu}+B_\nu n_\mu Z_B B^{\mu\nu})+\int_{r\rightarrow\infty} dx^3 \sqrt{-\gamma} (2\mathcal{K}-4),
\end{split}
\end{equation}
where $\gamma_{\mu\nu}$ is the induced metric on the boundary $r\rightarrow\infty$, $\mathcal{K}_{\mu\nu}$ is the associated extrinsic curvature and $n^\mu$ is the outward pointing unit normal vector to the boundary. The last term includes the Gibbons-Hawking term for a well-defined Dirichlet variational principle and further a surface counter term for removing divergence.\footnote{In principle, we should also consider the counter term for the scalar $\chi$. However, since we always deal with solutions without source, this term makes no contribution to the free energy.} The middle term is supplemented because we fix the charge density instead of chemical potential.

The temperature of the black brane is given by
\begin{equation}
T=\frac{g'(r_h)e^{-\xi(r_h)/2}}{4\pi},
\end{equation}
and the thermal entropy density is given by the Bekenstein-Hawking entropy in terms of the area of the horizon, i.e.,
\begin{equation}
s=\frac{2\pi}{\kappa^2_N}\frac{\text{Area}}{V}=\frac{2\pi}{\kappa^2_N} r_h^2,
\end{equation}
where we have defined $V=\int dx dy$ which is the spatial volume of the Minkowski space.

Employing the equations of motion, the on-shell action reduces to
\begin{equation}
-2\kappa^2_N S_{Euclidean}=\lim_{r\rightarrow\infty} [2\beta V r\sqrt{g}e^{-\xi/2}(r\mathcal{K}-2r-\sqrt{g})-\beta V r^2 e^{\xi/2}(A_t A_t'+B_t B_t')],
\end{equation}
with $\beta=1/T$. After substituting the UV expansion~\eqref{scuvexpan}, one can obtain the free energy
\begin{equation}
F=T S_{Euclidean}=\frac{V}{2\kappa^2_N}(g_v+\mu\rho+\mu_B\rho_B).
\end{equation}
According the above equation, the free energy for the normal phase~\eqref{RNads} is given by
\begin{equation}
F_{RN}=\frac{V}{2\kappa^2_N}\left[-r_h^3+\frac{3}{4}\frac{\rho^2+\rho_B^2}{r_h}\right].
\end{equation}

The expectation value of the energy-momentum tensor of the dual field theory is given by~\cite{Balasubramanian:1999re}
\begin{equation}
T_{ij}=\lim_{r\rightarrow\infty}\left[r(\mathcal{K}\gamma_{ij}-\mathcal{K}_{ij}-2\gamma_{ij})\right],
\end{equation}
with $i, j={t, x, y}$. By substituting the UV expansion~\eqref{scuvexpan}, we find
\begin{eqnarray}
\mathcal{E}=T_{tt}& = &\frac{1}{2\kappa^2_N} (-2g_v), \\
\mathcal{P}=T_{xx}& = &T_{yy}=\frac{1}{2\kappa^2_N}(-g_v+3\xi_v),
\end{eqnarray}
with vanishing non-diagonal components. Here $\mathcal{E}$ is the energy density and $\mathcal{P}$ is the pressure. Since we are considering a dual conformal field theory at the AdS boundary, the dual energy-momentum tensor is traceless, which in turn demands $\xi_v=0$. We have numerically checked that $\xi_v$ indeed vanishes within numerical errors. Note that the grand canonical potential $\Omega=F-\frac{V}{2\kappa^2_N}(\mu\rho+\mu_B\rho_B)=\frac{V}{2\kappa^2_N} g_v$, which matches the general expectation for a (2+1) CFT that the grand canonical potential is $\Omega=-\mathcal{E} V/2$. Using the conserved quantity~\eqref{appconserved}, we obtain the relation between IR and UV data,
\begin{equation}
-2\kappa_N^2 T s=\mu\rho+\mu_B\rho_B+3g_v.
\end{equation}
Hence, the system satisfies the expected thermodynamical relation
\begin{equation}
\mathcal{E}+\mathcal{P}=T s+\frac{1}{2\kappa^2_N}(\mu\rho+\mu_B\rho_B).
\end{equation}

\subsection{Condensate and free energy}

Since the differential equations depend on the precise form of the couplings, the condensate as well as the free energy should also depend on the details of the theory. To make it more clear, we shall consider two representative couplings,
\begin{eqnarray}
&&(a):  Z_A=1-5\chi^2,\quad Z_B=1-\frac{2}{3}\chi^2, \quad V_{int}=-\frac{5}{8}\chi^2,\label{sccasea}\\
&&(b): Z_A=\text{sech}(\sqrt{10}\chi),\, Z_B=\text{sech}\left(\sqrt{\frac{4}{3}}\chi\right),\, V_{int}=6-6\cosh\left(\sqrt{\frac{5}{24}}\chi\right)\label{sccaseb},
\end{eqnarray}
with $Z_{AB}=7 \chi^2/3$ and $\mathcal{F}=\chi^2/2$ the same.\footnote{Observe that in case (a) we must have $|\chi|<1/\sqrt{5}$. The coupling function in which a field has some special value restriction can also be found in some top-down theories~\cite{Gauntlett:2009bh}. An alternative way is to take the absolute value of $Z_A$ and $Z_B$ in~\eqref{sccasea}, which can result in the non-analyticity of the action. However, for all the examples we shall study, the solutions remain well inside the region of $Z_A>0$ and $Z_B>0$ for all radial positions.} We further choose $Q_A=1$ and $Q_B=0$ as before. One can check that two kinds of couplings give the same small $\chi$ expansion~\eqref{vpossi} with $a_2=-10, b_2=-4/3, c_2=14/3$ and $m^2=-5/4$, thus a linear analysis yields identical equation of motion for $\delta\chi$~\eqref{scpertub}.

We are interested in black brane solutions with regular event horizon. The VEV of the dual operator $\left<\mathcal{O}_\chi\right>=\chi_v$ can be read off from the asymptotical expansion of $\chi$ in the UV,
\begin{equation}
\chi=\frac{\chi_v}{r^{5/2}}+\cdots,
\end{equation}
and the free energy is given by
\begin{equation}
F=\frac{V}{2\kappa^2_N}(g_v+\mu\rho+\mu_B\rho_B),
\end{equation}
where $V$ is a volume of spatial region and $(g_v, \mu,\rho,\mu_B,\rho_B)$ are given in~\eqref{scuvexpan}. In the following, we will provide the plot of the condensate and compare the free energy between the normal phase and superconducting phase for each theory.

\subsubsection{Case $(a)$}
\begin{figure}[ht!]
\begin{center}
\includegraphics[width=.46\textwidth]{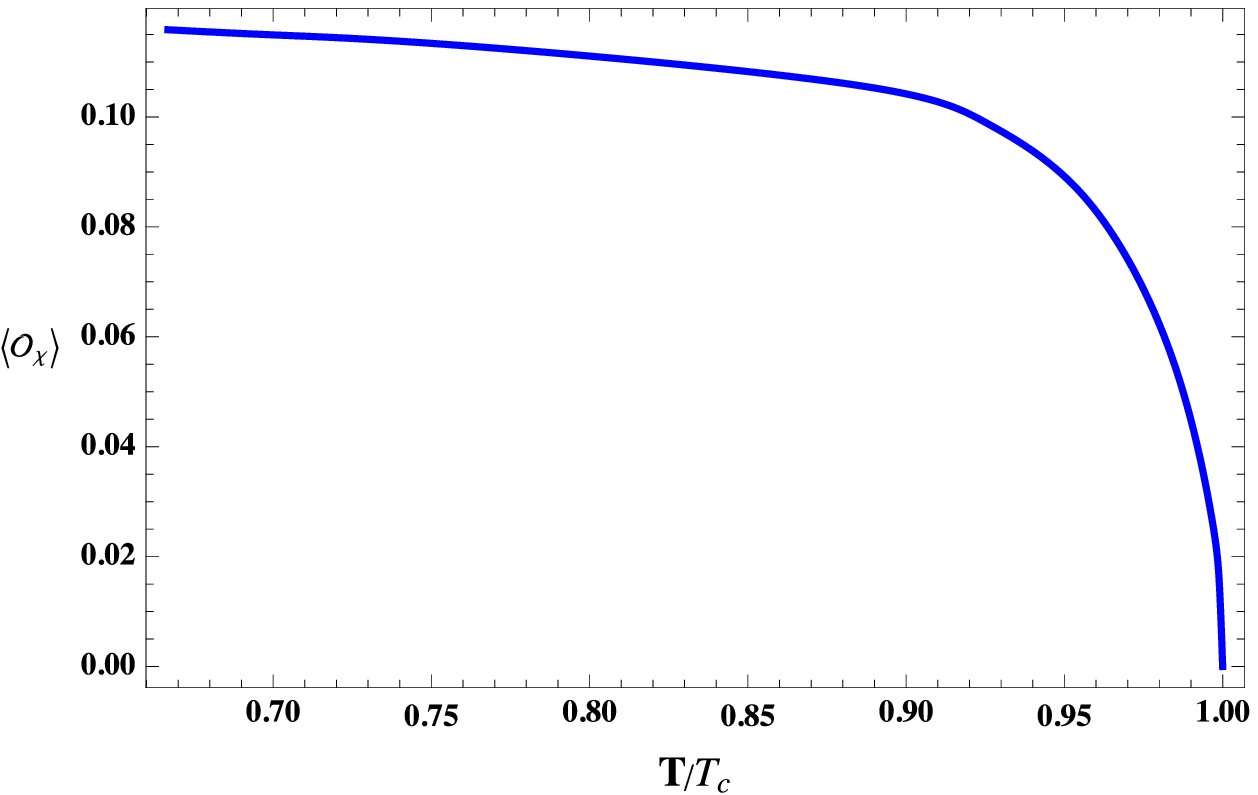}
\quad
\includegraphics[width=.50\textwidth]{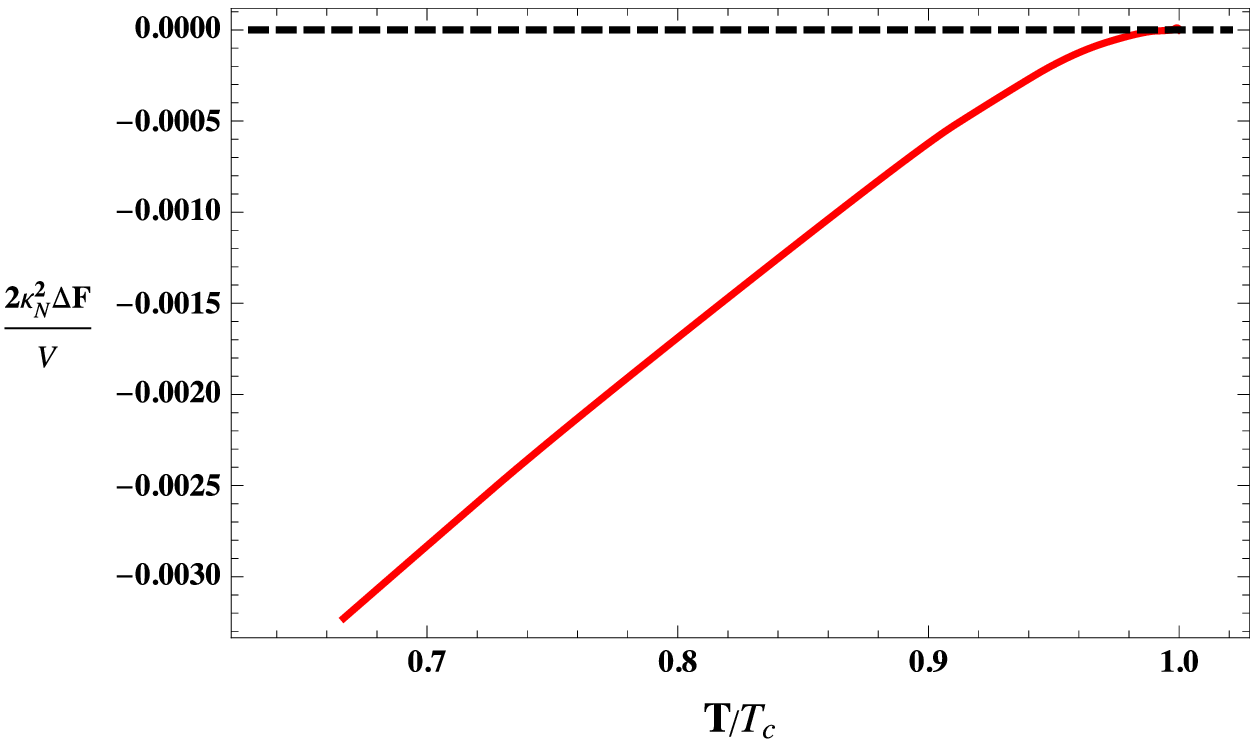}
\caption{The superconducting condensate $\left<\mathcal{O}_\chi\right>$ (blue line in the left) and the free energy difference $\Delta F=F-F_{RN}$ (red line in the right) for a fixed branch $\textbf{x}=2.4$ of theory $(a)$. The condensate appears continuously below a critical temperature $T_c\approx0.04597$. It is a second order phase transition from the RN branch (black dashed line). We have worked in the unites with $\rho=L=1$.}
\label{fig:casesca}
\end{center}
\end{figure}

The condensate as a function of temperature is presented in the left plot of figure~\ref{fig:casesca}. It is clear that as one lowers the temperature, the normal phase with vanishing charged scalar becomes unstable to developing scalar hair breaking the U(1) symmetry spontaneously in the dual field theory. By fitting the data near $T_c$, we find that for small condensate there is a critical behaviour with critical exponent $1/2$, suggesting a continuous phase transition.

From the data presented in the right plot of figure~\ref{fig:casesca}, one can ensure that the broken phase is thermodynamically preferred at the normal phase. This phase transition is a second order one. Therefore, our previous analysis by linear perturbation in the vicinity of critical temperature does work for case $(a)$. As have been expected, we indeed obtain a superconducting dome shown in figure~\ref{fig:scphase}. In particular, our numerical calculation suggests the tendency that the lower the effective mass squared on $AdS_2$, the higher the critical temperature for the broken phase. We stress that this empirical rule can only hold qualitatively.

\subsubsection{Case $(b)$}
We now comment on the second bulk theory $(b)$. For comparison with the result of case $(a)$ we also focus on the branch with $\textbf{x}=2.4$. Unsurprisingly, we obtain the same critical temperature at which the scalar hair begins to appear. However, we find drastically different behaviour from the first case, once we move away from the ``critical" temperature.

\begin{figure}[ht!]
\begin{center}
\includegraphics[width=.46\textwidth]{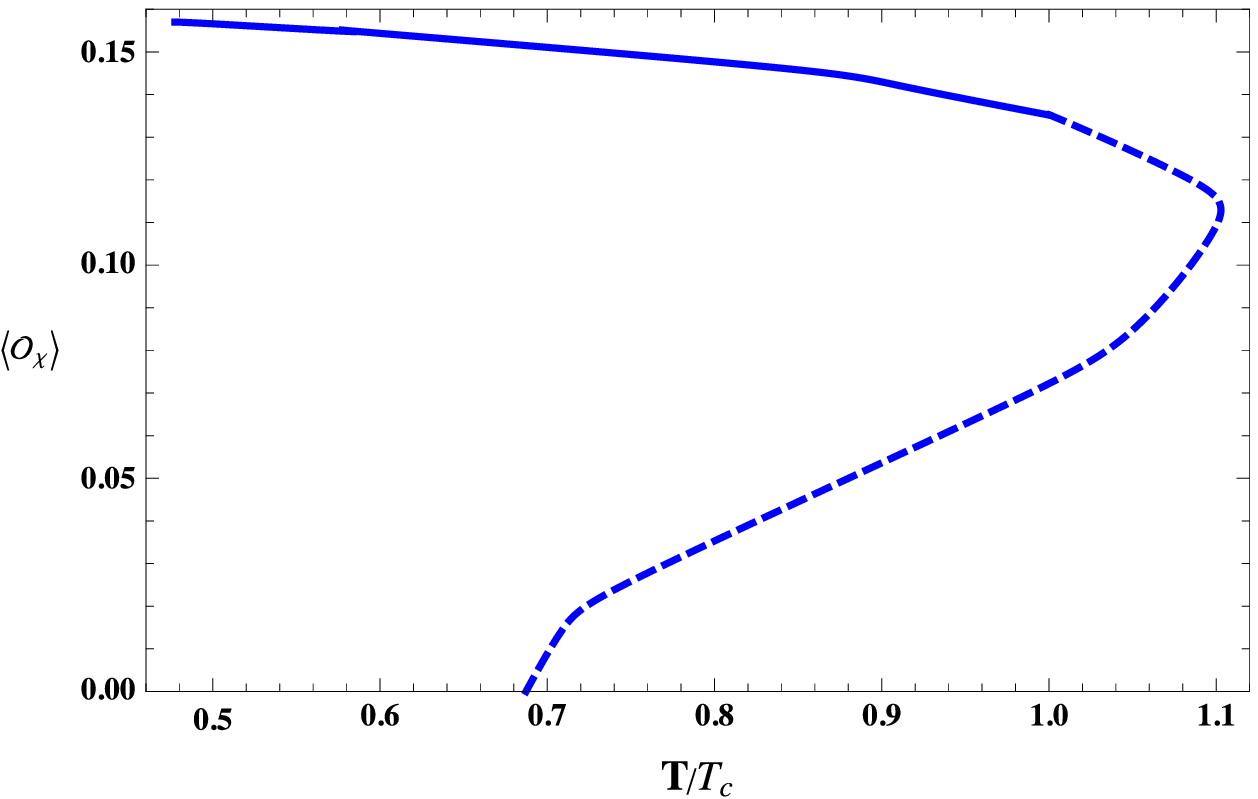}
\quad
\includegraphics[width=.50\textwidth]{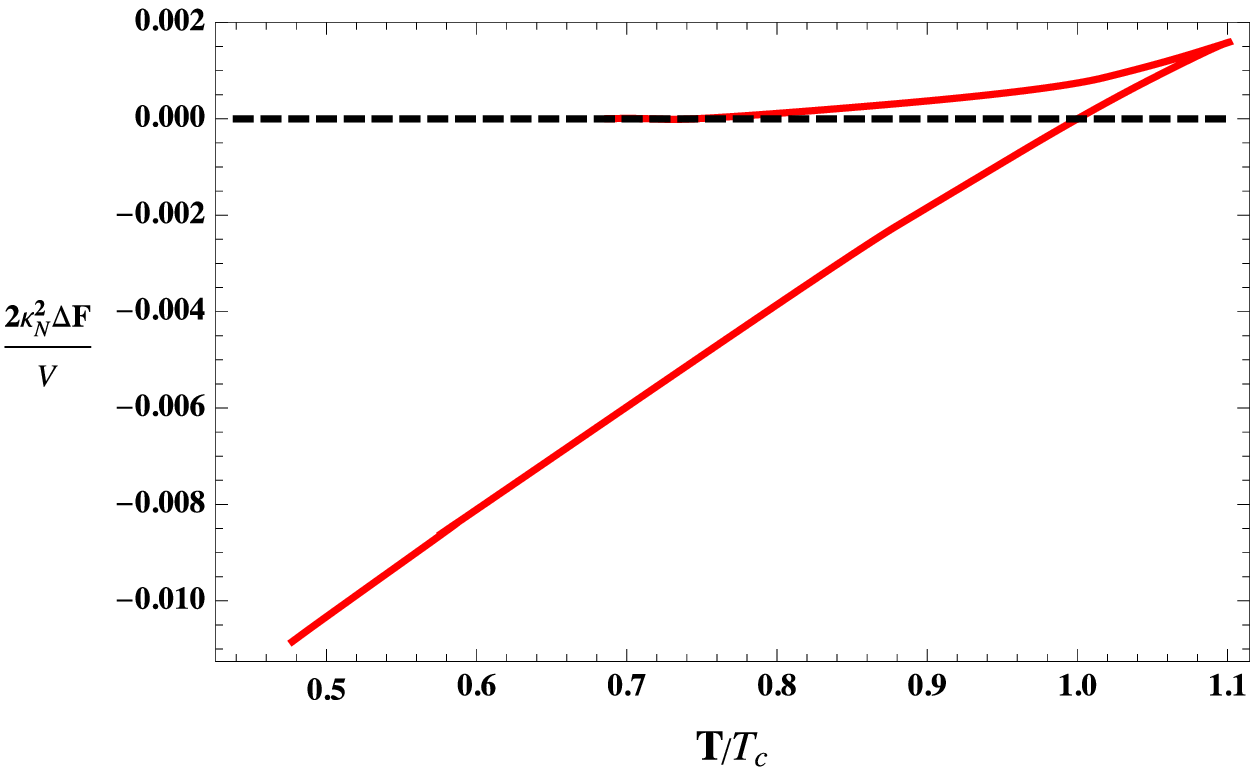}
\caption{The superconducting condensate $\left<\mathcal{O}_\chi\right>$ (blue line in the left) and the free energy difference $\Delta F=F-F_{RN}$ (red line in the right) for a fixed branch $\textbf{x}=2.4$ of theory $(b)$. The critical temperature is  $T_c\approx0.06689$ at which the condensate has a sudden jump from zero to the upper solid blue curve. It is a first order phase transition from the RN branch (black dashed line). We have worked in the unites with $\rho=L=1$.}
\label{fig:casescb}
\end{center}
\end{figure}

The condensate $\left<\mathcal{O}_\chi\right>$ versus temperature is presented in figure~\ref{fig:casescb}. One can find that the condensate becomes multi-valued and there are two new sets
of solutions with non-vanishing $\left<\mathcal{O}_\chi\right>$  developing below a particular temperature, involving an upper-branch with large $\left<\mathcal{O}_\chi\right>$ and a down-branch with small $\left<\mathcal{O}_\chi\right>$. Therefore, there are three states that are available to the system at some temperature, i.e., one is for $\left<\mathcal{O}_\chi\right>=0$ and two for $\left<\mathcal{O}_\chi\right>\neq 0$. To determine which state is the physical one, we draw the free energy in the right plot of figure~\ref{fig:casescb}. There is a characteristic ``swallow tail", suggesting that the normal phase is thermodynamically favoured at higher temperatures while the upper-branch of broken phase dominates as the temperature is lowered down to $T_c$ by a first order phase transition.

In case $(b)$ the critical temperature $T_c$ at a fixed $\textbf{x}$ is larger than the one in case $(a)$, because the real phase transition always occurs before the perturbative instability is reached. So the phase diagram in figure~\ref{fig:scphase} obtained by linear analysis should be modified. The SC phase will occupy a much larger region than case $(a)$ in the $(T,\textbf{x})$ plane. In particular, the boundary of the superconducting region now becomes the line of the first order superconducting phase transition. However, this modification is only quantitative and will not change the key feature, i.e., the existence of a superconducting dome.~\footnote{There is also a phenomenon called ``retrograde condensation" that was reported in some holographic theories~\cite{Charmousis:2010zz,Buchel:2009ge,Aprile:2011uq,Donos:2011ut,Cai:2013aca2014ija}. In that case, a hairy black brane solution exists only for temperatures above a critical value with the free energy much larger than the black brane without hair. However, for the theory  parameters we have checked, we do not observe this kind of behaviour.}

By comparing two cases, we see that the condensate can appear continuously at a particular temperature, but it might be thermodynamically unstable in the vicinity of that temperature. The order of the phase transition depends closely on the nonlinear details of the theory.  The critical temperature from perturbative analysis may be the point of an actually continuous phase transition, as it was in case (a), or it may label a spinodal point in a first order phase transition as in case (b). Nevertheless, our analysis shows that the existence of a superconducting dome is a rather generic feature of our theory provided the theory parameters are within the region~\eqref{sccondition}, independent of the choice of the coupling functions.

\section{The Antiferromagnetic Phase}
\label{sec:afphase}

In nature, at the corner near zero doping the antiferromagnetic phase dominates the $(T, \textbf{x})$ phase diagram. However, as the doping grows, the antiferromagnetic phase will disappear. For simplicity, we consider the case in which the critical temperature of antiferromagnetic phase transition, $T_N$, decreases monotonically as doping increases and then vanishes at a certain value of doping.\footnote{In principle, there is a possibility that the transition temperature $T_N$ first increases and then monotonically decreases to zero as the doping grows. See case 8 in subsection~\ref{alldiagrams}.} To consider such a kind of diagram one is suggested to choose theory parameters within the region~\eqref{afcondition}.

We first suppose that the transition from the normal phase to antiferromagnetic phase is continuous, something we will check explicitly later for the theory parameters we choose.  To determine the boundary of the AF phase in the $(T, \textbf{x})$ plane, which is the line of the antiferromagnetic phase transition, we can start in the normal phase, at larger values of temperature, and determine when it becomes unstable towards formation of the scalar hair of $\phi$.

\begin{figure}[ht!]
\begin{center}
\includegraphics[width=.6\textwidth]{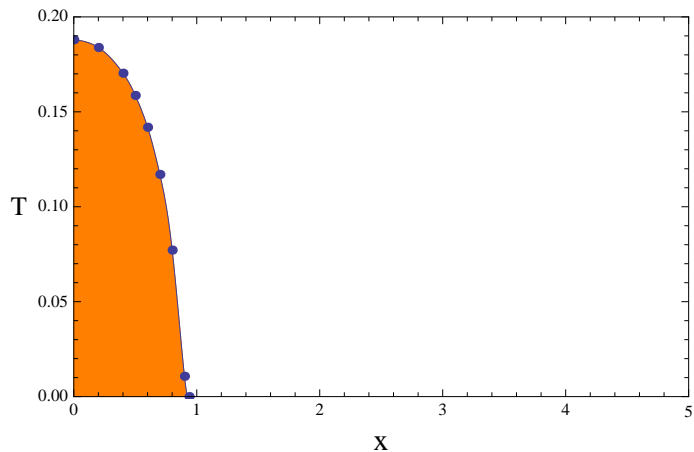}
\caption{ The critical temperature $T_N$ as a function of $\textbf{x}$. In the orange region the system is in the antiferromagnetic phase while in the white region it is in the normal phase. Blue dots are determined by numerically solving~\protect\eqref{afpertub}. We have chosen the theory parameters as $M^2=-2, a_1=6, b_1=-7, c_1=0$ and worked in the unites with $\rho=L=1$. }
\label{fig:afphase}
\end{center}
\end{figure}

The perturbation equation for $\phi$ in the vicinity of the phase transition point reads
\begin{equation}\label{afpertub}
\delta \phi''+\left(\frac{2}{r}+\frac{g'}{g}\right)\delta \phi'-\frac{1}{g}\left(M^2-\frac{b_1\textbf{x}^2+2 c_1 \textbf{x}+a_1}{2r^4}\right)\delta \phi=0,
\end{equation}
where we have used the expansion in~\eqref{vpossi} and set $\rho=1$. Note that the value of $T_{N}$ will be determined once the quadratic coefficients appearing in the expansion~\eqref{vpossi} are given, if the phase transition is continuous. Following the same procedure as before, we look for the static zero mode by solving~\eqref{afpertub} with regular condition in the IR and source free condition in the UV. The critical temperature $T_N$ versus $\textbf{x}$ is plotted in figure~\ref{fig:afphase}. We see that the critical temperature decreases monotonically as $\textbf{x}$ is increased. However, before moving on, we should first check that the phase transition is continuous and the AF phase is thermodynamically favoured at least for the particular couplings we are considering.

\begin{figure}[ht!]
\begin{center}
\includegraphics[width=.46\textwidth]{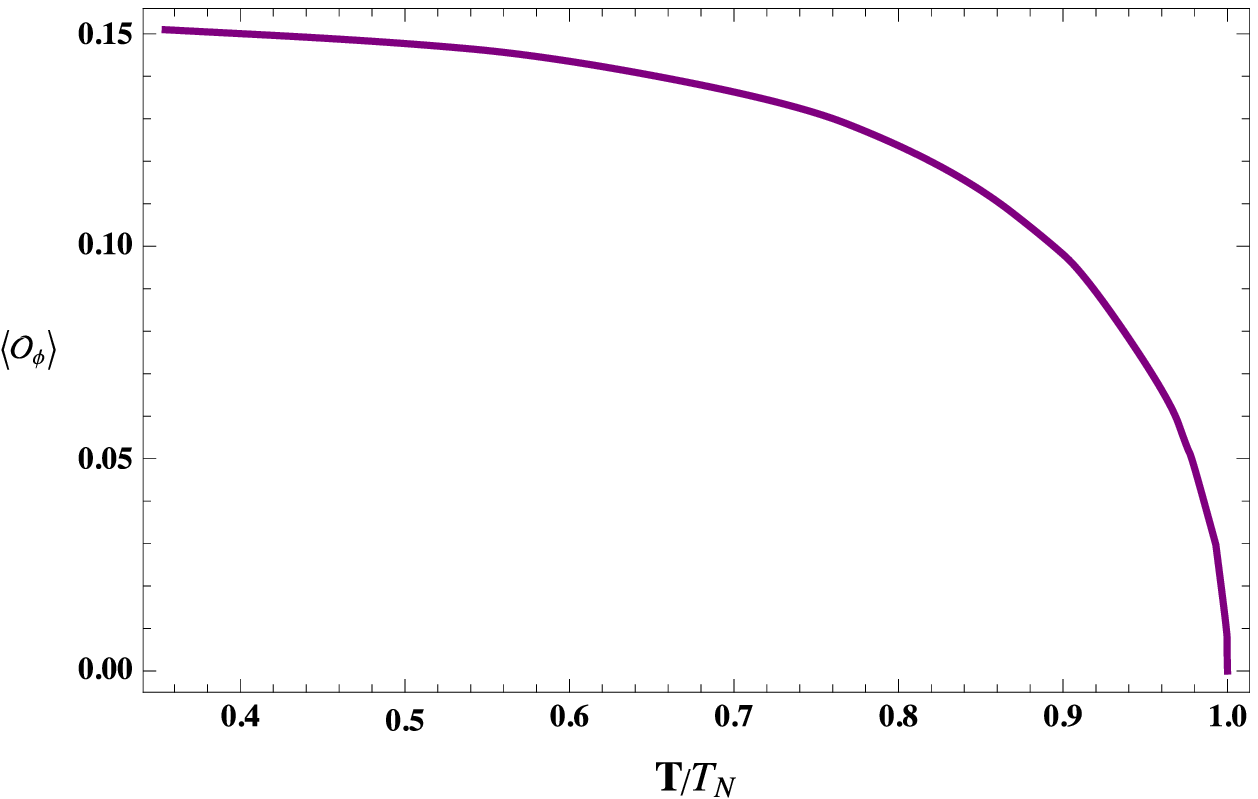}
\quad
\includegraphics[width=.50\textwidth]{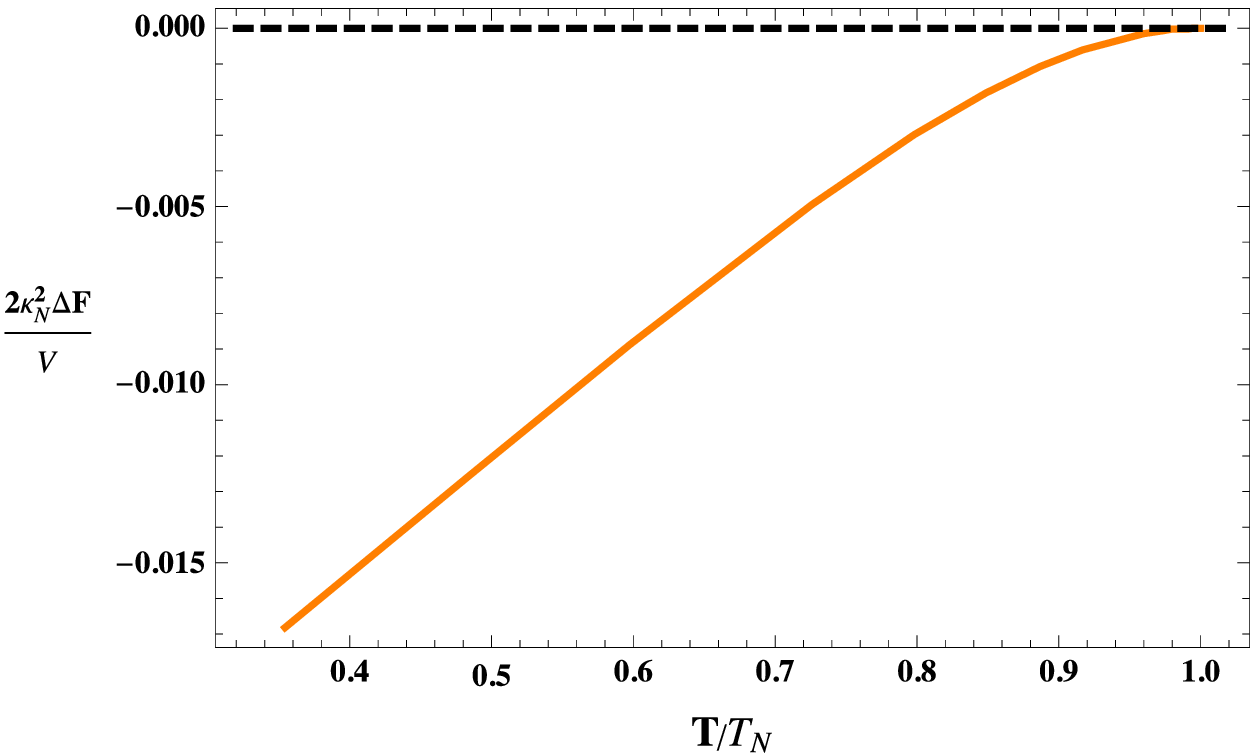}
\caption{The condensate $\left<\mathcal{O}_\phi\right>$ (purple line in the left) and the free energy difference $\Delta F=F-F_{RN}$ (orange line in the right) for a fixed branch $\textbf{x}=0.5$. The condensate begins to arise below a critical temperature $T_N\approx0.15863$ smoothly. It is a second order phase transition from the RN branch (black dashed line). We have worked in the unites with $\rho=L=1$.}
\label{fig:caseaf}
\end{center}
\end{figure}

For concreteness, we consider the theory by fixing the form of all couplings as follows,
\begin{equation}\label{afcoupling}
Z_A=1+\frac{a_1}{2}\phi^2,\quad Z_B=1+\frac{b_1}{2}\phi^2, \quad Z_{AB}=0,\quad V_{int}=\frac{1}{2}M^2\phi^2,
\end{equation}
without higher order corrections.  We consider similar ansatz adopted by superconducting case in subsection~\ref{scandfree}, which reads
\begin{equation}\label{afansatz}
\begin{split}
ds^2=-g(r) e^{-\xi(r)}d t^2+\frac{1}{g(r)} d r^2+r^2(dx^2+dy^2),\\
\phi=\phi(r),\quad A_t=A_t(r), \quad B_t=B_t(r),\quad \chi(r)=\alpha(r)=0.
\end{split}
\end{equation}
The independent equations are much simpler than the superconducting case and we give the precise form here.
\begin{eqnarray}
\phi''+\left(\frac{g'}{g}+\frac{2}{r}-\frac{\xi'}{2}\right)\phi'-\frac{1}{g}\left[M^2-\frac{1}{2}e^\xi (a_1 A_t'^2+b_1 B_t'^2)\right]\phi & = & 0, \\
\left(1+\frac{a_1}{2}\phi^2\right)A_t''+\left[\frac{2}{r}+\frac{\xi'}{2}+a_1\left(\frac{1}{r}\phi+\frac{\xi'}{4}\phi+\phi'\right)\phi\right] A_t'& = & 0,\\
\left(1+\frac{b_1}{2}\phi^2\right)B_t''+\left[\frac{2}{r}+\frac{\xi'}{2}+b_1\left(\frac{1}{r}\phi+\frac{\xi'}{4}\phi+\phi'\right)\phi\right] B_t'& = & 0,\\
\phi'^2+\frac{4g'}{rg}+\frac{e^\xi}{g}\left[\left(1+\frac{a_1}{2}\phi^2\right)A_t'^2+\left(1+\frac{b_1}{2}\phi^2\right)B_t'^2\right]+M^2\frac{\phi^2}{g}-\frac{12}{g}+\frac{4}{r^2}&=&0,\\
\frac{1}{2}\phi'^2+\frac{\xi'}{r}&=&0.
\end{eqnarray}
Adopting the same procedure as shown in subsection~\ref{scandfree}, one can obtain the condensate as well as the free energy by solving above coupled equations with regular condition in the IR and source free condition in the UV. To match figure~\ref{fig:afphase}, the theory parameters in~\eqref{afcoupling} are chosen to be $M^2=-2, a_1=6, b_1=-7$. Our numerical data is presented in figure~\ref{fig:caseaf}. As anticipated, we find that the phase transition from the normal phase to AF phase is second order. The AF phase has a lower free energy than the normal phase, and thus is thermodynamically stable. Therefore, the phase diagram shown in figure~\ref{fig:afphase} can be actually realised.~\footnote{Since $b_1$ we are taking is negative, one may worry about that the coefficient $Z_B$  can be negative for sufficiently large value of $\phi$ of our numerical solutions. In our calculation, we checked that this situation does not happen.}

We can not exclude the possibility that for other particular form of couplings which is different from~\eqref{afcoupling} the antiferromagnetic phase transition would become no longer continuous. In that case, the value of $T_N$ can not be determined by linear analysis from~\eqref{afpertub}. One has to find the real $T_N$ by comparing free energy between the AF phase and the normal phase.
However, the sharp of the region for the broken phase should not change qualitatively. In practice, we have a great deal of freedom to choose an appropriate theory.

We should stress out that the form of coupling functions, $Z_A$ and $Z_B$, adopted in figure~\ref{fig:scphase} and figure~\ref{fig:afphase} might not be a good choice from theoretical point of view, as it does not ensure positivity of the kinetic terms for an arbitrary value of fields. One should not take those coupling functions so seriously and they are only chosen for illustration. However, as discussed above, one can modify the coupling functions in order to keep the positivity of $Z_A$ and $Z_B$ in~\eqref{Smatter} without changing the phase diagram qualitatively. Indeed, we have explicitly checked that the theory  is almost the same as the one considered in figure~\ref{fig:afphase}. The only difference is that we add a higher order term into the coupling $Z_B$. The new function is given by

\begin{equation}\label{afcoupling2}
Z_B=1+\frac{b_1}{2}\phi^2+\tau \phi^4,\quad \tau>\frac{b_1^2}{16},
\end{equation}
thus is positive definite. As a typical example, considering $b_1=-7$ and $\tau=4$, we have numerically solved the resulted equations and find a second order phase transition from the normal phase to the AF phase. Therefore we can obtain the same critical temperature as shown in figure~\ref{fig:afphase} with both coupling functions $Z_A$ and $Z_B$ positive definite. A similar discussion can be, in principle, applied for case (a)~\eqref{sccasea} in the previous section by choosing some configuration of $Z_A$ and $Z_B$ which is positive definite and can give the same expansion coefficients $a_1$ and $b_1$ as $\chi\rightarrow 0$. Actually, case (b)~\eqref{sccaseb} is in fact a particular kind of modification of case (a). It is clear that $Z_A$ and $Z_B$ in~\eqref{sccaseb} are both positive definite. A dramatic change after such modification is that the superconducting phase transition becomes first order. But the superconducting dome is still preserved after modification.


\section{The Striped Phase}
\label{sec:spatial}
In the previous two sections, we have constructed desirable phase diagrams in the $(T,\textbf{x})$ plane for superconducting phase and antiferromagnetic phase. Looking at the phase diagram in figure~\ref{fig:cuprates},  between the two phases there is a so called ``pseudogap" region on which people have no good understanding so far.  One representative order in the pesudogap region is known as striped order~\cite{review,zaanen1}. Such a spatially modulated order is the consequence of spontaneously breaking of space translational symmetry.  As a replacement of the real ``pesudogap" region, our next goal is to investigate the presence of a spatially modulated phase which breaks translational invariance spontaneously. Then we will try to find a proper parameter space such that the striped phase is located between SC and AF phases.

Our attention mainly focuses on the striped instabilities corresponding to black brane solutions that are modulated in only one spatial direction. It is referred to as striped black branes, dual to the striped phases on the field theory side. This kind of solutions have been numerically constructed in~\cite{Rozali:2012es,Donos:2013wia,Withers:2013kva,Rozali:2013ama,Ling:2014saa,Withers:2013loa} by solving a nonlinear system of partial differential equations (PDEs). In most cases that have been studied, there is a second order phase transition to the striped solution. One exception is the theory in~\cite{Withers:2013loa} where the author reported the existence of a branch of striped solutions but no continuous phase transition. However, one can also obtain a second order transition by modifying the form of couplings in that theory. Another possibility corresponds to the spontaneous breaking of translational invariance in both boundary spatial directions, referred to as checkerboard phases with periodic modulation in two spatial directions. Checkerboard solutions have been numerically constructed in~\cite{Withers:2014sja} where it was shown that the checkerboard phase has lower free energy than the normal phase, while it may or may not be thermodynamically preferred from the striped phase, depending on the theory one considered.

Instead of dealing with PDEs, to identify the spatially modulated phase we use a perturbative analysis around the normal solution, namely, the AdS-RN solution~\eqref{RNads}. More precisely, we need to look for the spatially modulated static mode in the spectrum of fluctuations around the normal background. The presence of these modes would mark the instability of the homogeneous phase and the onset of the spatially modulated phases.

The strategy, at first, is to consider the zero temperature case in which there is an emergent electrically charged $AdS_2\times \mathbb{R}^2$ solution in the far IR. We try to identify possible spatially modulated modes that violate the $AdS_2$ BF bound. If such BF-violating mode exists, analogous modes would appear as well as in AdS-RN black brane away from zero temperature.
Then we consider finite temperature background~\eqref{RNads} to fix the critical temperature $T^*$ at which the spatially modulated phase begins to appear. We will give the behaviour of critical temperature $T^*$ with respect to $\textbf{x}$ for striped phases and briefly discuss the instabilities towards checkerboard structures.

\subsection{Striped instabilities of $AdS_2\times \mathbb{R}^2$ geometry}\label{stripads2}

We want to investigate striped instabilities of the electrically charged AdS-RN black brane~\eqref{RNads}. The starting point is to consider linearised perturbations in the $AdS_2\times \mathbb{R}^2$ background arising as the IR limit of the AdS-RN geometry at zero temperature,
\begin{equation}\label{ads2bk2}
\begin{split}
ds^2=\frac{1}{6 r^2 }dr^2-6 r^2 dt^2+r_h^2(dx^2+dy^2),\quad r_h^2=\frac{\sqrt{1+\textbf{x}^2}}{2\sqrt{3}}\rho,\\
A_t=\frac{2\sqrt{3}\rho}{\sqrt{\rho^2+\rho_B^2}}r=\frac{2\sqrt{3}}{\sqrt{1+\textbf{x}^2}}r,\quad
B_t=\frac{2\sqrt{3}\rho_B}{\sqrt{\rho^2+\rho_B^2}}r=\frac{2\sqrt{3}\textbf{x}}{\sqrt{1+\textbf{x}^2}}r,
\end{split}
\end{equation}
where we have set the AdS radius $L=1$ and omitted the tilde of $r$ for convenience. We try to turn on the following perturbations,~\footnote{One can easily find that the perturbation of $\theta$ does not appear at linear order.}
\begin{equation}\label{pertubzero}
\begin{split}
&\delta g_{ty}= \lambda\, r_h\, h_{ty}(r)\sin(kx),\quad \delta A_{y}= \lambda\, r_h\, a_{y}(r)\sin(kx),\\
&\delta B_{y}= \lambda\, r_h\, b_{y}(r)\sin(kx),\quad \delta \chi=  \lambda\, \upsilon(r)\cos(kx),\\
&\delta \phi= \lambda\, \varphi(r)\cos(kx),\quad \delta \alpha= \lambda\, w(r)\cos(kx),
\end{split}
\end{equation}
in which $\lambda$ is a formal expansion parameter that will be eventually set to unity. What we are really expanding in is the smallness of all perturbations to the background solution~\eqref{ads2bk2}.

By substituting into the equations of motion and working at linear level, we find that the above fluctuations satisfy
\begin{eqnarray}
6\partial_r(r^2 \partial_r \varphi)-\left(M_{(2)}^2+\frac{k^2}{r_h^2}\right)\varphi& = & 0,\label{stripads2af}\\
6\partial_r(r^2 \partial_r \upsilon)-\left(m_{(2)}^2+\frac{k^2}{r_h^2}\right)\upsilon& = & 0,\label{stripads2sc} \\
6\partial_r(r^2 \partial_r w)-\left(\widetilde{m}_{(2)}^2+\frac{k^2}{r_h^2}\right)w-\frac{4\sqrt{3}(2n_1+n_2\textbf{x})}{\sqrt{1+\textbf{x}^2}}\frac{k}{r_h}a_y-\frac{4\sqrt{3}n_2}{\sqrt{1+\textbf{x}^2}}\frac{k}{r_h}b_y& = & 0, \\
6\partial_r(r^2 \partial_r a_y)+\frac{2\sqrt{3}}{\sqrt{1+\textbf{x}^2}}h_{ty}'-\frac{4\sqrt{3}(2n_1+n_2\textbf{x})}{\sqrt{1+\textbf{x}^2}}\frac{k}{r_h}w-\frac{k^2}{r_h^2}a_y&=&0,\\
6\partial_r(r^2 \partial_r b_y)+\frac{2\sqrt{3}\textbf{x}}{\sqrt{1+\textbf{x}^2}}h_{ty}'-\frac{4\sqrt{3}n_2}{\sqrt{1+\textbf{x}^2}}\frac{k}{r_h}w-\frac{k^2}{r_h^2}b_y&=&0,\\
-3r^2 h_{ty}''-\frac{6\sqrt{3}r^2}{\sqrt{1+\textbf{x}^2}}a_y'-\frac{6\sqrt{3}\textbf{x}r^2}{\sqrt{1+\textbf{x}^2}}b_y'+\frac{1}{2}\frac{k^2}{r_h^2}h_{ty}&=&0,\label{stripads2hty}
\end{eqnarray}
where $n_1$ and $n_2$ denote two Chern-Simons couplings given in~\eqref{vpossi}, and $M_{(2)}^2$, $m_{(2)}^2$ and $\widetilde{m}_{(2)}^2$  have been defined in equations~\eqref{ads2M},~\eqref{ads2m} and~\eqref{ads2axi}, respectively.  It is clear that the equations of motion for $\varphi$ and $\upsilon$ decouple from other modes. Furthermore, after turning on momentum dependence there is an additional positive contribution, $k^2/r_h^2$, to the effective mass squared of $\varphi$  and $\upsilon$, see~\eqref{stripads2af} and~\eqref{stripads2sc}, and thus will stabilise the normal background. The striped instabilities can only happen for other mixed modes including $w, a_y, b_y, h_{ty}$.

To move forward, we further introduce the field redefinition
\begin{equation}
\eta=-\frac{1}{2}h_{ty}'-\frac{\sqrt{3}}{\sqrt{1+\textbf{x}^2}}a_y-\frac{\sqrt{3}\textbf{x}}{\sqrt{1+\textbf{x}^2}}b_y,
\end{equation}
and define the four vector $\mathbb{V}^T=(\eta, a_y, w, b_y)$. By differentiating equation~\eqref{stripads2hty}, we can rewrite the last four equations of motion in a compact form, which reads
\begin{equation}\label{stripeads2}
6\partial_r(r^2\partial_r\mathbb{V})-\mathbb{M}^2\mathbb{V}=0,
\end{equation}
with the mass matrix
\begin{equation}
\mathbb{M}^2=\left[\begin{array}{cccc}\frac{k^2}{r_h^2} & \frac{\sqrt{3}}{\sqrt{1+\textbf{x}^2}}\frac{k^2}{r_h^2} & 0 & \frac{\sqrt{3}\textbf{x}}{\sqrt{1+\textbf{x}^2}} \frac{k^2}{r_h^2}\\ \frac{4\sqrt{3}}{\sqrt{1+\textbf{x}^2}} & \frac{k^2}{r_h^2}+\frac{12}{1+\textbf{x}^2} & \frac{4\sqrt{3}(2n_1+n_2\textbf{x})}{\sqrt{1+\textbf{x}^2}}\frac{k}{r_h}& \frac{12\textbf{x}}{1+\textbf{x}^2} \\
0 & \frac{4\sqrt{3}(2n_1+n_2\textbf{x})}{\sqrt{1+\textbf{x}^2}}\frac{k}{r_h} & \widetilde{m}_{(2)}^2(\textbf{x})+\frac{k^2}{r_h^2} & \frac{4\sqrt{3}n_2}{\sqrt{1+\textbf{x}^2}}\frac{k}{r_h} \\
\frac{4\sqrt{3}\textbf{x}}{\sqrt{1+\textbf{x}^2}} & \frac{12\textbf{x}}{1+\textbf{x}^2}& \frac{4\sqrt{3}n_2}{\sqrt{1+\textbf{x}^2}}\frac{k}{r_h} & \frac{k^2}{r_h^2}+\frac{12\textbf{x}^2}{1+\textbf{x}^2} \end{array}\right].
\end{equation}
Here $\widetilde{m}_{(2)}^2$ is defined in~\eqref{ads2axi}, and two Chern-Simons coupling constants $n_1, n_2$ are given in~\eqref{vpossi}.
Note that
\begin{equation}
\frac{k^2}{r_h^2}=\frac{k^2/ \rho}{r_h^2/\rho}=\frac{2\sqrt{3}(k/ \sqrt{\rho})^2}{\sqrt{1+\textbf{x}^2}},
\end{equation}
depends on $\textbf{x}$ and $k/\sqrt{\rho}$ is the dimensionless wave number.

This is the standard system of four mixed modes propagating at the $AdS_2$ background~\eqref{ads2bk2}. Thus a diagnostic for an instability is whether the mass matrix $\mathbb{M}^2$ has an eigenvalue that violates the $AdS_2$ BF bound.  More precisely, for fixed theory parameters as well as $\textbf{x}$ we look for a range of $k$ where the smallest eigenvalue, say $\mathbb{M}^2_{min}[k]$, among four eigenvalues of $4\times 4$ matrix $\mathbb{M}^2$ violates the BF bound, which means
\begin{equation}
\mathbb{M}^2_{min}[k]<-3/2,
\end{equation}
in our units with $L=1$. Such BF-violating modes are associated with striped phases with periodic modulation in the $x$ direction. Among those BF-violating modes there is a particular value $k_0$ which gives the smallest eigenvalue $\mathbb{M}^2_{min}[k_0]$ among all wave numbers. It is worth noting that the off-diagonal terms originating from $\mathcal{L}_{cs}$~\eqref{Smatter} in $\mathbb{M}^2$ provide a mechanism which drives down the smallest eigenvalue at a nonzero $k$. Indeed, if one sets $n_1=n_2=0$, one will find that the smallest eigenvalue can be only at $k=0$, and therefore there will be no striped instabilities.

The value of $\mathbb{M}^2_{min}[k]$ has to be determined numerically. However, before doing a numerical calculation, one can find some useful properties of $\mathbb{M}^2_{min}[k]$ from, for example, the characteristic polynomial of $\mathbb{M}^2$.
\begin{itemize}
\item 1)  $\mathbb{M}^2_{min}[-k]=\mathbb{M}^2_{min}[k]$, which means that if there is an unstable mode at $k$, there should be an unstable mode at $-k$. So we can only consider modes with positive $k$.
\item 2) $\mathbb{M}^2_{min}[k]$ is invariant under the transformation $(n_1, n_2)\rightarrow (-n_1,-n_2)$. It can effectively reduce the $(n_1, n_2)$ parameter space by half. As a consequence, when $n_2=0$, $\mathbb{M}^2_{min}[k]$ does not change  under the substitution $n_1\rightarrow -n_1$. Similarly, when $n_1=0$, $\mathbb{M}^2_{min}[k]$ is also independent of the sign of $n_2$.
\item 3) For the particular case $\textbf{x}=0$, $\mathbb{M}^2_{min}[k]$ is invariant under $(n_1, n_2)\rightarrow (n_1,-n_2)$ and $(n_1, n_2)\rightarrow (-n_1, n_2)$.
\end{itemize}
By using the above symmetries, the parameter space of BF-violating modes can be significantly reduced. In order to find the striped phase in the $(T,\textbf{x})$ plane, it is not necessary to know all the details of the parameter space of  BF-violating modes, although it will be interesting to work out the complete parameter space in terms of $(\widetilde{m}^2, a_3, b_3, c_3, n_1, n_2,\textbf{x})$ where the theory parameters are given in the expansion~\eqref{vpossi}.

\begin{figure}[ht!]
\begin{center}
\includegraphics[width=.55\textwidth]{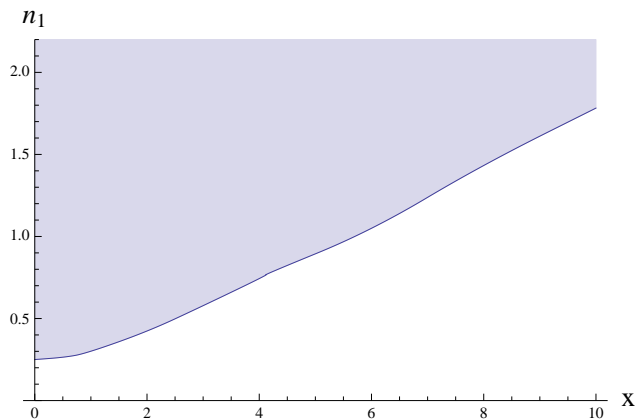}
\caption{ The shaded region in the $(\textbf{x}, n_1)$ plane includes unstable modes with non-vanishing wave number $k$ in the $AdS_2\times \mathbb{R}^2$ background. The phase boundary is given by $\mathbb{M}^2_{min}[k_0]=-3/2$ where $\mathbb{M}^2_{min}[k_0]$ denotes the most smallest eigenvalue of $\mathbb{M}^2$.
We chose $n_2=0$ and $ \widetilde{m}_{(2)}^2=0$, i.e., $\widetilde{m}^2=a_3=b_3=c_3=0$.}
\label{fig:cs1}
\end{center}
\end{figure}
\begin{figure}[ht!]
\begin{center}
\includegraphics[width=.55\textwidth]{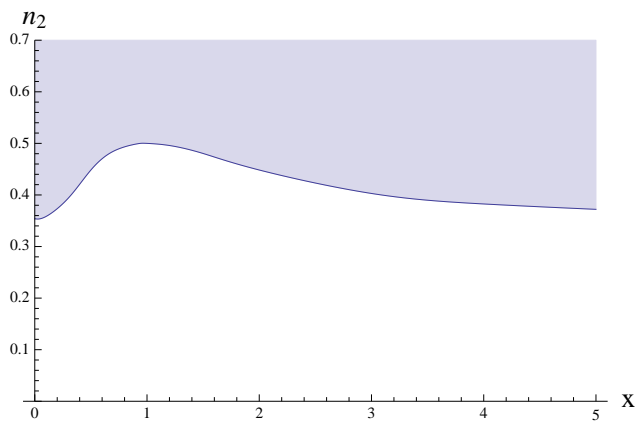}
\caption{ The shaded region in the $(\textbf{x}, n_2)$ plane includes unstable modes with non-vanishing wave number $k$ in the $AdS_2\times \mathbb{R}^2$ background. The phase boundary is determined by requiring $\mathbb{M}^2_{min}[k_0]=-3/2$. We chose $n_1=0$ and $ \widetilde{m}_{(2)}^2=0$, i.e., $\widetilde{m}^2=a_3=b_3=c_3=0$. Due to the exchange symmetry between $A_\mu$ and $B_\mu$ in this case, if there is an instability at $\textbf{x}$, there must have the same instability at $1/\textbf{x}$. }
\label{fig:cs2}
\end{center}
\end{figure}

In the present paper, we focus on a simple situation of the theory~\eqref{Smatter}, where $(Z_A, Z_B, Z_{AB})$ are independent of $\alpha$, $(\vartheta_1, \vartheta_2)$ depend on $\alpha$ linearly and $\alpha$ is massless. This situation with $\widetilde{m}^2=a_3=b_3=c_3=0$ is natural from string theory point of view. We are left with two Chern-Simons coupling constants $n_1$ and $n_2$ defined in~\eqref{vpossi}.

We show the the domain in which there exists a range of wave numbers violating the corresponding $AdS_2$ BF bound in figure~\ref{fig:cs1} for $n_2=0$ and figure~\ref{fig:cs2} for $n_1=0$, respectively. Both shaded regions include only unstable modes with non-vanishing wave number $k$. In the first case, as one increases $\textbf{x}$, a much larger value of $n_1$ is needed to ``excite" the striped instability. In contrast, in the latter case the boundary between unstable modes and stable modes first rises then falls as $\textbf{x}$ is increased. The maximum is precisely at $\textbf{x}=1$ and the critical value of $n_2$ is exactly the same at $\textbf{x}$ and $1/\textbf{x}$. This behaviour is a consequence of the exchange symmetry between two U(1) gauge fields when $\vartheta_1=0$. Actually, one can easily check that under the substitution $a_y\rightarrow b_y$ and $\textbf{x}\rightarrow 1/\textbf{x}$, the form of~\eqref{stripeads2} does not change when $n_1=0$.

\begin{figure}[ht!]
\begin{center}
\includegraphics[width=.58\textwidth]{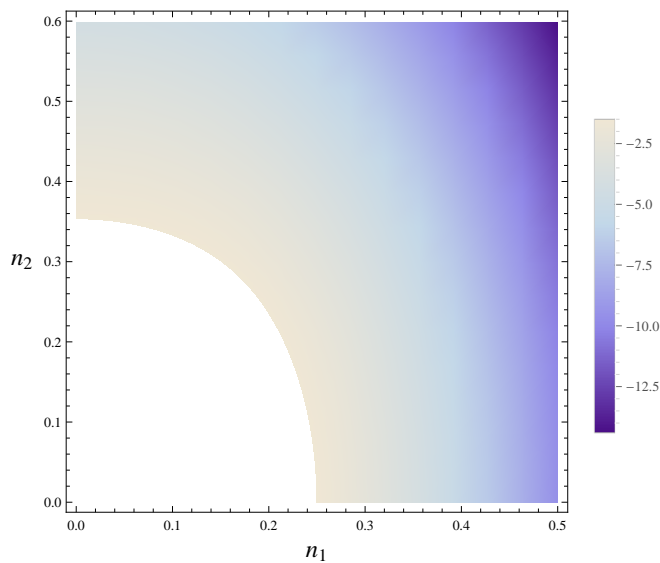}
\caption{ The density plot of the most smallest eigenvalue of $\mathbb{M}^2$, i.e., $\mathbb{M}^2_{min}[k_0]$  in the $(n_1, n_2)$ plane when $\textbf{x}=0$. The modes in the white region do not violate the $AdS_2$ BF bound, while the shaded region includes unstable modes with non-vanishing wave number in the $AdS_2$ background. Note that the BF-violating modes correspond to cases with $\mathbb{M}^2_{min}[k_0]<-3/2$.  We chose $\widetilde{m}^2=a_3=b_3=c_3=0$.}
\label{fig:csstripx0}
\end{center}
\end{figure}

In figure~\ref{fig:csstripx0} we show the density plot of $\mathbb{M}^2_{min}[k_0]$ in the $(n_1, n_2)$ plane when $\textbf{x}=0$. Remember that $k_0$ denotes the wave number at which $\mathbb{M}^2_{min}[k]$ as a function of $k$ takes the minimum value. We only give the case in the first quadrant. However, according to the argument in $\bullet \,3)$ one can easily obtain the results in the other three quadrants. One can see from figure~\ref{fig:csstripx0} that the larger the values of $n_1$ and $n_2$, the lower the effective mass squared of BF-violating modes. Therefore, to increase the critical temperature at the onset of striped instabilities, one is suggested to choose a larger value of $n_1$ or $n_2$.

\begin{figure}[ht!]
\begin{center}
\includegraphics[width=.55\textwidth]{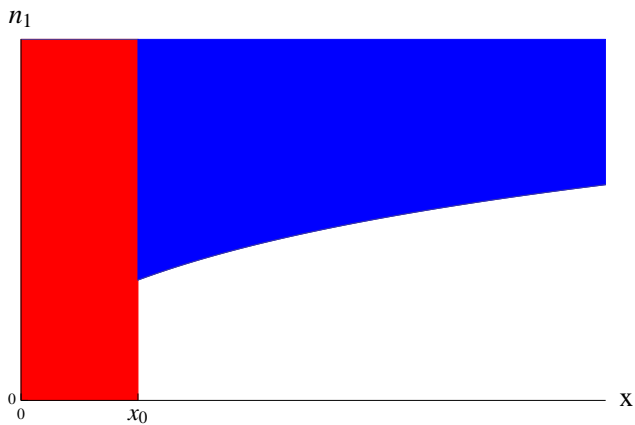}
\caption{The schematic plot for the case in which $\widetilde{m}_{(2)}^2(\textbf{x})<-3/2$ as $\textbf{x}<\textbf{x}_0$ (red region) and $\widetilde{m}_{(2)}^2(\textbf{x})>-3/2$ as $\textbf{x}>\textbf{x}_0$ (blue region). In the former case, $\widetilde{m}_{(2)}^2(\textbf{x})$ violates the $AdS_2$ BF bound. The red region has unstable modes including $k=0$, while the blue region only has unstable modes with $k\neq 0$. }
\label{fig:cs3}
\end{center}
\end{figure}

\begin{figure}[ht!]
\begin{center}
\includegraphics[width=.55\textwidth]{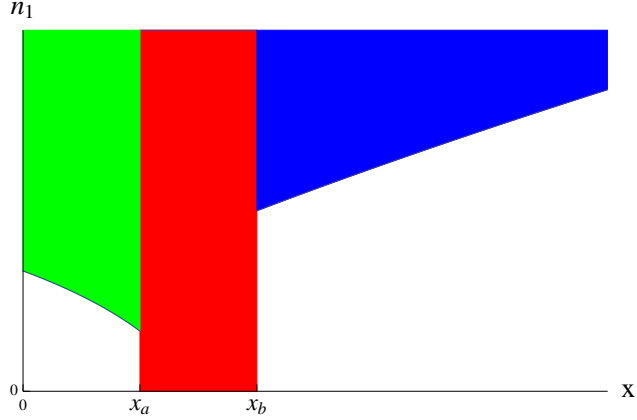}
\caption{ The schematic plot for the case in which $\widetilde{m}_{(2)}^2(\textbf{x})<-3/2$ as $\textbf{x}_a<\textbf{x}<\textbf{x}_b$ (red region) while $\widetilde{m}_{(2)}^2(\textbf{x})>-3/2$ as $\textbf{x}<\textbf{x}_a$ (green region) and $\textbf{x}>\textbf{x}_b$ (blue region). In the red region, $\widetilde{m}_{(2)}^2(\textbf{x})$ violates the $AdS_2$ BF bound, thus this region has unstable modes including $k=0$, while the green and blue regions only have unstable modes with $k\neq 0$. This plot corresponds to a dome of striped phase in the $(T, \textbf{x})$ phase diagram.}
\label{fig:cs4}
\end{center}
\end{figure}

In an effective theory, one can in principle choose  general coupling constants $(\widetilde{m}^2, a_3, b_3, c_3)$ of~\eqref{vpossi}. Correspondingly, the region of unstable modes towards striped solutions can be changed. Two representative plots are schematically shown in figure~\ref{fig:cs3} and figure~\ref{fig:cs4}.  In both cases, there is a red region where the $AdS_2$ BF bound is violated even at $k=0$. Nevertheless, after turning on Chern-Simons couplings $n_1$ and $n_2$ in~\eqref{vpossi}, the $k=0$ mode is in general not the one with the smallest $AdS_2$ mass squared. In the latter case of figure~\ref{fig:cs4}, there would be a dome for striped phases near the red region as $n_1$ is neither too large nor too small.

\subsection{Striped instabilities of AdS-RN black brane}
In the last subsection the spatially modulated instabilities that we analysed on the $AdS_2$ background happen at zero temperature. Nevertheless, it suggests that analogous instabilities would appear as well as in AdS-RN black brane~\eqref{RNads} at finite temperatures. We will then calculate the critical temperature at which the AdS-RN geometry becomes unstable as a function of $\textbf{x}$.

Motivated by the analysis in the $AdS_2$ geometry case, we only need to turn on fluctuations as~\footnote{We have refined our parameterisation of perturbations in~\eqref{pertubzero} to include some explicit $r$-dependent term, for numerical convenience.}
\begin{equation}\label{pertufir}
\begin{split}
&\delta g_{ty} =  \lambda\, r(r-r_h)h(r)\sin(kx),\quad \delta\alpha=  \lambda\, w(r)\cos(kx),\\
&\delta A_{y} =  \lambda\,  a_{y}(r)\sin(kx),\quad\quad\quad\quad \delta B_{y}= \lambda\, b_{y}(r)\sin(kx).
\end{split}
\end{equation}
By expanding around the AdS-RN background to linear order, one can obtain four coupled linear ODEs,
\begin{eqnarray}
\nonumber w''+\left(\frac{2}{r}+\frac{g'}{g}\right)w'-\frac{1}{g}\left(\widetilde{m}_\alpha^2+\frac{k^2}{r^2}-\frac{b_3\textbf{x}^2+2c_3\textbf{x}+a_3}{2r^4}\right)w-\\
\frac{2k(2n_1+n_2\textbf{x})}{r^4 g}a_y-\frac{2k n_2}{r^4 g}b_y& = & 0, \\
a_y''+\frac{g'}{g}a_y'+\frac{r-r_h}{r g}h'-\frac{2k (2n_1+n_2\textbf{x})}{r^2 g}w-\frac{k^2}{r^2 g}a_y+\frac{r_h}{r^2 g}h&=&0,\\
b_y''+\frac{g'}{g}b_y'+\frac{(r-r_h)\textbf{x}}{r g}h'-\frac{2k n_2}{r^2 g}w-\frac{k^2}{r^2 g}b_y+\frac{r_h \textbf{x}}{r^2 g}h&=&0,\\
h''+\frac{4r-2r_h}{r(r-r_h)}h'+\frac{a_y'+\textbf{x}b_y'}{r^3(r-r_h)}-\frac{k^2(r-r_h)-2r_h g}{r^2(r-r_h)g}h&=&0.
\end{eqnarray}
We have worked in units with $\rho=1$. The instabilities towards the formation of striped order correspond to the normalisable zero modes with non-vanishing wave number $k$ allowed by four coupled linearised equations. Those marginally-unstable zero modes have to be found using numerics.

 We demand the fluctuations to be regular on the horizon at $r=r_h$. Therefore, near the horizon, the four functions $h, a_y, b_y, w$ behave as
\begin{equation}
\begin{split}
&h(r) = h^h+\mathcal{O}(r-r_h),\quad\quad w(r)=w^h+\mathcal{O}(r-r_h),\\
&a_{y}(r) =a^h +\mathcal{O}(r-r_h),\quad\quad b_y(r)= b^h +\mathcal{O}(r-r_h).
\end{split}
\end{equation}
The most general asymptotical behaviour in the UV as $r\rightarrow \infty$ is given by
\begin{equation}\label{stripuvexp}
\begin{split}
&h(r) = h_s+\cdots+\frac{h_v}{r^3}+\cdots,\quad\quad w(r)=w_s+\cdots+\frac{w_v}{r^3}+\cdots,\\
&a_{y}(r) =a_s+\cdots+\frac{a_v}{r}+\cdots,\quad\quad b_{y}(r) =b_s+\cdots+\frac{b_v}{r}+\cdots.
\end{split}
\end{equation}
Since we are interested in the case that breaks translational invariance spontaneously, we  turn off the parameters $(h_s, w_s, a_s, b_s)$ which correspond to the sources of the dual operators in the boundary field theory.

We now have four coupled second order ODEs with regular condition in the IR and source free condition in the UV. In order to integrate from the outer horizon, $r=r_h$, to asymptotical boundary, $r\rightarrow \infty$, one needs to specify eight integration constants. For given wave number $k$ and doping parameter $\textbf{x}$, there are nine parameters $(r_h, h^h, w^h, a^h, b^h, h_v, w_v, a_v, b_v)$ entering the ODEs. Since the equations are linear, we can always scale one of these parameters to unity. Therefore, for given $k$ and $\textbf{x}$, we are expected to find a normalisable zero mode, if at all, appearing at a particular temperature.

We solve this problem numerically by using of the double side shooting method from both the IR and UV (see appendix~\ref{app:doushot} for more details). For a given set of theory parameters and $\textbf{x}$ that admit striped instabilities associated with BF-violating modes, we try to look for the normalisable zero modes and then find that these modes exist for a range of $k$ with the critical temperature depending on $k$. The curve is expected to give the usual ``bell curve" type behaviour: the critical temperature versus $k$ first increases and then decreases. Hence, there exists a particular wave number $k^*$ associated with the most highest critical temperature $T^*$. At $T=T^*$ a new branch of black branes with wave number $k^*$ will appear  breaking translation symmetry along the $x$ direction spontaneously. Note that the value of $T^*$ as well as $k^*$ depends on $\textbf{x}$.

\begin{figure}[ht!]
\begin{center}
\includegraphics[width=.65\textwidth]{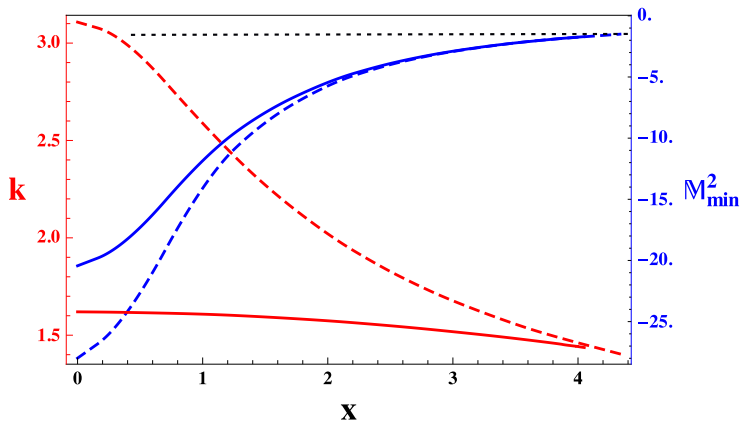}
\caption{ The red solid curve denotes $k^*$ versus $\textbf{x}$, while the red dashed curve gives the wave number $k_0$ at which the eigenvalue of $\mathbb{M}^2$ has a global smallest value for all $k$ at each $\textbf{x}$. The blue solid curve represents the smallest eigenvalue of $\mathbb{M}^2$ at $k^*$ and the blue dashed curve indicates the smallest eigenvalue of $\mathbb{M}^2$ at $k_0$. The horizontal black dotted line is the $AdS_2$ BF bound. Note that two blue curves are both below the horizontal line. We have chosen $n_1=0.8$ and $\widetilde{m}^2=a_3=b_3=c_3=n_2=0$.}
\label{fig:stripkm}
\end{center}
\end{figure}

\begin{figure}[ht!]
\begin{center}
\includegraphics[width=.55\textwidth]{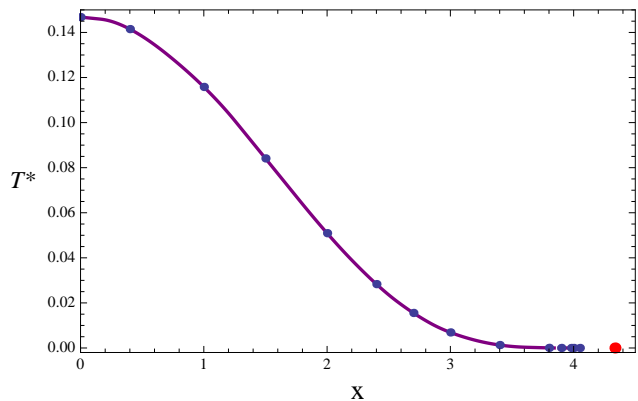}
\caption{The maximum critical temperature $T^*$ at which a striped solution appears as a function of $\textbf{x}$.  The blue dots are from numerical calculation in the AdS-RN black brane, while the red point $(4.335, 0)$ at far right is obtained from the $AdS_2$ geometry which is the external limit of AdS-RN background. We have chosen $n_1=0.8$ and $\widetilde{m}^2=a_3=b_3=c_3=n_2=0$.}
\label{fig:striptx}
\end{center}
\end{figure}

We consider a particular case with $n_1=0.8$ and $\widetilde{m}^2=a_3=b_3=c_3=n_2=0$, where all parameters are given in~\eqref{vpossi}. As one can see from figure~\ref{fig:cs1}, the striped instabilities will appear as $\textbf{x}<4.335$. One can see from figure~\ref{fig:stripkm} that although the mode at $k_0$ has the smallest eigenvalue $\mathbb{M}^2_{min}[k_0]$ that violates the $AdS_2$ BF bound, it is in general not the mode that appears first at finite temperatures. Nevertheless, the difference between $k^*$ (red solid) and $k_0$ (red dashed) is reduced as one increases $\textbf{x}$. This is reasonable since as $\textbf{x}$ is increased the range of $k$ that allows striped instabilities shrinks. At the particular value $\textbf{x}\simeq 4.335$ where the instability region shrinks to one point with $k_0\simeq 1.404$, $k^*$ and $k_0$ are expected to coincide with each other. The smallest eigenvalue of $\mathbb{M}^2$ at $k^*$ (blue solid)  is obviously above the one at $k_0$ (blue dashed). As shown in figure~\ref{fig:stripkm}, both increase monotonically by increasing $\textbf{x}$ and arrival at the BF bound from below at $\textbf{x}\simeq 4.335$. The critical temperature $T^*$ as a function of $\textbf{x}$ is given in figure~\ref{fig:striptx}. $T^*$ decreases with the increase of $\textbf{x}$ and vanishes as $\textbf{x}>4.335$.\footnote{Due to the lake of numerical control at sufficiently low temperatures, it is very difficult to calculate the values of $k^*$ and $T^*$ very closed to the critical point $x\simeq 4.335$. Nevertheless, we solve the theory up to the temperature as low as $10^{-7}$ and find good agreements with $AdS_2$ analysis.} The current result also suggests the tendency: the lower the effective mass squared $\mathbb{M}^2[k^*]$ on $AdS_2$, the higher the critical temperature for the broken phase. We also checked other values of $n_1$ while keeping $\widetilde{m}^2=a_3=b_3=c_3=n_2=0$. Just as expected, the critical temperature $T^*$ is below the curve in figure~\ref{fig:striptx} for a smaller value of $n_1$.


\subsection{Striped phase and charge density wave}
We have found the static normalisable zero modes in the electrically charged AdS-RN background by virtue of perturbation analysis. By working at leading order, for each doping $\textbf{x}$, one can obtain the critical temperature versus wave number at which the AdS-RN geometry becomes unstable to the formation of a spatially modulated phase. They form a typical ``bell curve" from which one can read off the highest critical temperature $T^*$ and the corresponding wave number $k^*$.

From the leading order perturbations we find that the spatially modulated solutions are associated with nonzero $h_v, a_v, b_v$ and $w_v$ of~\eqref{stripuvexp}. A non-vanishing $h_v$ means that there is momentum transfer in the $y$ direction. A non-vanishing $a_v$ ( $b_v$) implies that the dual current component $\langle\hat{J}_A^y\rangle$ ($\langle\hat{J}_B^y\rangle$) acquires a spatially modulated VEV, thus the dual system exhibits current density wave with spatial modulation $k^*$ given at $T^*$. Finally, a nonzero $w_v$ means that the scalar operator dual to $\alpha$ is acquiring a spatially modulated VEV which breaks the time-reversal and parity invariance spontaneously. Note that at the linear order, the charge density is still space independent. To see whether the spatially modulated charge density (CDW)  exists or not, we need to study our perturbation theory at next to leading order.

Working at  next to leading order, we find that the system is also accompanied with CDWs with a spatial modulation given by $2 k^*$ at $T^*$. More specifically, we turn on the following second order perturbations,
\begin{equation}\label{pertusec}
\begin{split}
\delta g_{tt}& =  \lambda^2 [h_{tt}^{(0)}(r)+h_{tt}^{(1)}(r)\cos(2kx)],\\
\delta g_{xx}& =  \lambda^2 [h_{xx}^{(0)}(r)+h_{xx}^{(1)}(r)\cos(2kx)],\\
\delta g_{yy}& =  \lambda^2 [h_{yy}^{(0)}(r)+h_{yy}^{(1)}(r)\cos(2kx)],\\
\delta A_{t}& =  \lambda^2 [a_{t}^{(0)}(r)+a_{t}^{(1)}(r)\cos(2kx)],\\
\delta B_{t}& =  \lambda^2 [b_{t}^{(0)}(r)+b_{t}^{(1)}(r)\cos(2kx)].
\end{split}
\end{equation}
Expending the total field fluctuations in~\eqref{pertusec} together with~\eqref{pertufir} up to order $\mathcal{O}(\lambda^2)$, we can obtain a closed set of inhomogeneous ODEs for  the perturbations above, which are sourced by the linear order zero mode solutions. Actually, there are thirteen equations for ten functions. However, the two equations from the $tt$ component of Einstein's equations are implied by the other equations. By using the equation from the $rx$ component and the one from the $rr$ component one finds that $h_{tt}^{(1)}$ satisfies an algebraic equation and hence can be eliminated from the system. Finally, we obtain nine independent differential equations in which the functions $(h_{xx}^{(0)}, h_{xx}^{(1)}, h_{yy}^{(0)}, h_{yy}^{(1)}, a_{t}^{(0)}, a_{t}^{(1)}, b_{t}^{(0)}, b_{t}^{(1)})$ satisfy second order equations and the function $h_{tt}^{(0)}$ satisfies a first order equation.

By imposing a regularity condition at the horizon and demanding the source free boundary condition in the UV, for given $k$ and $\textbf{x}$, we are left with one free parameter which can be chosen to be the temperature. Then from the asymptotically UV expansion of $a_{t}^{(1)}$ or $b_{t}^{(1)}$, one can see that there is a spatial modulation of charge density with wave number given by $2 k^*$ at $T^*$. So the spatially modulated phase is always associated with CDW order with half the period of the condensate of pseudo-scalar and current density waves.

To sum up, we demonstrated the instabilities towards the formation of CDWs which occur to  next to leading order in perturbation theory. Therefore, the new branch of spatially modulated black brane solutions, assuming that they are thermodynamically preferred at the normal solutions, will be dual to striped phases accompanying with both current density waves and charge density waves.

\subsection{Checkerboard phase}\label{subsec:check}
In the previous subsection  we have focused on the striped order where space translation symmetry is only spontaneously broken in one direction, resulting in a periodic phase. There is another possibility that the symmetry is broken in both directions and the configuration becomes periodic in two spatial directions. This is known as checkerboard phase. In this subsection, we shall briefly discuss the instability towards formation of checkerboard structures.

We take $AdS_2\times \mathbb{R}^2$ background~\eqref{ads2bk2} as our starting point. We try to turn on the following perturbations with momentum $k_x$ in the $x$ direction and $k_y$ in the $y$ direction,
\begin{equation}\label{checkerbordpub}
\begin{split}
&\delta g_{tx}= \lambda r_h h_{tx}(r)\cos(k_x\,x)\sin(k_y\,y),\quad \delta g_{ty}= \lambda r_h h_{ty}(r)\sin(k_x \,x)\cos(k_y\,y),\\
&\delta A_{x}= \lambda r_h a_{x}(r)\cos(k_x\,x)\sin(k_y\,y),\quad\;  \delta A_{y}= \lambda r_h a_{y}(r)\sin(k_x\, x)\cos(k_y\,y),\\
&\delta B_{x}= \lambda r_h b_{x}(r)\cos(k_x\,x)\sin(k_y\, y),\quad\;  \delta B_{y}= \lambda r_h b_{y}(r)\sin(k_x \,x)\cos(k_y\,y),\\
&\delta \alpha= \lambda w(r)\cos(k_x \,x)\cos(k_y \,y),
\end{split}
\end{equation}
with $\lambda$ a formal expansion parameter as before.  We find that the finally independent equations are given by
\begin{eqnarray}
\nonumber 6\partial_r(r^2 \partial_r w)-\left(\widetilde{m}_{(2)}^2+\frac{k_x^2+k_y^2}{r_h^2}\right)w-\frac{4\sqrt{3}(2n_1+n_2\textbf{x})}{\sqrt{1+\textbf{x}^2}}\frac{k_x^2+k_y^2}{k_x r_h}a_y\\
-\frac{4\sqrt{3}n_2}{\sqrt{1+\textbf{x}^2}}\frac{k_x^2+k_y^2}{k_x r_h}b_y& = & 0, \\
6\partial_r(r^2 \partial_r a_y)+\frac{2\sqrt{3}}{\sqrt{1+\textbf{x}^2}}h_{ty}'-\frac{4\sqrt{3}(2n_1+n_2\textbf{x})}{\sqrt{1+\textbf{x}^2}}\frac{k_x}{r_h}w-\frac{k_x^2+k_y^2}{r_h^2}a_y&=&0,\\
6\partial_r(r^2 \partial_r b_y)+\frac{2\sqrt{3}\textbf{x}}{\sqrt{1+\textbf{x}^2}}h_{ty}'-\frac{4\sqrt{3}n_2}{\sqrt{1+\textbf{x}^2}}\frac{k_x}{r_h}w-\frac{k_x^2+k_y^2}{r_h^2}b_y&=&0,\\
-3r^2 h_{ty}''-\frac{6\sqrt{3}r^2}{\sqrt{1+\textbf{x}^2}}a_y'-\frac{6\sqrt{3}\textbf{x}r^2}{\sqrt{1+\textbf{x}^2}}b_y'+\frac{1}{2}\frac{k_x^2+k_y^2}{r_h^2}h_{ty}&=&0,
\end{eqnarray}
together with algebraic relation $k_x a_x+k_y a_y=k_x b_x+k_y b_y=k_x h_{tx}+k_y h_{ty}=0$. For more details one can consult appendix~\ref{app:checkerboard}.

\begin{figure}[ht!]
\begin{center}
\includegraphics[width=.55\textwidth]{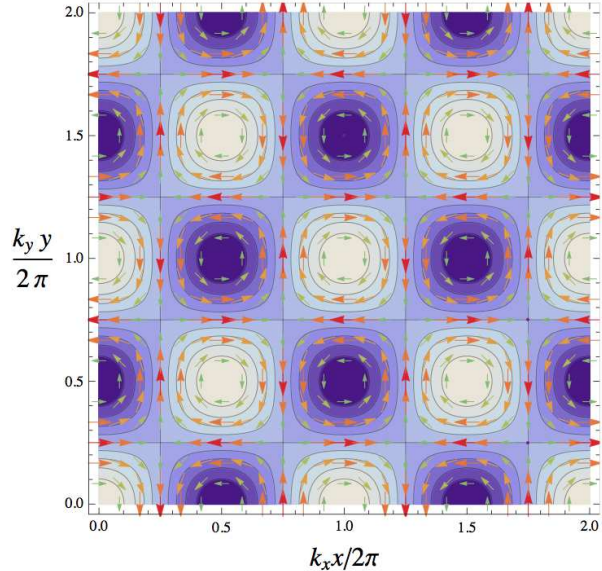}
\caption{A square checkerboard in the vicinity of the critical temperature.  We show a contour plot of the VEV of the pseudo-scalar operator, $\left<\mathcal{O}_\alpha\right>$, together with some integral curves of the dual current $(\langle\hat{J}_A^x \rangle,\langle\hat{J}_A^y\rangle)$.}
\label{fig:checker}
\end{center}
\end{figure}

As a consistent check, by setting $k_y=0$ and $k_x=k$, one recovers the linear equations for striped perturbations in subsection~\ref{stripads2} exactly. Those equations above should give very similar unstable modes that break the $AdS_2$ BF bound as discussed in the striped case previously. They will also give similar marginally-unstable modes on the finite temperature background, giving rise to black brane solutions which are spatially modulated in both spatial directions.

Since the expectation values of the operator $\mathcal{O}_\alpha$ dual to $\alpha$ and the current $\hat{J}_A^\mu$ dual to $A_\mu$ are read off from the coefficients in the UV, they should have the same spatial dependence on two spatial coordinates $x$ and $y$ as their corresponding bulk field configurations near the boundary. As a consequence, $\left<\mathcal{O}_\alpha\right>\propto \cos(k_x\,x)\cos(k_y\,y)$ as well as $(\langle\hat{J}_A^x \rangle,\langle\hat{J}_A^y\rangle)\propto ( \cos(k_x\,x)\sin(k_y\,y), -\sin(k_x\,x)\cos(k_y\,y))$ when the temperature is sufficiently close to the critical temperature at which the checkerboard instability starts to appear. A schematic plot for a square pattern $k_x=k_y$ is drawn in figure~\ref{fig:checker}. The current circulates with the sense alternating between adjacent plaquettes of the checkerboard. We obtain this picture very close to the critical temperature. Nevertheless, this behaviour was observed at temperatures far below the critical temperature in a similar theory~\cite{Withers:2014sja}.


\section{The Global Phase Diagram}
\label{sec:diagram}
The precise form of phase diagram depends critically on the specific choice of the theory parameters as discussed above. A generic case allows for a much larger zoo of phase diagrams. In this section we will first classify possible phase diagrams for a broken phase in the homogeneous case. Then we choose theory parameters to arrive at the expected phase diagram.

\subsection{Classification of possible phase diagrams for broken phase \label{alldiagrams}}
In this subsection we will determine the phase boundaries in the $(T, \textbf{x})$ phase diagram for each broken phase.
We will restrict ourselves to the homogeneous case. This kind of phase transition should be triggered by an instability in the normal state.

The staring point is to consider fluctuations of each scalar around the IR background of the extremal AdS-RN black brane. As one can see from subsection~\ref{sub:ads2}, all equations can be regarded as for a scalar in the $AdS_2$ background with the form of effective mass squared given by
\begin{equation}
\mathcal{M}^2(\textbf{x})=\mathbf{m}^2-\frac{b\textbf{x}^2+2c\textbf{x}+a}{1+\textbf{x}^2}.
\end{equation}
In out units with $L=1$ the $AdS_2$ BF bound is given by $\mathcal{M}_{BF}^2=-3/2$. The value of $\mathcal{M}^2$ can be tuned by changing $\textbf{x}$. The instability towards the formation of nontrivial scalar hair will appear if $\mathcal{M}^2(\textbf{x})<-3/2$. The type of $(T, \textbf{x})$ phase diagram can be inferred from the instability region of $\mathcal{M}^2(\textbf{x})$. Our numerical experience suggests that the lower the value of $\mathcal{M}^2(\textbf{x})$, the higher the critical temperature we will obtain.

Taking advantage of above argument, we can classify all possible kinds of $(T, \textbf{x})$ phase diagram by finding the region in which $\mathcal{M}^2(\textbf{x})<-3/2$. The BF-violating region depends on four parameters $(\mathbf{m}^2, a, b ,c)$. After a through analysis, we find ten different phase diagrams. We will give the parameter space and a schematic figure for each phase diagram one by one.

\begin{figure}[ht!]
\begin{center}
\includegraphics[width=.46\textwidth]{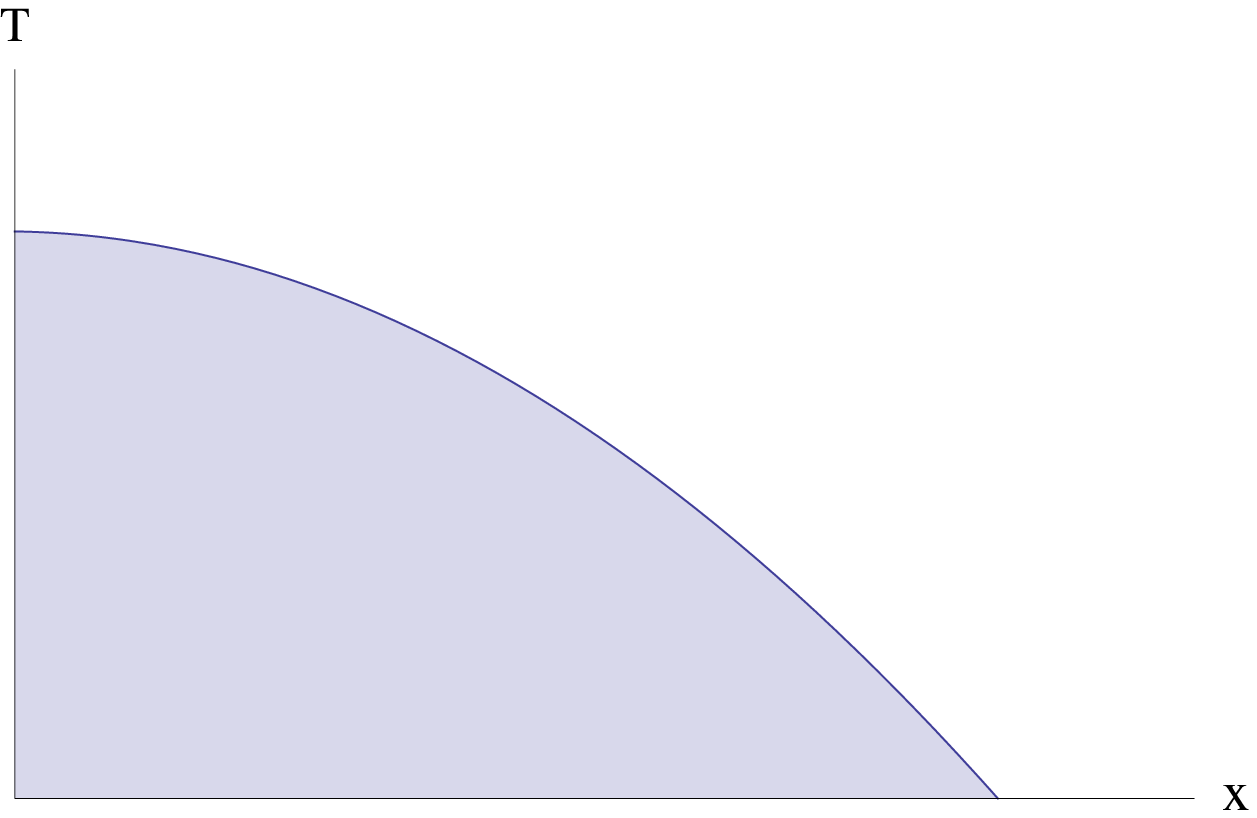}
\quad\quad
\includegraphics[width=.46\textwidth]{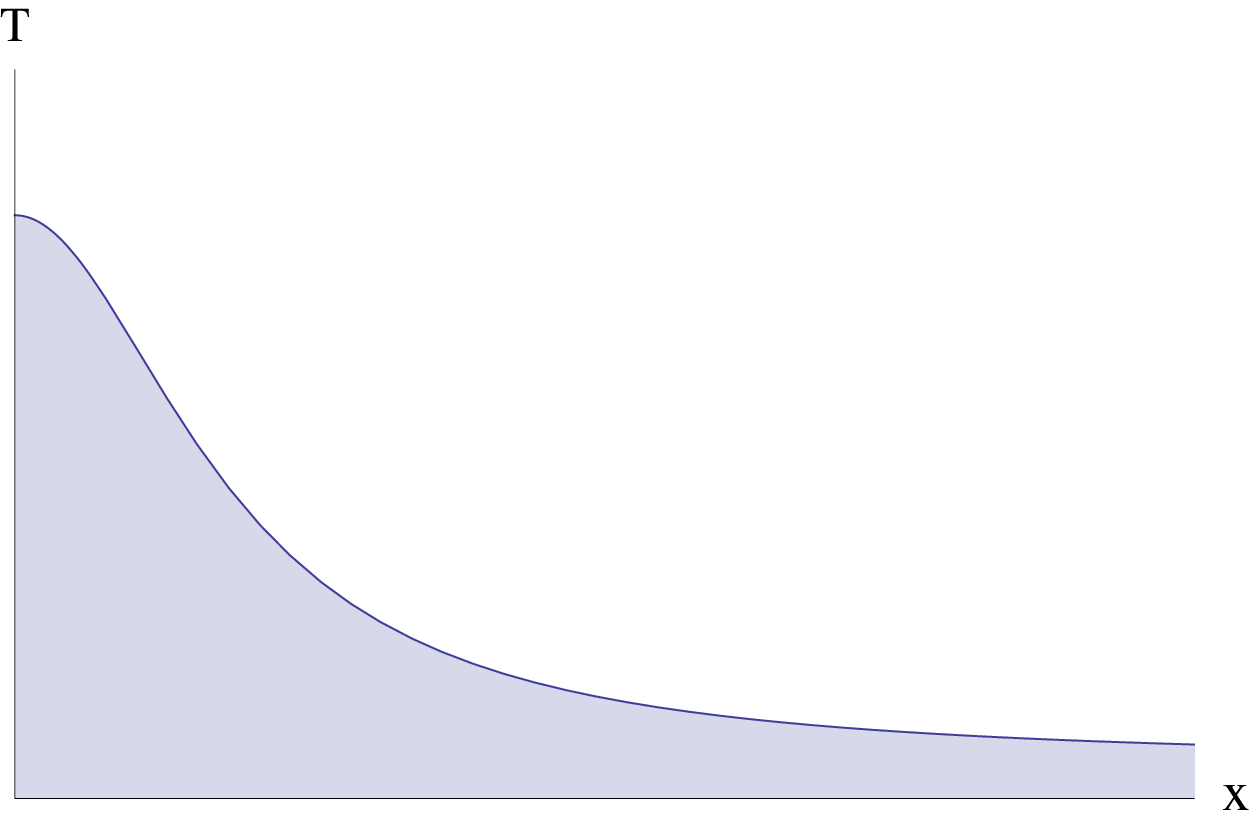}
\caption{The schematic plots for case 1 (left) and case 2 (right). The shadow region corresponds to the broken phase. In case 1 the critical temperature becomes zero at a particular value of $\textbf{x}$, while in the later case the critical temperature never vanishes.}
\label{fig:case12}
\end{center}
\end{figure}

\subsubsection{Case 1}
In this case, the critical temperature decreases monotonically with respect to $\textbf{x}$ and finally vanishes at a certain value of $\textbf{x}$, see the left plot of figure~\ref{fig:case12}. This kind of phase diagram is adopted to realise the antiferromagnetic region in the global $(T, \textbf{x})$ phase diagram.

The theory parameters satisfy the following conditions,
\begin{equation}
c=0,\quad \textbf{m}^2-a<-3/2< \textbf{m}^2-b,
\end{equation}
or
\begin{equation}
c<0,\quad \textbf{m}^2-a<-3/2< \textbf{m}^2-b.
\end{equation}
In the former one $\mathcal{M}^2$ is a monotonically increasing function of $\textbf{x}$ in the positive $\textbf{x}$ direction, while the latter is non-monotonic.

\subsubsection{Case 2}
In case 2, the critical temperature also decreases monotonically with respect to $\textbf{x}$. The difference from case 1 is that  the critical temperature keeps a finite value even for very large value of  $\textbf{x}$, see the right plot in figure~\ref{fig:case12}.

The theory parameters satisfy
\begin{equation}
c=0, \quad b<a, \quad \textbf{m}^2-b<-3/2.
\end{equation}
\begin{figure}[ht!]
\begin{center}
\includegraphics[width=.46\textwidth]{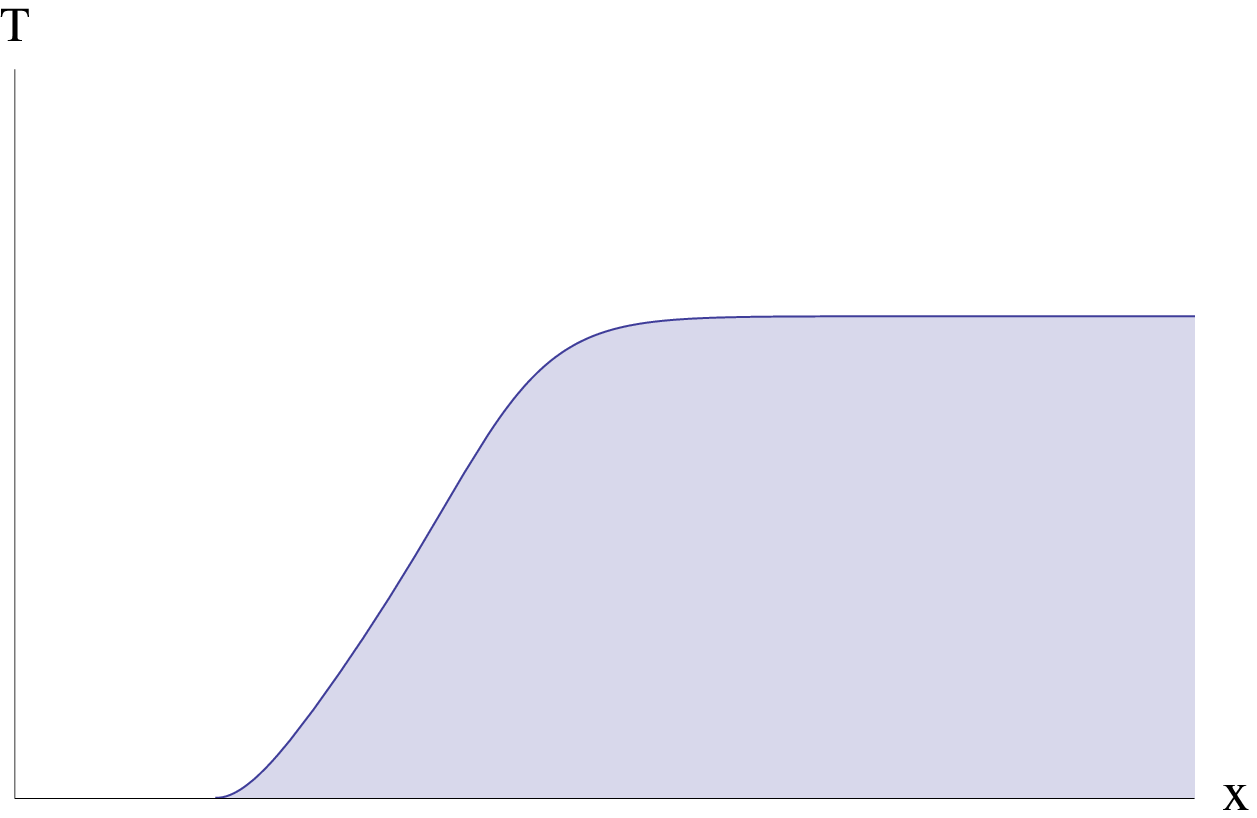}
\quad\quad
\includegraphics[width=.46\textwidth]{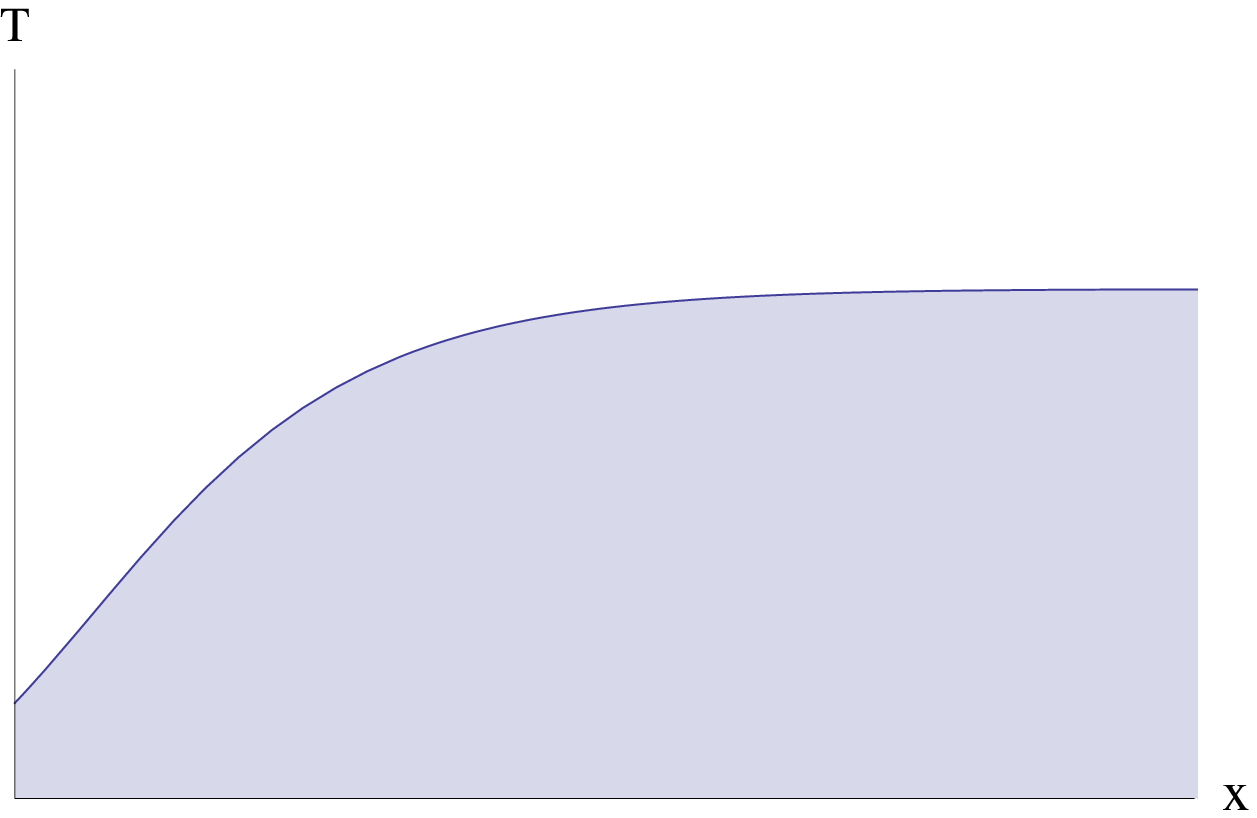}
\caption{The schematic plots for case 3 (left) and case 4 (right). The shadow region corresponds to the broken phase. In case 3 the critical temperature appears at a nonzero $\textbf{x}$, while in the later case the critical temperature develops at the origin $\textbf{x}=0$.}
\label{fig:case34}
\end{center}
\end{figure}

\subsubsection{Case 3}
In case 3, the critical temperature is nonzero only when $\textbf{x}$ is larger than a certain finite value. Then it increases as one increases the value of $\textbf{x}$. See the right plot in figure~\ref{fig:case34}.

There are two sets of  parameters supporting this kind of phase diagram,
\begin{equation}\label{appcsae3a}
c=0, \quad \textbf{m}^2-b<-3/2<\textbf{m}^2-a,
\end{equation}
as well as
\begin{equation}
c<0, \quad  \textbf{m}^2-b<-3/2<\textbf{m}^2-a.
\end{equation}
The difference is that $\mathcal{M}^2$ is a monotonically decreasing function in the positive $\textbf{x}$ direction in the first case~\eqref{appcsae3a}, while it is non-monotonic in the latter one.

\subsubsection{Case 4}
Case 4 in the right plot of figure~\ref{fig:case34} is similar as case 3. The only difference is that the critical temperature is nonzero at $\textbf{x}=0$. Then it increases monotonically as a function of $\textbf{x}$.

The parameter space that supports this phase diagram reads
\begin{equation}
c=0, \quad a<b,\quad \textbf{m}^2-a<-3/2.
\end{equation}
\begin{figure}[ht!]
\begin{center}
\includegraphics[width=.46\textwidth]{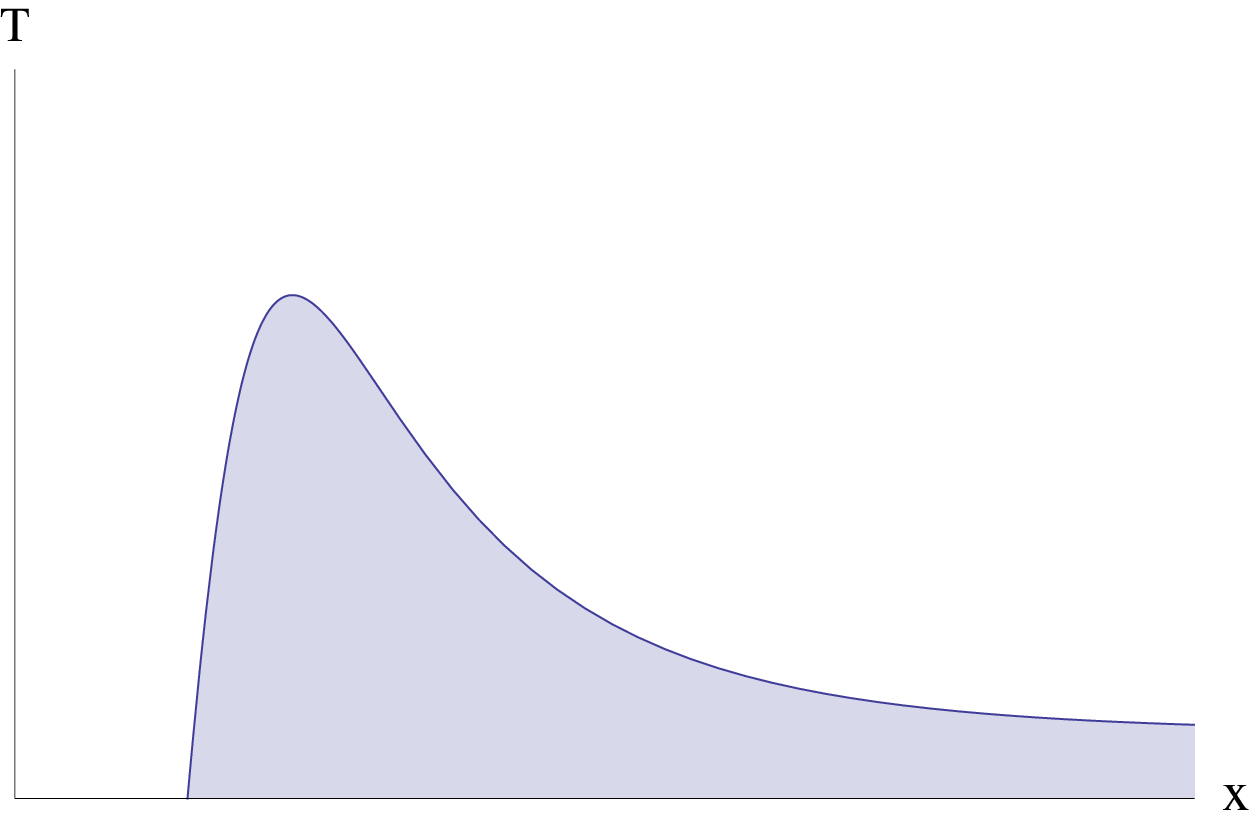}
\quad\quad
\includegraphics[width=.46\textwidth]{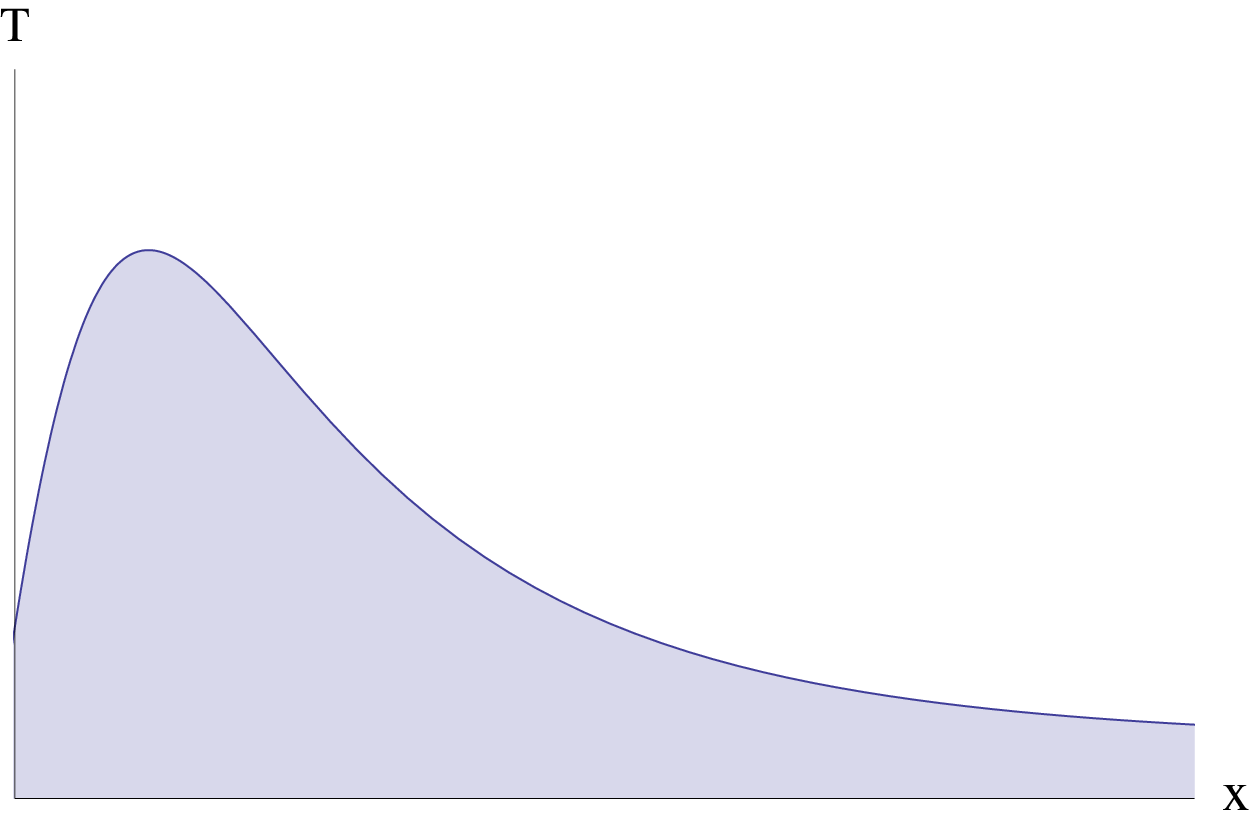}
\caption{The schematic plots for case 5 (left) and case 6 (right). The shadow region denotes the broken phase. In case 5 the critical temperature appears at a nonzero $\textbf{x}$, while in case 6 the critical temperature appears at the origin $\textbf{x}=0$.}
\label{fig:case56}
\end{center}
\end{figure}

\subsubsection{Case 5}
For case 5, the critical temperature begins to appear above a nonzero value of $\textbf{x}$, then rises as the value of $\textbf{x}$ is increased, arrives at its maximum at a certain $\textbf{x}$,
and finally decreases monotonously. See the left plot in figure~\ref{fig:case56}.

The parameter space for this type of phase diagram is given by
\begin{equation}
c>0, \quad \textbf{m}^2-b<-3/2,\quad \textbf{m}^2-a>-3/2.
\end{equation}

\subsubsection{Case 6}
We show the behaviour of critical temperature as a function of $\textbf{x}$ in the right plot of figure~\ref{fig:case56}. The non-monotonic behaviour of the critical temperature is similar as the previous case. However, in case 6 the critical temperature does not vanish even at $\textbf{x}=0$.

The theory parameters that can give  this kind of phase diagram satisfy
\begin{equation}
c>0,\quad \textbf{m}^2-b<-3/2,\quad \textbf{m}^2-a<-3/2.
\end{equation}

One should note that the critical temperature never vanishes for sufficiently large $\textbf{x}$ in case 5 and case 6.

\begin{figure}[ht!]
\begin{center}
\includegraphics[width=.46\textwidth]{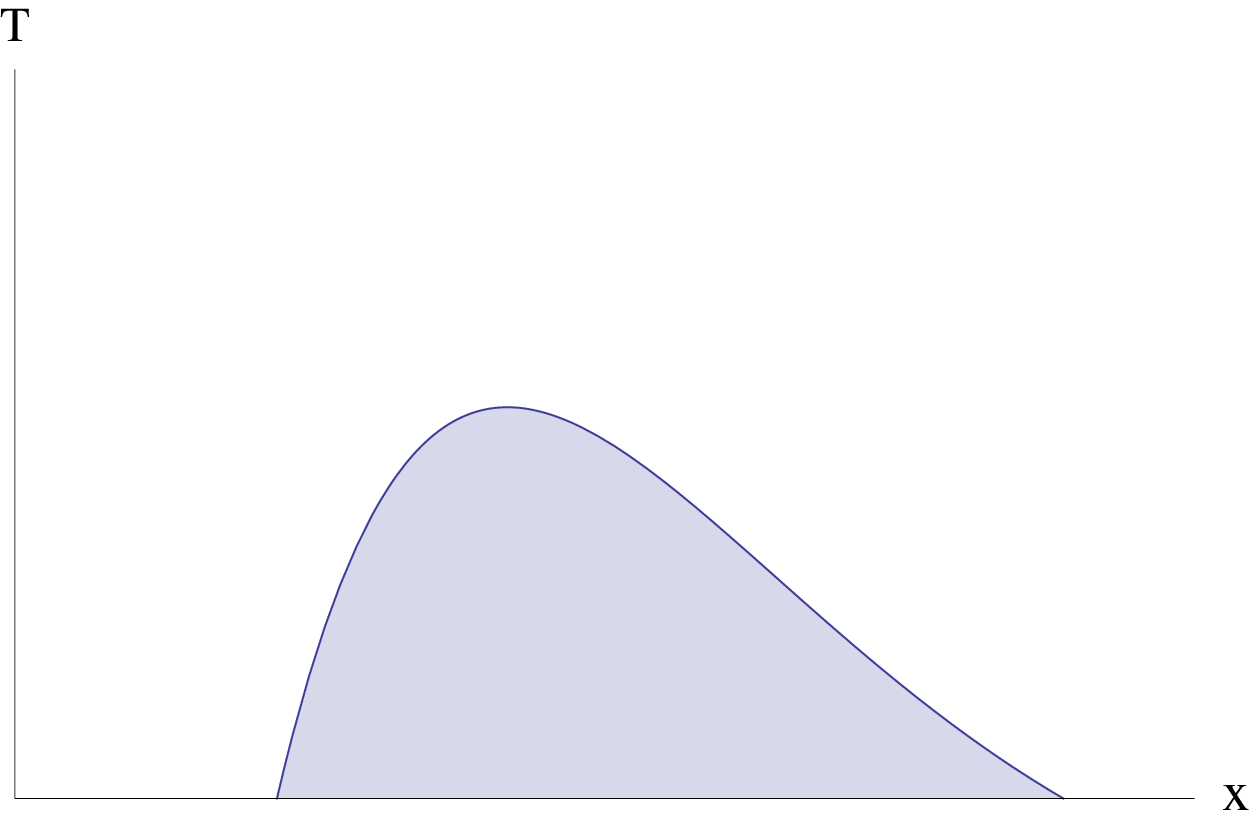}
\quad\quad
\includegraphics[width=.46\textwidth]{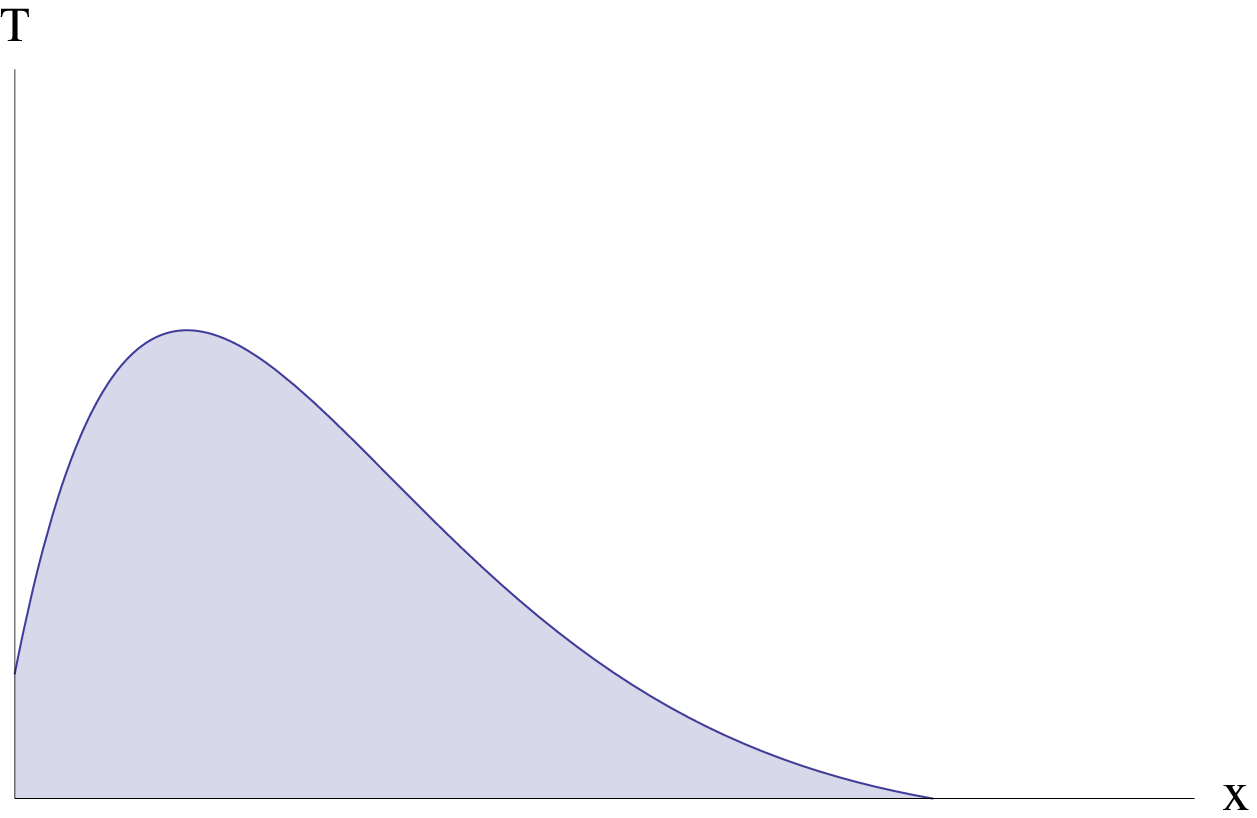}
\caption{The schematic plots for case 7 (left) and case 8 (right). The shadow region denotes broken phase. In both cases, there exists a dome for broken phase. In the former case the critical temperature appears at a nonzero $\textbf{x}$ and then vanishes at a larger value, while in case 8 the critical temperature appears at the origin $\textbf{x}=0$.}
\label{fig:case78}
\end{center}
\end{figure}

\subsubsection{Case 7}
There is a dome for case 7 shown in the left plot of figure~\ref{fig:case78}. The broken phase appears at a nonzero $\textbf{x}$ and then disappears above a certain larger value of $\textbf{x}$. This kind of phase diagram is used to construct the superconducting dome in this paper.

To obtain the dome we should choose the theory parameters as follows.
\begin{equation}
c>0,\quad \mathcal{M}^2(\textbf{x}^+_{min})<-3/2<\textbf{m}^2-b,\quad \textbf{m}^2-a>-3/2.
\end{equation}
When $c>0$, there is a minimum denoted as $\textbf{x}^+_{min}$ in the positive $\textbf{x}$ direction. Two critical points at which the broken phase disappears can be uniquely obtained by solving the algebraic equation $\mathcal{M}^2(\textbf{x})=-3/2$.

\subsubsection{Case 8}
From the right plot of figure~\ref{fig:case78}, one can see that  as we increase the doping parameter $\textbf{x}$ the transition temperature to the broken phase first increases, arrivals at a maximum value and then monotonically decreases to zero as $\textbf{x}$ grows. This is still a kind of dome.

The parameter space for case 8 is given by
\begin{equation}
c>0,\quad \textbf{m}^2-b>-3/2,\quad \textbf{m}^2-a<-3/2.
\end{equation}
\begin{figure}[ht!]
\begin{center}
\includegraphics[width=.46\textwidth]{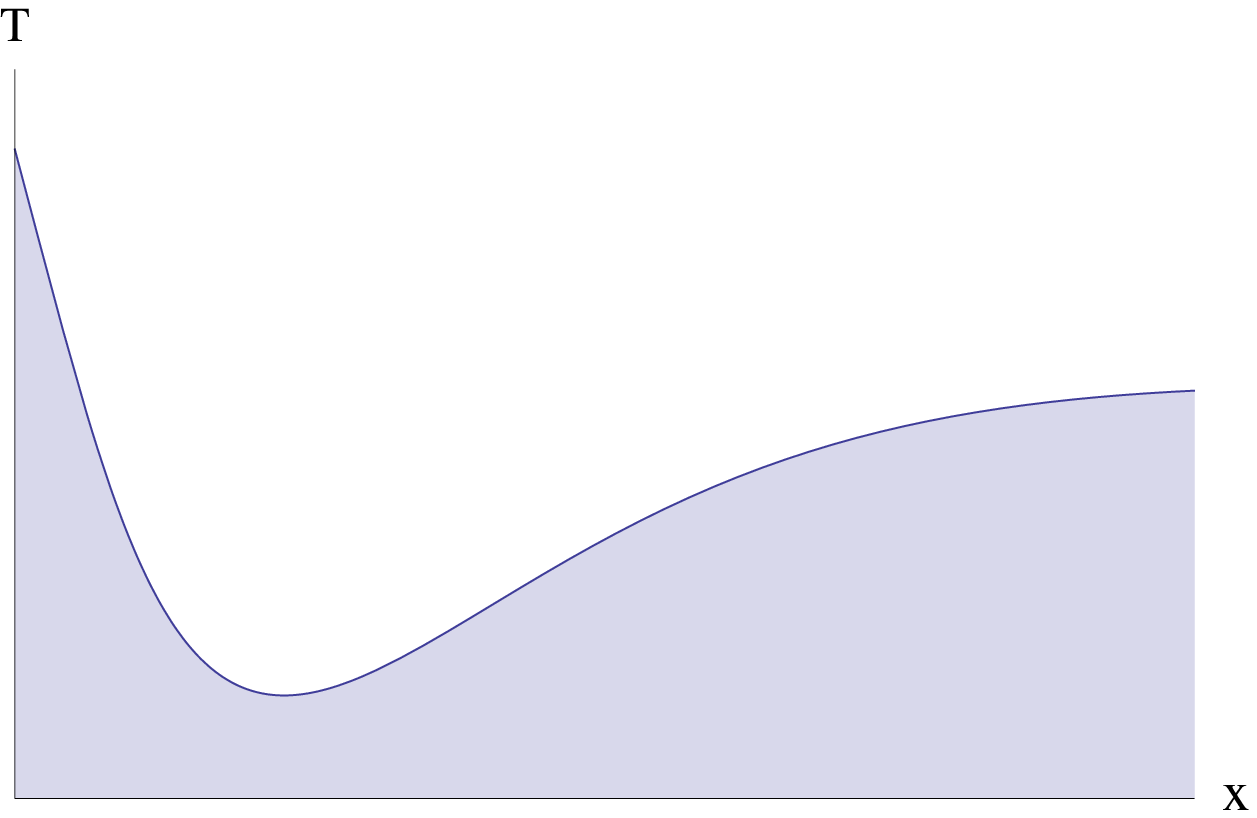}
\quad\quad
\includegraphics[width=.46\textwidth]{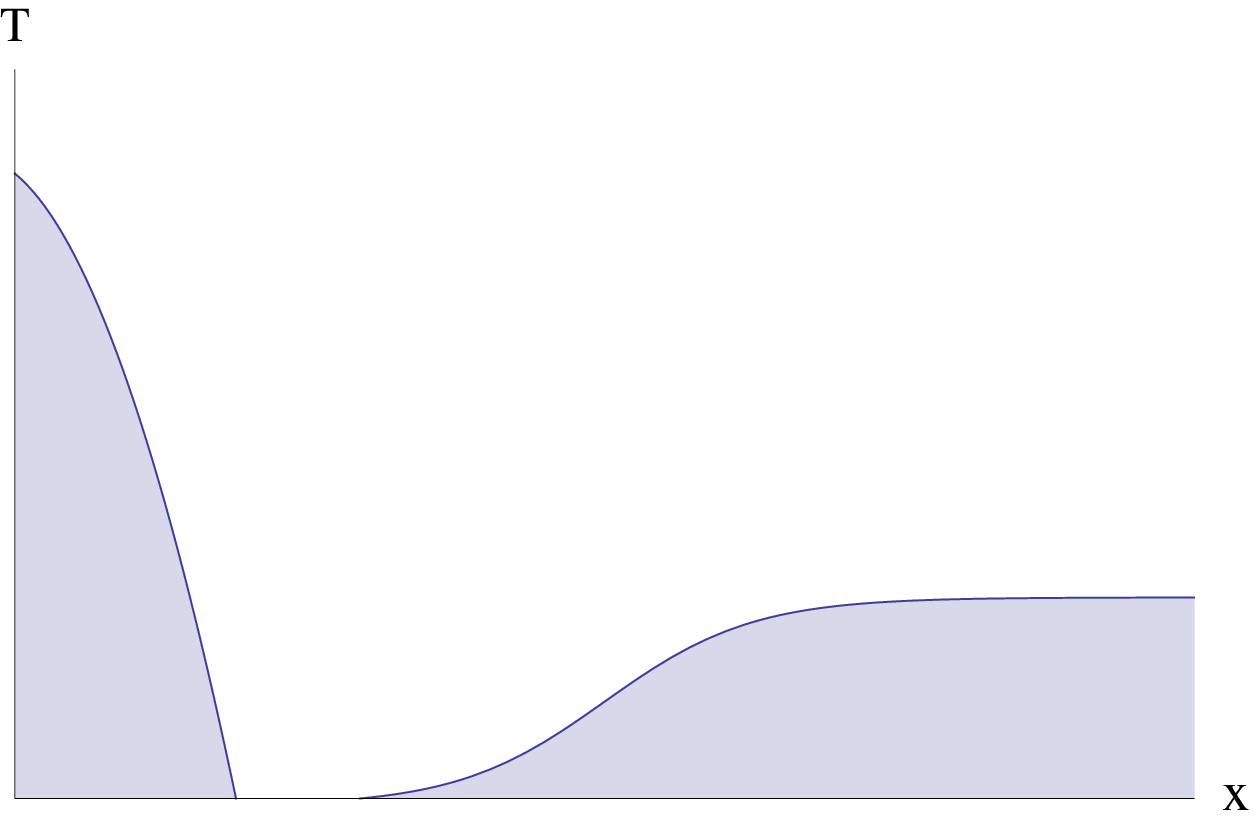}
\caption{The schematic plot for case 9 (left) and case 10 (right). The shadow region denotes the broken phase. In case 9 the critical temperature has a non-vanishing minimum value. There are two disconnected shadow regions for the right one.}
\label{fig:case910}
\end{center}
\end{figure}

\subsubsection{Case 9}
In this case the broken phase can always develop for any positive value of $\textbf{x}$. As one can see from the left plot of figure~\ref{fig:case910}, the critical temperature first decreases and then increases.

The parameter space that supports this phase diagram is given by
\begin{equation}
c<0,\quad \mathcal{M}^2(\textbf{x}^-_{max})<-3/2.
\end{equation}
Here the finite value $\textbf{x}^-_{max}$ denotes the maximum in the positive $\textbf{x}$ direction when $c<0$.

\subsubsection{Case 10}
The last case is shown in the left plot of figure~\ref{fig:case910}, from which one can find two disconnected regions for the broken phase. One is near the origin, and the other one happens at larger values of $\textbf{x}$.

The parameter space for case 10 is given by
\begin{equation}
c<0,\quad \textbf{m}^2-b<-3/2<\mathcal{M}^2(\textbf{x}^-_{max}),\quad \textbf{m}^2-a<-3/2.
\end{equation}
Here $\textbf{x}^-_{max}$ is the location of the unique maximum in the positive $\textbf{x}$ direction when $c<0$.

Before the end of this subsection, we point out that the above analyses are based on the BF-violating linear modes on the $AdS_2$ background. However, those kinds of phase diagrams are expected to be realised by choosing specific models within their corresponding parameter spaces given above. Indeed, case 1 and case 7 are explicitly constructed in section~\ref{sec:afphase} and section~\ref{sec:scphase}, respectively. Nevertheless, to obtain a particular type of phase diagram quantitatively, one should solve the coupled equations for the fully back-reacted geometry and compare its free energy to the normal unbroken phase.

\subsection{Towards cuprates and strange metals}
\begin{figure}[ht!]
\begin{center}
\includegraphics[width=.58\textwidth]{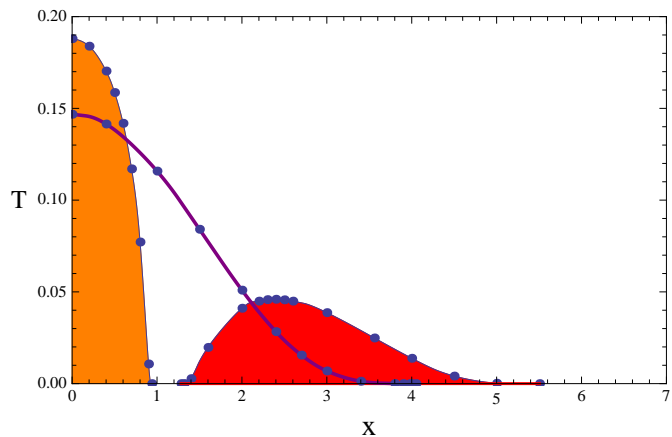}
\caption{ A typical phase diagram for our holographic theory. This phase diagram is the combination of figure~\protect\ref{fig:scphase}, figure~\protect\ref{fig:caseaf} and figure~\protect\ref{fig:striptx}. The orange region denotes the antiferromagnetic phase and the red region denotes the superconducting phase. The region below the purple curve indicates the possible striped phase associated with CDWs. The white region above the purple curve represents the normal unbroken phase.}
\label{fig:fullphase}
\end{center}
\end{figure}

We specify to the theory parameters that can arrive at the phase diagram we are interested. As one can see our study so far provides several ingredients that are necessary for a more realistic model, including a superconducting dome, a corner of antiferromagnetic phase and a region with striped phase or checkerboard phase. By putting them together we arrive at the global phase diagram shown in the temperature-doping plane in figure~\ref{fig:fullphase}. To our knowledge this is the first determination including all three ingredients. In particular, the critical temperatures for all broken phases are comparable in figure~\ref{fig:fullphase}. This figure is very reminiscent of the phase diagram of high-$T_c$ cuprates in figure~\ref{fig:cuprates}.

However, that is not the whole story. There is an overlap between the homogeneous AF phase and the striped phase. Something additional would happen within this overlap region. One possibility is that two orders will compete with each  other and only one of them can survive, resulting in a first order phase transition. The other possibility is that both orders can coexist and there is a new phase with both AF order and striped order. To see this more clearly, we can study the possible spatially modulated static mode in the spectrum of fluctuations around the condensed phase of antiferromagnetism~\eqref{afansatz}.  If no such kind of mode can exist, the homogeneous AF background is stable and one should compare the free energy between AF phase and striped phase to find the real phase boundary between them. There is a first order phase transition occurring at this phase boundary. In contrast, if one can find that kind of zero mode, there might be a possibility that the AF order is spatially modulated and can coexist with CDWs. In a very simple case, the striped zero mode around the condensed phase of the antiferromagnetism is analysed in appendix~\ref{app:stripedonaf}. It is shown that such zero mode can be developed at some temperature range lower than the critical temperature $T_N$. However, it does not guarantee that a striped AF order can really appear in the global phase diagram.\footnote{The coexisting phase would be thermodynamically stable or unstable, depending on the nonlinear details of the theory~\cite{BauandCai,NishidaLi:2014,Nie:2014qma}. In principle, we can add higher order corrections to the theory without changing our perturbative analysis, but the thermodynamics (free energy) of the coexisting phase can change significantly.} On should construct this coexisting phase and compare its free energy to other phases. All that can be done by solving nonlinear PDEs which is numerically challenging and is very sensitive to the exact form of the theory we are considering. Similar story will  also hold for the superconducting case. We shall study the competition and coexistence of different order parameters in the future.

\begin{figure}[ht!]
\begin{center}
\includegraphics[width=.55\textwidth]{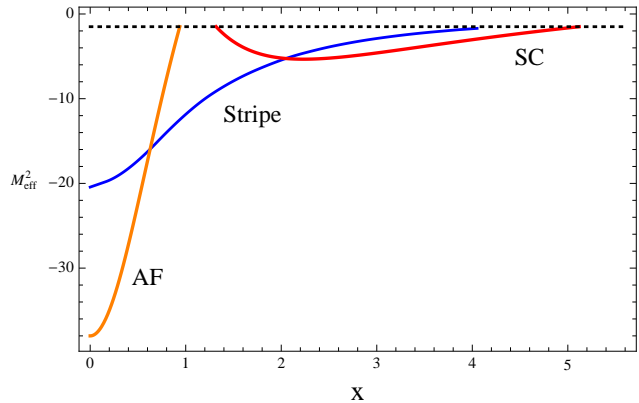}
\caption{ The effective mass squared for superconducting phase (red), antiferromagnetic phase (orange) and striped phase at $k^*$ (blue) on the $AdS_2$ background. The horizontal dotted line denotes the $AdS_2$ BF bound. The theory parameters we choose are the same as the previous figure~\protect\ref{fig:fullphase}.}
\label{fig:masseff}
\end{center}
\end{figure}

Nevertheless, we can obtain some intuition by comparing the effective mass squared for each broken phase on the $AdS_2$ background. From figure~\ref{fig:masseff} one can see that at small doping region on the far left, the effective mass squared of the AF phase (orange line) is much smaller than the one of striped phase (blue line), which suggests that the mode of antiferromagnetism around the ground state of the normal phase is much more unstable and thus the AF phase would first develop and dominate the phase diagram at least when the doping is sufficiently small. In contrast, when $\textbf{x}$ is large the SC phase  dominates the phase diagram. Therefore, we can propose a simple phase diagram shown before in figure~\ref{fig:holoydiagram}. To arrival at that diagram we have assumed that all broken solutions continue down to $T=0$ without sprouting additional branches, which seems reasonable since all broken solutions arise as a consequence of breaking the BF bound of the ground state of the normal, unbroken phase. We also assume that either the possible coexisting phases will not develop, or they can develop but are not thermodynamically favoured.

\begin{figure}[ht!]
\begin{center}
\includegraphics[width=.55\textwidth]{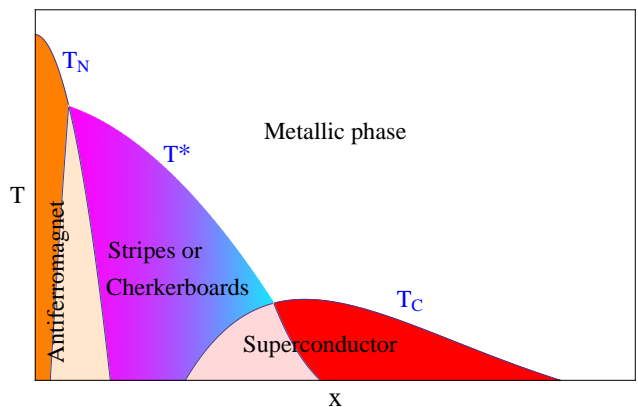}
\caption{ A plausible phase diagram in the temparature-doping plane after including the possible spatially modulated AF order and SC order. The light orange region on the left denotes the phase with spatially modulated AF condensate, i.e., an AF+CDW phase. The light red region on the right corresponds to the phase with striped SC condensate, namely, a SC+CDW phase.}
\label{fig:holoydiagramfull}
\end{center}
\end{figure}

The key features in the phase diagram of figure~\ref{fig:holoydiagram} are as follows. At low doping level, the phase diagram is captured by the AF phase. By increasing doping $\textbf{x}$, some kind of spatially modulated phase becomes dominant. One particularly interesting point is that there is a dome-shaped superconducting region in the $(T,\textbf{x})$ plane. For sufficiently high doping, the superconductivity can be destroyed and be replaced by the normal phase. Notice that even the phase with striped superconducting order or antiferromagnetic order develops, the main structure of the phase diagram will not change a lot. We still have a superconducting dome and a corner region of antiferromagnetism but with spatially modulated condensates. A schematic phase diagram is shown in figure~\ref{fig:holoydiagramfull}. Compared with figure~\ref{fig:holoydiagram}, there are two new phases appearing in the $(T,\textbf{x})$ plane. The left one near the homogeneous AF phase is an inhomogeneous AF phase known as AF+CDW phase in which both the AF condensate and the charge density are spatially modulated. Very similarly, the right one near the homogeneous SC region is an inhomogeneous SC phase, namely, the SC+CDW phase with spatially modulated SC condensate and charge density wave.

To establish the exact phase diagram of a given theory, one should know, in principle, all of the black brane solutions that exist at low temperatures, including the full back-reacted spatially modulated black branes we mentioned above. Carrying this out in full detail is thus potentially very involved. In the present paper, we do not address this issue as it is not very essential for the key features of the phase diagram and too technically involved.  It would be helpful to construct a full phase diagram by finding all broken phases numerically in the future.

Besides cuprates, iron-based superconductors have become the second high temperature superconductor families. Iron-based superconductors share many characteristics with the cuprates. Both of them are layered materials and both have similar phase diagrams: superconductivity only emerges after doping an antiferromagnetic parent state.
However, the parent state of the iron-based superconductors is a conductor, whereas the parent state in cuprates is an antiferromagnetic insulator. Hole doped cuprates exhibit the ``pseudogap" region which is absent in iron-based superconductors. By turning off the pseudo-scalar $\alpha$ as well as Chern-Simons couplings, the striped phase will disappear. We can then obtain a system in which superconducting dome locates near an AF phase with possible coexistence of the superconducting and antiferromagnetic orders at certain doping levels~\cite{Pratt:2009}, which is reminiscent of the phase diagram of iron-based superconductors. Our theory allows for systematic improvements of the applied approximations.  Actually, as it has been shown in subsection~\ref{alldiagrams}, our theory allows phase diagrams that are significantly different from figure~\ref{fig:holoydiagram}. Nevertheless, we pay much more attention to the phase diagram as figure~\ref{fig:holoydiagram} and we shall study other kinds of phase diagrams in more details elsewhere.


\section{Conclusion and Outlook}
\label{sec:outlook}

Progress in applying the holographic duality to condensed matter physics has recently moved towards more concrete holographic theory building for strongly correlated systems.
As a step towards realising a holographic theory of the doped high-temperature superconductors, we analysed a very general bulk theory which has the ability to produce interesting phase diagrams as functions  of temperature $T$ and doping-like parameter $\textbf{x}$. All couplings specifying the action are a priori general, provided that they admit an asymptotically expansion as in~\eqref{vpossi}. The theory exhibits a rich phase structure: depending on the values of the temperature and the doping the boundary system can be in superconducting phase, antiferromagnetic phase, normal metallic phase or spatially modulated phase with striped order or checkerboard order.

The sharp transition as well as its position for each broken phase in the global $(T, \textbf{x})$ phase diagram can be found by changing theory parameters. We classify possible phase diagrams for each broken phase that can be realised in our theory in subsection~\ref{alldiagrams}. One particularly interesting point is that the existence of a superconductivity dome is a rather general feature in our theory provided that the theory parameters are within the region~\eqref{sccondition}. We are aware of one holographic theory that can give a superconducting dome with respect to a theory parameter, i.e., the magnitude of the translational symmetry breaking~\cite{Baggioli:2015zoa}.\footnote{The first attempt towards finding a superconducting dome from holography has been done in the Abelian-Higgs model~\cite{Hartnoll:2008vx} by introducing a modulating chemical potential~\cite{Ganguli:2013oya}. It was shown that there is a critical modulation wave number above which the critical temperature will vanish or nearly vanish.} In the previous work,  the height of the dome was very limited: the critical temperature was very small (of the order of $10^{-8}$ in units of charge density) and thus not accessible through stable numerical analysis. In our setup, it is very natural to obtain a superconducting dome with comparable transition temperature to other phases, see figure~\ref{fig:fullphase}.

Most of our attention is to construct a $(T, \textbf{x})$ phase diagram that is reminiscent of  real doped high-temperature superconductors as shown schematically in figure~\ref{fig:cuprates}.
To find the desirable patten of $(T, \textbf{x})$ phase diagram, including a superconducting dome, a corner of antiferromagnetic phase as well as a spatially modulated region between them,  the instability of the extremal AdS-RN black brane plays an important role. The near horizon geometry features an $AdS_2$ factor. Therefore, to trigger the phase transition to a certain kind of broken phase can be related to the violation of the BF bound of the corresponding dual IR operator. There might be another black brane background that describes the normal phase much better than the AdS-RN we are considering. One can then, in principle, perform similar analysis on that background, although it is expected that it is more involved. The holographic phase diagram in figure~\ref{fig:holoydiagram} (as well as figure~\ref{fig:holoydiagramfull}) can be a step towards realising the phase diagram of high-temperature superconductivity. Nevertheless, there are still many features we need to identify and study in the future.

It should be very interesting to study the transport properties, especially the conductivity. We leave this to future work, as future improvements of the framework as mechanisms for momentum dissipation (see, for example,~\cite{masssivegrav,Donos:2013eha,Andrade:2013gsa}) were not included here.

There are several phase transitions in the theory. First there are three quantum phase transitions as one changes the doping parameter, one to go from AF to striped phases, another to go from striped to SC phases and another from SC phase to metal. As they are all triggered by violations of the BF bound, they are in the BKT universality class~\cite{son,karch,liu,matti}.
Then there are several finite temperature phase transitions from the high temperature  metallic phase to the AF, striped or SC phase. The theory can in principle admit mixed phases in which more than one orders can coexist, for example, an AF+SC state. These can be first or second order phase transition as a function of the detailed form of the holographic theory.

 It would be interesting to compare our phase diagram via holographic construction with other theoretical approaches. For example, by analysing SO(5) quantum nonlinear $\sigma$ model, the author of~\cite{zhangso5} has classified four types of transitions from the AF phase to the SC phase, see phase diagrams in figure~1 of~\cite{zhangso5}. The topology of the phase diagram depends on the strength of the quantum fluctuation. Similar kind of phase diagrams can be in principle realised in our holographic theory with a proper choice of parameters. Furthermore, there is a new ingredient, i.e., striped order in our holographic phase diagrams. Another interesting theory is the phase string theory~\cite{weng:2007} for doped antiferromagnets, which is described by two bosonic degrees of freedom, say the spinon and the holon, rather than a quasiparticle theory of interacting electrons. In the phase string picture, the doping is the hole concentration with the hole number conserved. There are two emergent U(1) gauge fields and a possible external electromagnetic field. Therefore, in a sense, the phase string theory shares some similarity with our holographic theory. The global phase diagram in the minimal phase string model can be found in figure~31 of~\cite{weng:2007}. The phase diagram covers many interesting regimes of the cuprate high-temperature superconductors. However, it needs to be modified to describe physics for the very underdoped regime as well as the overdoped regime. Our holographic theory, in some sense, gives the expected phase behaviours in those two doping regions.

As we can see, the precise form of the global phase diagram depends on the nonlinear details of the theory: the phase boundary can be second order or first order and  there also might be phases with coexisting orders. Coupling of the SC mode to other degrees of freedom such as AF excitation and striped order is expected to give rise to a rich phase diagram of physical phenomena.
It would certainly be interesting to study whether one can obtain a thermodynamic stable solution in which the superconducting condensate becomes to be spatially modulated spontaneously, i.e., striped superconductors. Similarly, there also might be a phase with the striped antiferromagnetic order. With a proper choice of parameters, the theory should admit coexisting phase with AF and SC condensates. It will be interesting to construct the superconducting-antiferromagnetic-superconducting junction and the superconducting vortices with antiferromagnetic cores by virtue of such kind of mixing phase and compare the holographic results, for example, with the SO(5) theory of high-temperature superconductivity~\cite{demler2004}. Finally, one of the remaining questions is how the details of the coupling functions are mapped to the dual boundary theory, which could shed light on the quest for the origin of high-temperature superconductors.

The class of effective holographic theories we analysed here however is expected upon improvement to provide phase diagrams for a larger variety of strongly correlated systems. There are several ingredients however that should be included in order to make it more realistic:
\begin{itemize}

\item A regular strange-metal phase should be included by modifying the effective action to include the most relevant scalar operator along the lines of \cite{Charmousis:2010zz}. In such a case there are several non-fermi liquid ground states possible to choose from.

\item Agents for momentum dissipation should be added to make conductivity realistic. Except for the striped phase there is now ballistic conductivity  in the rest of phases as there is translation invariance and finite density. Ways to do this in holographic theories are known and they can be incorporated.

\item The two U(1) charge densities were taken to be independently conserved. This will cease to be true if charge of one kind could be converted to the other. This can be modeled by giving a field dependent mass to a linear combination of the two U(1) vector fields.

\end{itemize}

A further problem that can be studied along similar lines is how a layered structure of strongly coupled systems can affect the competition of phases using the formalism that was developed in~\cite{jo}.

Quark matter or QCD matter is also believed to display a rich phase structure at finite temperature and baryon density~\cite{Fukushima:2010bq}. The techniques of this paper could be also applicable in such a context. It was already shown in~\cite{Alho:2012mh} that in holographic models in the class of V-QCD a superconducting dome can appear in their phase diagrams for the axial global symmetry. It would be interesting to explore further exotic phases in this context.

\addcontentsline{toc}{section}{Acknowledgments}
\acknowledgments\label{ACKNOWL}

We would like to thank  M.~Ammon, J.~Erdmenger,  Z.~Y.~Weng, H.~Yao and J. Zaanen for helpful discussions.

This work was supported in part by European Union's Seventh Framework Programme under grant agreements (FP7-REGPOT-2012-2013-1) no 316165, the EU-Greece program ``Thales" MIS 375734 and was also co-financed by the European Union (European Social Fund, ESF) and Greek national funds through the Operational Program ``Education and Lifelong Learning" of the National Strategic Reference Framework (NSRF) under ``Funding of proposals that have received a positive evaluation in the 3rd and 4th Call of ERC Grant Schemes". LL is also grateful to KITPC for its hospitality and its partial support during the completion of this work.

\newpage
\appendix
\renewcommand{\theequation}{\thesection.\arabic{equation}}
\addcontentsline{toc}{section}{Appendix}
\section*{Appendices}


\section{{Generalised Effective Holographic Superconductor Theory}}\label{app:stuck}

The holographic superconducting phase was first constructed by the Abelian-Higgs model~\cite{Hartnoll:2008vx} in which a complex scalar $\psi$ is minimally coupled with a U(1) gauge field in the bulk. The key ingredient is that the U(1) gauge symmetry can be spontaneously broken below a critical temperature via a charged scalar condensate slightly outside the event horizon~\cite{Gubser:2008px}.
What we do here is to write the most general two-derivative holographic action that may break U(1) spontaneously. A mild generalisation was presented in~\cite{Franco:2009yz}, and the most general action in~\cite{gk2}.

To illustrate this case, we consider a holographic theory with a complex scalar $\psi$ that is charged under the U(1) gauge symmetry,
\begin{equation}
\label{modelchi}
\mathcal{L}_\psi^0=-\frac{G(\psi^*\psi)}{2}|(\partial_\mu-iqA_\mu)\psi|^2-V(\psi^*\psi)-\frac{Z(\psi^*\psi)}{4}H^2,
\end{equation}
with $H=d A$. We can innocuously rewrite this theory by changing variables to $\psi=|\psi|e^{i\theta}$, then the theory becomes
\begin{equation}
\label{modelchi}
\mathcal{L}_\psi^0=-\frac{G(|\psi|^2)}{2}(\partial_\mu|\psi|)^2-\frac{G(|\psi|^2)}{2}|\psi|^2(\partial_\mu\theta-q A_\mu)^2-V(|\psi|^2)-\frac{Z(|\psi|^2)}{4}H^2,
\end{equation}
where $|\psi|$ and $\theta$ are real scalars. The gauge symmetry becomes $A_\mu\rightarrow A_\mu+\partial_\mu\beta/q$ together with $\theta\rightarrow \theta+\beta$. Remember that $|\psi|$  is the absolute value of a complex field and hence positive definite.

So far we have done nothing but just rewrite the theory using different variables. Nevertheless, it is straightforward to generalise the theory as broad as possible while preserving gauge invariance. Notice that one can change variable $|\psi|\rightarrow\chi$ such that the kinetic term of the new scalar $\chi$ takes the standard form, $-\frac{1}{2}(\partial_\mu\chi)^2$. Therefore, the generalised one reads
\begin{equation}
\label{modelchi}
\mathcal{L}_\chi=-\frac{1}{2}(\partial_\mu\chi)^2-\mathcal{F}(\chi)(\partial_\mu\theta-q A_\mu)^2-V(\chi)-\frac{Z(\chi)}{4}H^2,
\end{equation}
where $\mathcal{F}$, $V$ and $Z$ are three arbitrary functions of real scalar $\chi$ that takes in general any real value. One demands $\mathcal{F}$  and $Z$ to be positive, at least for a solution one is considering,  to ensure positivity of the kinetic term for $\theta$ and $A_\mu$, respectively. We will take this generalised theory in our present study to describe holographically a superconductor phase transition.


\section{Equations of Motion \label{app:eoms}}

In this appendix we provide the detailed equations of motion under the ansatz~\eqref{fullansatz}, which read
\begin{equation}\label{appalph}
\theta'=0,
\end{equation}
\begin{equation}\label{appaxion}
\frac{1}{\sqrt{EDC^2}}\left(\sqrt{\frac{DC^2}{E}}\alpha'\right)'-\partial_\alpha V_{int}+\frac{\partial_\alpha Z_A A_t'^2+\partial_\alpha Z_B B_t'^2+2\partial_\alpha Z_{AB} A_t' B_t'}{2ED}=0,
\end{equation}
\begin{equation}\label{appchi}
\begin{split}
\frac{1}{\sqrt{EDC^2}}\left(\sqrt{\frac{DC^2}{E}}\chi'\right)'-\partial_\chi V_{int}+&\frac{(q_A A_t+q_B B_t)^2}{D}\partial_\chi \mathcal{F}\\
+&\frac{\partial_\chi Z_A A_t'^2+\partial_\chi Z_B B_t'^2+2\partial_\chi Z_{AB} A_t' B_t'}{2ED}=0,
\end{split}
\end{equation}
\begin{equation}\label{appphi}
\frac{1}{\sqrt{EDC^2}}\left(\sqrt{\frac{DC^2}{E}}\phi'\right)'-2\partial_\Phi V_{int}\phi+\frac{\partial_\Phi Z_A A_t'^2+\partial_\Phi Z_B B_t'^2+2\partial_\Phi Z_{AB} A_t' B_t'}{ED}\phi=0,
\end{equation}
\begin{equation}\label{appA}
\left(Z_A\sqrt{\frac{C^2}{ED}}A_t'\right)'+\left(Z_{AB}\sqrt{\frac{C^2}{ED}}B_t'\right)'=\sqrt{\frac{EC^2}{D}} 2\mathcal{F}(\chi)q_A(q_A A_t+q_B B_t),
\end{equation}
\begin{equation}\label{appB}
\left(Z_B\sqrt{\frac{C^2}{ED}}B_t'\right)'+\left(Z_{AB}\sqrt{\frac{C^2}{ED}}A_t'\right)'=\sqrt{\frac{EC^2}{D}} 2\mathcal{F}(\chi)q_B(q_A A_t+q_B B_t),
\end{equation}
\begin{eqnarray}\label{appmetric1}
\nonumber\frac{D'C'}{DC}+\frac{1}{2}\frac{C'^2}{C^2}+\frac{Z_A A_t'^2+2Z_{AB}A_t' B_t'+Z_B B_t'^2}{2D}
-\frac{1}{2}(\phi'^2+\chi'^2+\alpha'^2)\\-\frac{ \mathcal{F}(q_A A_t+q_B B_t)^2}{D}E+ V_{int}E-\frac{6}{L^2}E=0,
\end{eqnarray}
\begin{eqnarray}\label{appmetric2}
\frac{C''}{C}-\frac{1}{2}\left(\frac{C'}{C}+\frac{D'}{D}+\frac{E'}{E}\right)\frac{C'}{C}+\frac{1}{2}(\phi'^2+\chi'^2+\alpha'^2)+\frac{\mathcal{F}(q_A A_t+q_B B_t)^2}{D}E=0,
\end{eqnarray}
\begin{eqnarray}\label{appmetric3}
\nonumber2\frac{D''}{D}-2\frac{C''}{C}+\left(\frac{C'}{C}-\frac{D'}{D}-\frac{E'}{E}\right)\frac{D'}{D}+\frac{C' E'}{C E}-\frac{4 \mathcal{F} (q_A A_t+q_B B_t)^2}{D}E\\
-\frac{2(Z_A A_t'^2+2Z_{AB}A_t' B_t'+Z_B B_t'^2)}{D}=0,
\end{eqnarray}
where we use primes to denote radial derivatives. Note that only two of the last three equations are independent.

Making use of equations~\eqref{appA},~\eqref{appB} and~\eqref{appmetric3}, we can obtain the following conserved quantity
\begin{equation}\label{appconserved}
Q=\frac{C}{\sqrt{DE}}\left[Z_A A_t A_t'+Z_B B_t B_t'+Z_{AB}(A_t B_t)'-C\left(\frac{D}{C}\right)'\right],
\end{equation}
which can be used to connect IR to boundary data.
If there is a black brane horizon at $r=r_h$, then the temperature and entropy density are given by
\begin{eqnarray}
T=\frac{1}{4\pi}\sqrt{\frac{d_h}{e_h}},\quad s=\frac{2\pi}{\kappa^2}C(r_h),
\end{eqnarray}
where $D=d_h(r-r_h)+\cdots$ and $E=e_h/(r-r_h)+\cdots$ near the horizon. In order for $A_t\,dt$ and $B_t\,dt$ to be well defined at the horizon as a one form, $A_t(r_h)$ and $B_t(r_h)$ have to be zero with their first derivatives finite. As a consequence, it is straightforward to check that
\begin{eqnarray}
Q=-2\kappa^2 T s.
\end{eqnarray}
Therefore $Q=0$ signals extremity, which implies that either there is no horizon, or that if there is one it must have $T s=0$.


\section{Antiferromagnetism  and Spin Waves \label{app:afspin}}
In the low-energy limit, spin rotations should  be thought as a global SU(2) symmetry, and then antiferromagnetic ordering corresponds to a spontaneous breaking from SU(2) to U(1) which corresponds to rotations about a single axis.   Although the background value of spin density is vanishing, there is a staggered spin order parameter $\phi^a$ transforming as a triplet under spin rotations. To construct the gravity dual of AF phase, one needs to introduce a real scalar $\Phi^a$ which transforms as a triplet under SU(2) in the bulk. A bulk solution with vanishing SU(2) gauge fields but with a normalisable $\Phi^a$ in one direction is then holographically dual to an AF phase or spin density wave phase in the boundary theory~\cite{Iqbal:2010eh}.

In this appendix we will briefly show that the gravity description in terms of the spontaneous breaking of SU(2) symmetry is reminiscent of the N\'eel phase in a spin system and can display the associated spin wave with linear dispersion. These can be done by studying the linear perturbations around the background of symmetry broken phase. To begin with, we will explicitly work out that the linear fluctuations of SU(2) gauge field and triplet scalar in two symmetry breaking directions form a closed system even one considers Einstein's equations.

\subsection{Linearised equations of motion}
We start from the theory~\eqref{action1} and turn on the most general perturbations around the background~\eqref{fullansatz},
\begin{eqnarray}\label{afmpertu}
\begin{split}
ds^2=g_{\mu\nu}(r) dx^\mu dx^\nu+\lambda h_{\mu\nu} dx^\mu dx^\nu,\\
g_\mu^a=0+\lambda f_\mu^a,\quad\quad \Phi=(0+\lambda\varphi_1, 0+\lambda\varphi_2, \phi(r)+\lambda \varphi),\\
A=A_t(r)dt+\lambda a_\mu dx^\mu,\quad B=B_t(r)dt+\lambda b_\mu dx^\mu,\\
\quad \chi=\chi(r)+\lambda \upsilon,\quad \theta=\theta(r)+\lambda \vartheta,\quad \alpha=\alpha(r)+\lambda w,
\end{split}
\end{eqnarray}
where $\lambda$ is a formal expansion parameter. At the beginning, all perturbative fields can depend on general coordinates $(t, r, x, y)$.

Since we are interested in the physics of AF phase, we focus on the equations of motion for perturbations of SU(2) gauge field and the triplet scalar. The linearised equations for $f_\mu^a$ can be obtained from~\eqref{eomgg}, which read
\begin{eqnarray}\label{eomff}
\begin{split}
&\nabla^\nu f_{\nu\mu}^1-\phi^2 f_\mu^1=+\varphi_2 \nabla_\mu \phi-\phi \nabla_\mu \varphi_2,\\
&\nabla^\nu f_{\nu\mu}^2-\phi^2 f_\mu^2=-\varphi_1 \nabla_\mu \phi+\phi \nabla_\mu \varphi_1,
\end{split}
\end{eqnarray}
and
\begin{equation}
\nabla^\nu f_{\nu\mu}^3=0,
\end{equation}
where we have set the coupling constant $Z_G=1$ for simplicity and defined the field strength for $f_\mu^a$ as $f_{\mu\nu}^a=\partial_\mu f_\nu^a-\partial_\nu f_\mu^a$. The covariant derivative $\nabla$ corresponds to the background metric, i.e., $\nabla_\beta g_{\mu\nu}=0$. Note that due to the symmetry breaking in $a=1, 2$ direction, the gauge fields $f_\mu^1$ and $f_\mu^2$ obtain an effective mass  with its value set by the condensate $\phi$ in the third direction, while $f_\mu^3$ is still massless since the symmetry in this direction is unbroken.

The equations of motion for $\varphi_1$ and $\varphi_2$ can be derived from~\eqref{eomphi},
\begin{eqnarray}\label{eomvpsi}
\begin{split}
&\nabla^\mu(\nabla_\mu \varphi_1+\phi f_\mu^2)+f_\mu^2\nabla^\mu \phi-M^2_{eff}\varphi_1=0,\\
&\nabla^\mu(\nabla_\mu \varphi_2+\phi f_\mu^1)+f_\mu^1\nabla^\mu \phi-M^2_{eff}\varphi_2=0,\\
\end{split}
\end{eqnarray}
where the effective mass $M^2_{eff}$ is provided by background fields. The precise form of $M^2_{eff}$ is not important here. Note that under the ansatz~\eqref{fullansatz} the equation of motion for $\phi$ in terms of $M^2_{eff}$ can be written as
\begin{equation}\label{eombkphi}
\nabla^\mu\nabla_\mu\phi-M^2_{eff}\phi=0.
\end{equation}

Thanks to the SU(2) symmetry which demands each coupling in~\eqref{action1} to be analytic functions of $\Phi^a\Phi^a$,  after a simple analysis one can find that $f_\mu^1,f_\mu^2, \varphi_1, \varphi_2$ do not enter into other equations of motion at linear order.  Hence, as long as one knows the profile of background condensate $\phi$, the equations of ($f_\mu^1,f_\mu^2, \varphi_1, \varphi_2$) can be solved. Actually, equations~\eqref{eomff} and~\eqref{eomvpsi} can be rewritten as a much more compact form. We define two new functions $\Pi^1$ and $\Pi^2$ by
\begin{equation}
\varphi_2=\Pi^1 \phi, \quad  \varphi_1=-\Pi^2 \phi.
\end{equation}
Then~\eqref{eomff} and~\eqref{eomvpsi} in terms of $\Pi^1$ and $\Pi^2$ become
\begin{eqnarray}
\label{spineom1}\nabla^\nu f_{\nu\mu}^1+\phi^2 (\nabla_\mu \Pi^1-f_\mu^1)& = & 0, \\
\label{spineom2}\nabla^\nu[\phi^2(\nabla_\mu \Pi^1-f_\mu^1)] &= & 0,
\end{eqnarray}
and
\begin{eqnarray}
\nabla^\nu[\phi^2(\nabla_\mu \Pi^2-f_\mu^2)] &= & 0,\\
\nabla^\nu f_{\nu\mu}^2+\phi^2 (\nabla_\mu \Pi^2-f_\mu^2)& = & 0.
\end{eqnarray}
To obtain above equations we have used the background equation~\eqref{eombkphi}.

It is obviously that two sets of perturbations decouple with each other and have the same form. Hence we can only focus on one of them and drop the upper index for simplicity. They are the full set of equations for this system. Actually, the perturbation equations are the same as the one discussed in~\cite{Iqbal:2010eh}, in which $\Pi^1$ and $\Pi^2$ are known as bulk Goldstone fields. In the next subsection, following~\cite{Iqbal:2010eh}, we will briefly show that the system displays spin waves and their velocity has the form as expected from the standard theory for a quantum antiferromagnet.\footnote{One can refer to~\cite{Iqbal:2010eh} for more details.} We choose the symmetry breaking background as~\eqref{fullansatz}, where the AF order parameter dual to $\phi$ acquires a nonzero VEV spontaneously.\footnote{We only consider the standard quantisation for $\phi$, which means that for large $r$ the normalisable solution behaves as $\phi\sim r^{-\Delta^{\phi}_+}$ with $\Delta^{\phi}_+=\frac{3+\sqrt{9+4M^2}}{2}$.} We restrict ourselves to the finite temperature case in which $g_{tt}=-D$ ($g_{rr}=E$) has a simple zero (pole) at the horizon $r=r_h$.

\subsection{Hydrodynamic limit and spin wave velocity}
To do hydrodynamics analysis, for definiteness and without loss of generality, we demands that all fluctuations have a small frequency $\omega$ and momentum $k$ in the $x$ direction. More precisely, we consider the fluctuations as follows,
\begin{equation}
\Pi=\pi(r)e^{-i\omega t+i kx},\quad f_t=f_t(r)e^{-i\omega t+i kx},\quad f_x=f_x(r)e^{-i\omega t+i kx}.
\end{equation}
As $\omega=k=0$, there is a simple solution to the equations~\eqref{spineom1} and~\eqref{spineom2}: a constant Goldstone field $\Pi(r)=\pi_0$ with vanishing gauge profile $f_\mu=0$. Note that it is a normalisable solution existing in the limit $\omega\rightarrow 0$, thus a gapless Goldstone mode. Actually, this solution is the consequence of the global SU(2) gauge rotation of the background order parameter $\Phi=(0,0,\phi)$. To find general Goldstone mode which is a normalisable solution to~\eqref{spineom1} and~\eqref{spineom2} with ingoing or regular condition at the horizon, one makes an expansion in powers of $\omega$ and $k$ and then finds that
\begin{eqnarray}
&&\Pi(r) =  \pi_0+\omega^2\pi_0\, \mathcal{C}^t(r)+k^2\pi_0\, \mathcal{C}^x(r), \\
&&f_t  =  -i\omega\pi_0(1-\zeta_t(r)),\quad f_x  = -ik\pi_0(1-\zeta_x(r)).
\end{eqnarray}
Here $\zeta_t$ and $\zeta_x$ are solutions of
\begin{eqnarray}
\partial_r(\sqrt{-g}g^{rr}g^{tt}\partial_r \zeta_t)-\sqrt{-g}g^{tt}\phi^2\zeta_t& = & 0, \\
\partial_r(\sqrt{-g}g^{rr}g^{xx}\partial_r \zeta_x)-\sqrt{-g}g^{xx}\phi^2\zeta_x& = & 0,
\end{eqnarray}
with ingoing conditions at the horizon and Dirichlet conditions $\zeta_t|_{r\rightarrow\infty}=\zeta_x|_{r\rightarrow\infty}=1$ in the UV to make $f_t$ and $f_x$ be normalisable.
The equations of $\mathcal{C}^t$ and $\mathcal{C}^x$ come from~\eqref{spineom1}, reading
\begin{eqnarray}
&\partial_r\alpha^t=0,\quad  \alpha^t= -\sqrt{-g}\phi^2 g^{rr}\partial_r \mathcal{C}^t-\sqrt{-g}g^{rr}g^{tt}\partial_r\zeta_r, \\
&\partial_r\alpha^x=0,\quad  \alpha^x= -\sqrt{-g}\phi^2 g^{rr}\partial_r \mathcal{C}^x+\sqrt{-g}g^{rr}g^{xx}\partial_r\zeta_x.
\end{eqnarray}
The last constraint is the $r$ component of the gauge equations~\eqref{spineom1} from which one finds the relation
\begin{equation}
\omega^2=v_s^2\, k^2,\quad v_s^2=\alpha_x/\alpha_t.
\end{equation}
Thus the dynamical bulk equations of motion admit Goldstone mode, i.e., normalisable ingoing solution, only when $\omega$ and $k$ obey a linear dispersion relation with the spin wave velocity $v_s$.
It has been shown that the constants $\alpha_t$ and $\alpha_x$ are directly related to the $tt$ and $xx$ components of the retarded function $G^R_{\mu\nu}$ of spin current $j_s^\mu$ along a symmetry breaking direction, i.e.,
\begin{equation}
\alpha_t=-G^R_{tt}(\omega=0,k=0),\quad \alpha_x=G^R_{xx}(\omega=0,k=0).
\end{equation}
By further recognising
\begin{equation}
\rho_s=G^R_{xx}(\omega=0,k=0), \quad \chi_{\bot}=-G^R_{tt}(\omega=0,k=0),
\end{equation}
with $\rho_s$ the spin stiffness and $\chi_{\bot}$ the transverse magnetic susceptibility, one finds that
\begin{equation}
v_s=\sqrt{\frac{G^R_{xx}(\omega=0,k=0)}{-G^R_{tt}(\omega=0,k=0)}}=\sqrt{\frac{\rho_s}{\chi_{\bot}}},
\end{equation}
which is  the expected expression for the spin velocity for a quantum antiferromagnet.


\section{Normalisable Static Modes by Shooting Method \label{app:doushot}}

In this appendix we will introduce the method we use to determine numerically the existence of normalisable static modes.

As mentioned in section~\ref{sec:spatial},  a nontrivial bulk solution with desirable boundary conditions at particular value of $k\neq 0$ means that a static spatially modulated mode exists, leading to a spatially modulated phase. The quantitative value of $k$ can be determined by the shooting method numerically. A typical shooting strategy is to impose boundary conditions on the one side and to check the boundary values on the other side. However, since the system in AdS spacetime typically has singularity at the horizon $r=r_h$ as well as the asymptotical boundary $r\rightarrow \infty$, it would generally develop  divergent mode in this single side shooting process.  To bypass such undesirable singular modes, one can use a double side shooting approach, in which one prepares a pair of solutions with desirable boundary conditions independently at each side and then connects them smoothly at a particular point in the interior. Similar method was adopted to look for the spatially modulated mode in a holographic d-wave superconductor~\cite{Krikun:2013iha}.

Specific to the system we are considering, i.e., the linear ODEs for four functions $\eta_J=(h, a_y, w, b_y)$, the desirable solution can be obtained by a linear superposition of four independent solutions $\xi^j_J, (j=1, 2, 3, 4)$ of the ODEs. The solution satisfying the IR boundary condition can be expressed as
\begin{equation}
\eta_{\text{ir}J}=\sum_{j=1}^4 s_j \xi_{\text{ir}J}^j,
\end{equation}
while the solution satisfying the UV boundary condition can be expressed as
\begin{equation}
\eta_{\text{uv}J}=\sum_{j=1}^4 t_j \xi_{\text{uv}J}^j,
\end{equation}
where we have used subscript $``\text{ir}"$ and $``\text{uv}"$ to highlight the boundary conditions satisfied by the solution, $s_j$ and $t_j$ are all constants. To make it more clear, the four sets of solutions $\xi_{\text{ir}J}^j$ are independent solutions that only satisfy the regular condition on the horizon, while $\xi_{\text{uv}J}^j$ are four sets of independent solutions that only satisfy the source free condition in the UV.

A smooth connection demands that the values of the functions and their first derivates should coincide at a particular interior point, say $r_0$. Thus we obtain eight constraints,
\begin{eqnarray}
\sum_{j=1}^4 s_j \xi_{\text{ir}J}^j(r_0)& = &\sum_{j=1}^4 t_j \xi_{\text{uv}J}^j(r_0), \\
\sum_{j=1}^4 s_j \partial_r\xi^{j}_{\text{ir}J}(r_0)& = &\sum_{j=1}^4 t_j \partial_r\xi_{\text{uv}J}^j(r_0).
\end{eqnarray}
To have nontrivial value for eight coefficients $(s_j, t_j)$, the Wronskian
\begin{equation}\label{appwroski}
W(r) = \begin{vmatrix}
\xi_{\text{ir}\textbf{1}}^1(r) & \dots & \xi_{\text{ir}\textbf{1}}^4(r) & \xi_{\text{uv}\textbf{1}}^1(r) &\dots &\xi_{\text{uv}\textbf{1}}^4(r) \\
\vdots & \ddots & \vdots & \vdots & \ddots & \vdots \\
\xi_{\text{ir}\textbf{4}}^1(r) & \dots & \xi_{\text{ir}\textbf{4}}^4(r) & \xi_{\text{uv}\textbf{4}}^1(r) &\dots & \xi_{\text{uv}\textbf{4}}^4(r)\\
\partial_r\xi_{\text{ir}\textbf{1}}^1(r) & \dots & \partial_r\xi_{\text{ir}\textbf{1}}^4(r) &\partial_r \xi_{\text{uv}\textbf{1}}^1(r) &\dots &\partial_r\xi_{\text{uv}\textbf{1}}^4(r) \\
\vdots & \ddots & \vdots & \vdots & \ddots & \vdots \\
\partial_r\xi_{\text{ir}\textbf{4}}^1(r) & \dots & \partial_r\xi_{\text{ir}\textbf{4}}^4(r) &\partial_r\xi_{\text{uv}\textbf{4}}^1(r) &\dots & \partial_r\xi_{\text{uv}\textbf{4}}^4(r)\\
\end{vmatrix}
\end{equation}
should be zero at $r=r_0$. Note that the vanishing of Wronskian does not depend on the choose of $r_0$ and independent solutions $\xi_{\text{ir}J}^j$ and $\xi_{\text{uv}J}^j$, as long as the normalisable solution we are looking for does exist at a given wave number.


\section{Analysis on Checkerboard Instabilities}\label{app:checkerboard}
In this appendix  we will show how one can obtain the perturbation equations for the checkerboard structure.

To simplify our analysis, we focus on the $AdS_2\times \mathbb{R}^2$ background~\eqref{ads2bk2} as an illustrative example. Substituting the general perturbations~\eqref{checkerbordpub} into the equations of motion and working at linear order of $\lambda$, we obtain six ODEs from two gauge field equations, three from gravity equations and one from the pseudo-scalar equation,
\begin{eqnarray}
6\partial_r(r^2 \partial_r w)-\left(\widetilde{m}_{(2)}^2+\frac{k_x^2+k_y^2}{r_h^2}\right)w-\frac{4\sqrt{3}(2n_1+n_2\textbf{x})}{\sqrt{1+\textbf{x}^2}}(\frac{k_x}{r_h}a_y-\frac{k_y}{r_h}a_x)\\ \nonumber
-\frac{4\sqrt{3}n_2}{\sqrt{1+\textbf{x}^2}}(\frac{k_x}{r_h}b_y-\frac{k_y}{r_h}b_x)& = & 0, \\
6\partial_r(r^2 \partial_r a_x)+\frac{2\sqrt{3}}{\sqrt{1+\textbf{x}^2}}h_{tx}'+\frac{4\sqrt{3}(2n_1+n_2\textbf{x})}{\sqrt{1+\textbf{x}^2}}\frac{k_y}{r_h}w-\frac{k_y^2}{r_h^2}a_x+\frac{k_x k_y}{r_h^2}a_y&=&0,\\
6\partial_r(r^2 \partial_r a_y)+\frac{2\sqrt{3}}{\sqrt{1+\textbf{x}^2}}h_{ty}'-\frac{4\sqrt{3}(2n_1+n_2\textbf{x})}{\sqrt{1+\textbf{x}^2}}\frac{k_x}{r_h}w+\frac{k_x k_y}{r_h^2}a_x-\frac{k_x^2}{r_h^2}a_y&=&0,\\
6\partial_r(r^2 \partial_r b_x)+\frac{2\sqrt{3}\textbf{x}}{\sqrt{1+\textbf{x}^2}}h_{tx}'+\frac{4\sqrt{3}n_2}{\sqrt{1+\textbf{x}^2}}\frac{k_y}{r_h}w-\frac{k_y^2}{r_h^2}b_x+\frac{k_x k_y}{r_h^2}b_y&=&0,\\
6\partial_r(r^2 \partial_r b_y)+\frac{2\sqrt{3}\textbf{x}}{\sqrt{1+\textbf{x}^2}}h_{ty}'-\frac{4\sqrt{3}n_2}{\sqrt{1+\textbf{x}^2}}\frac{k_x}{r_h}w+\frac{k_x k_y}{r_h^2}b_x-\frac{k_x^2}{r_h^2}b_y&=&0,\\
-3r^2 h_{tx}''-\frac{6\sqrt{3}r^2}{\sqrt{1+\textbf{x}^2}}a_x'-\frac{6\sqrt{3}\textbf{x}r^2}{\sqrt{1+\textbf{x}^2}}b_x'+\frac{1}{2}\frac{k_y^2}{r_h^2}h_{tx}-\frac{1}{2}\frac{k_x k_y}{r_h^2}h_{ty}&=&0,\\
-3r^2 h_{ty}''-\frac{6\sqrt{3}r^2}{\sqrt{1+\textbf{x}^2}}a_y'-\frac{6\sqrt{3}\textbf{x}r^2}{\sqrt{1+\textbf{x}^2}}b_y'-\frac{1}{2}\frac{k_x k_y}{r_h^2}h_{tx}+\frac{1}{2}\frac{k_x^2}{r_h^2}h_{ty}&=&0,\\
k_x a_x'+k_y a_y'+\frac{k_x}{\sqrt{3}\sqrt{1+\textbf{x}^2} r^2}h_{tx}+\frac{k_y}{\sqrt{3}\sqrt{1+\textbf{x}^2} r^2}h_{ty}&=&0,\\
k_x b_x'+k_y b_y'+\frac{k_x \textbf{x}}{\sqrt{3}\sqrt{1+\textbf{x}^2} r^2}h_{tx}+\frac{k_y \textbf{x}}{\sqrt{3}\sqrt{1+\textbf{x}^2} r^2}h_{ty}&=&0,\\
-\frac{k_x}{2}h_{tx}'-\frac{k_y}{2}h_{ty}'+\frac{k_x}{r}h_{tx}+\frac{k_y}{r}h_{ty}&=&0,
\end{eqnarray}
where $\widetilde{m}_{(2)}^2$ is defined in~\eqref{ads2axi}, and two Chern-Simons coupling constants $n_1, n_2$ are given in~\eqref{vpossi}.

The last three equations are first order and therefore can be thought of as constraint equations, from which we obtain three constraints,
\begin{eqnarray}
k_x h_{tx}+k_y h_{ty} & =& c\, r^2,\\
k_x a_x+k_y a_y& = & a+\frac{c}{\sqrt{3}\sqrt{1+\textbf{x}^2}},\\
k_x b_x+k_y b_y& = &b+ \frac{c\,\textbf{x}}{\sqrt{3}\sqrt{1+\textbf{x}^2}},
\end{eqnarray}
with $a, b, c$ three integration constants. By using those three constraints, one can easily check that the second order equations for $(a_x, b_x, h_{tx})$ and $(a_y, b_y, h_{ty})$ are not independent and can be derived from each other. This feature is a consequence of the background  rotational symmetry in the $(x, y)$ plane. If we consider AdS-RN background, three constants $a, b, c$ should vanish because we are looking for marginally-unstable perturbations without turning on any source.  Although what we are considering here is the $AdS_2$ background which is the near horizon geometry of the extremal AdS-RN solution, we still set $a=b=c=0$ so as to be self consistent with finite temperature analysis. By using this condition, the checkerboard perturbations in subsection~\ref{subsec:check} can be deduced straightforwardly.


\section{Striped Instabilities around the Antiferromagnetic Phase}\label{app:stripedonaf}

In this appendix we analyse in more detail the striped instabilities in the vicinity of the AF phase.

Since there is an overlap between AF phase and striped phase in figure~\ref{fig:fullphase}, those two phases could compete or coexist with each other. It is interesting to investigate the interplay between the inhomogeneous instabilities and the antiferromagnetic instabilities. A complete study needs one to solve the full coupled equations of motion numerically. In particular, one needs to check whether the solution with coexisting AF order and striped order can exist or not and whether such solution is thermodynamically favoured. One should solve a set of coupled PDEs which is much more numerical challenge. Nevertheless, one can obtain some useful hints from perturbative analysis. More precisely, one can search for the possible spatially modulated zero mode in the spectrum of fluctuations around the AF phase~\eqref{afansatz}.

The background solution of AF phase is sensitive to the details of the action. As a concrete example, we will consider the theory~\eqref{afcoupling} with ansatz given by~\eqref{afansatz} in section~\ref{sec:afphase}. For the pseudo-scalar $\alpha$, we consider the same setup as the case in figure~\ref{fig:striptx}, i.e., $n_1=0.8$ and $\widetilde{m}^2=a_3=b_3=c_3=n_2=0$.
A set of consistent perturbations with momentum $k$ in the $x$ direction are given by
\begin{equation}\label{stripedaf}
\begin{split}
&\delta g_{ty}= \lambda\, r (r-r_h)h_{ty}(r)\sin(kx)+\mathcal{O}(\lambda^2),\quad \delta \alpha= \lambda\, w(r)\cos(kx)+\mathcal{O}(\lambda^2),\\
&\delta B_{y}= \lambda\, b_{y}(r)\sin(kx)+\mathcal{O}(\lambda^2), \quad \delta A_{y}= \lambda\, a_{y}(r)\sin(kx)+\mathcal{O}(\lambda^2),\\
&\delta \phi= 0+\mathcal{O}(\lambda^2),
\end{split}
\end{equation}
while other perturbations can be consistently set to zero at linear order of $\lambda$.

The resulted linearised equations around the background~\eqref{afansatz} have a similar form as the case in AdS-RN background.
\begin{eqnarray}
 w''+\left(\frac{2}{r}+\frac{g'}{g}-\frac{\xi}{2}\right)w'-\frac{k^2}{r^2g}w-\frac{4k n_1 e^{\xi/2} A_t'}{r^2 g}a_y& = & 0, \\
\nonumber a_y''+\left(\frac{g'}{g}-\frac{\xi'}{2}\right)a_y'+\frac{r(r-r_h)e^{\xi}A_t'}{g}h_{ty}'-\frac{8k n_1 e^{\xi/2}A_t'}{g (2+a_1\phi^2)}w-\frac{k^2}{r^2 g}a_y\\+\frac{r_h e^{\xi}A_t'}{ g}h_{ty}&=&0,\\
b_y''+\left(\frac{g'}{g}-\frac{\xi'}{2}\right)b_y'+\frac{r(r-r_h)e^{\xi}B_t'}{g}h_{ty}'-\frac{k^2}{r^2 g}b_y+\frac{r_h e^{\xi}B_t'}{ g}h_{ty}&=&0,\\
\nonumber h_{ty}''+\left(\frac{4r-2r_h}{r(r-r_h)}+\frac{\xi'}{2}\right)h_{ty}'+\frac{(2+a_1\phi^2)A_t'}{2r(r-r_h)}a_y'+\frac{(2+b_1\phi^2)B_t'}{2r(r-r_h)}b_y'\\
-\left(\frac{k^2}{r^2g}-\frac{r_h(4+r \xi')}{2r^2(r-r_h)}\right)h_{ty}&=&0.
\end{eqnarray}
Therefore, for a given background profiles $(g, \xi, \phi, A_t, B_t)$ in~\eqref{afansatz}, one can check whether the system admits normalisable zero modes with non-vanishing wave number $k$. Note that the above coupled equations are unchanged under the transformation $k\rightarrow-k$ and $(w, a_y, b_y, h_{ty})\rightarrow(w, -a_y, -b_y, -h_{ty})$. So we can only focus on the positive value of $k$.

The fluctuations should be regular on the horizon at $r=r_h$. Therefore, the four functions $h_{ty}, a_y, b_y, w$ behave as
\begin{equation}
\begin{split}
&h_{ty}(r) = h^h+\mathcal{O}(r-r_h),\quad\quad w(r)=w^h+\mathcal{O}(r-r_h),\\
&a_{y}(r) =a^h +\mathcal{O}(r-r_h),\quad\quad b_y(r)= b^h +\mathcal{O}(r-r_h),
\end{split}
\end{equation}
near the horizon. On the other hand, to break translational invariance spontaneously we should not turn on any source. So the asymptotical behaviour in the UV as $r\rightarrow \infty$ is given by
\begin{equation}
\begin{split}
&h_{ty}(r) =\frac{h_v}{r^3}+\cdots,\quad\quad w(r)=\frac{w_v}{r^3}+\cdots,\\
&a_{y}(r) =\frac{a_v}{r}+\cdots,\quad\quad b_{y}(r) =\frac{b_v}{r}+\cdots,
\end{split}
\end{equation}
in which the parameters $(h_v, w_v, a_v, b_v)$ correspond to the response of the dual operators on the field theory side.

To solve this problem we take advantage of the double side shooting method from both the IR and UV (see appendix~\ref{app:doushot}). We first prepare four independent sets of $(h^h, w^h, a^h, b^h)$ as initial boundary conditions in the IR as well as four independent sets of $(h_v, w_v, a_v, b_v)$ as initial boundary conditions in the UV. After obtaining eight solutions we can generate the Wronskian~\eqref{appwroski}. The value of the Wronskian depends one the location of the connection point, say $r=r_0$. However, if the zero mode indeed exists, say, at $k=k_0$, the Wronskian $W(r_0)$ at $k=k_0$ should be zero for any choice of connection point $r_0$.

\begin{figure}[ht!]
\begin{center}
\includegraphics[width=.45\textwidth]{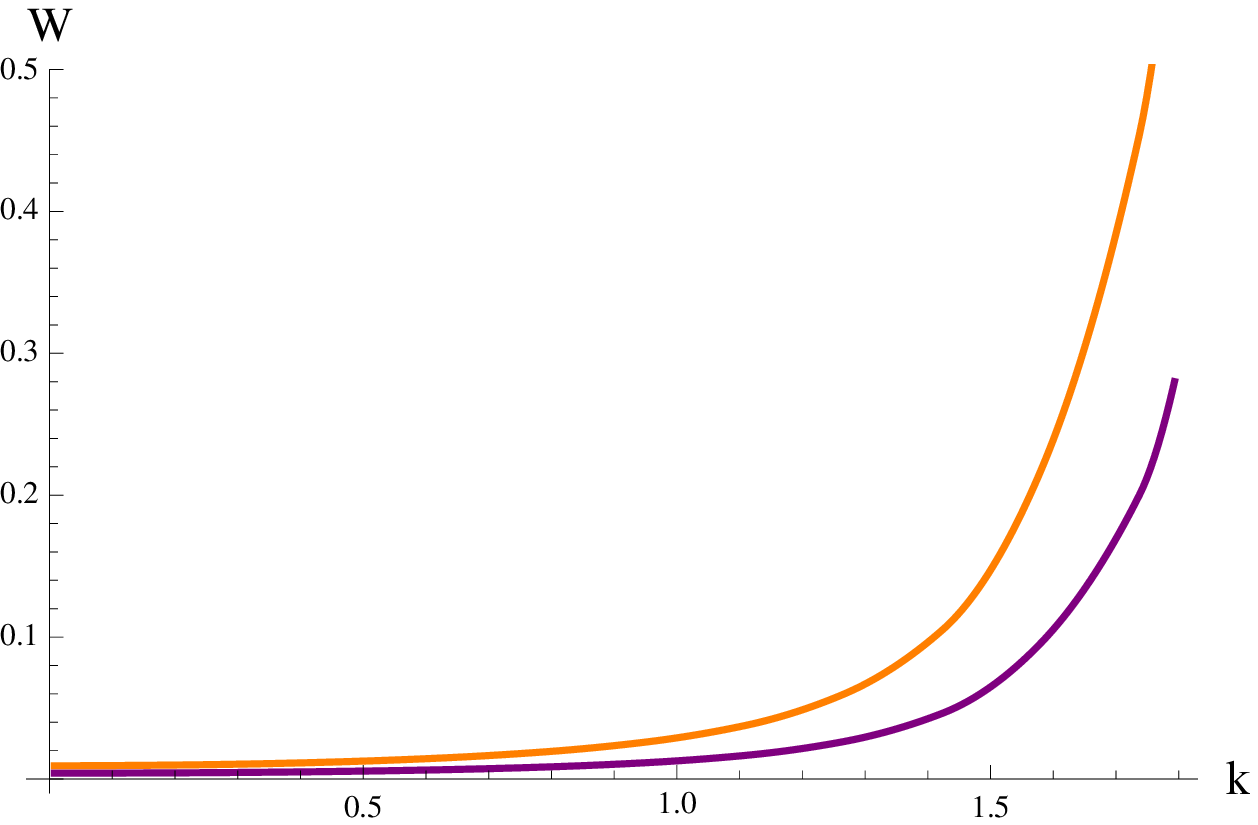}
\quad\quad
\includegraphics[width=.47\textwidth]{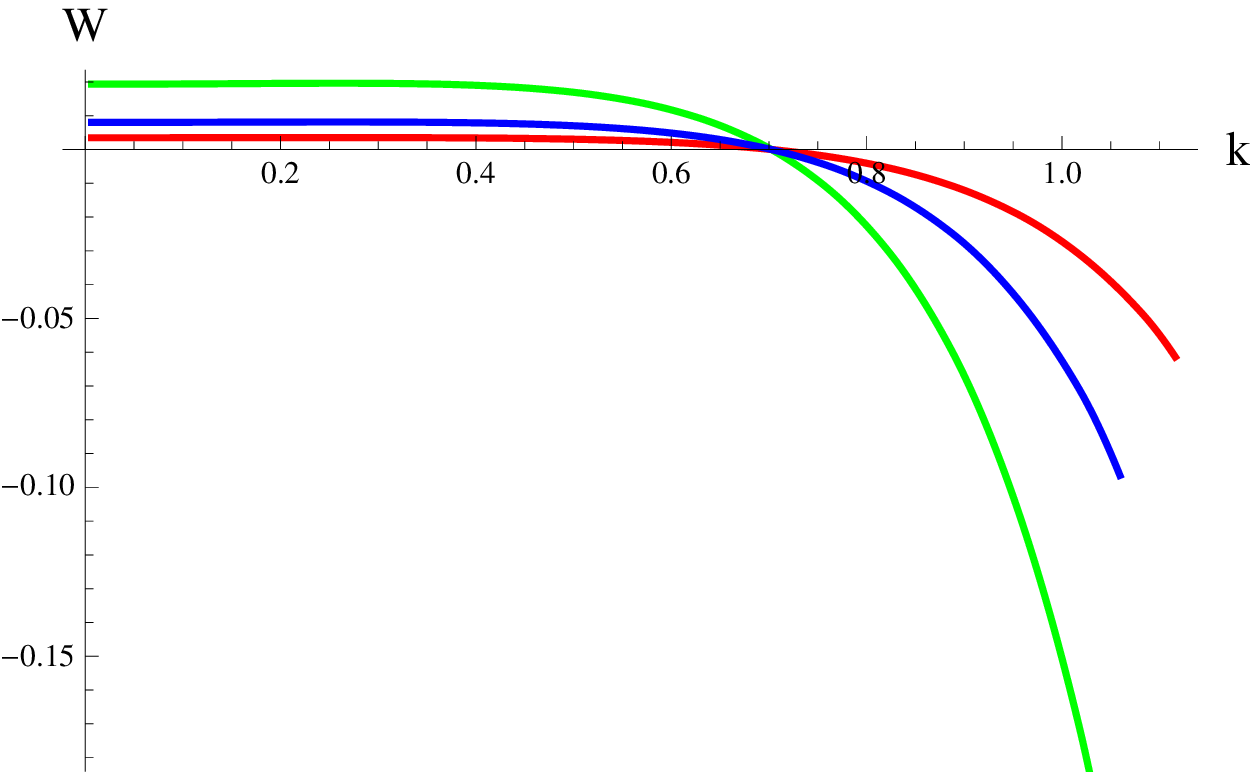}
\caption{The behaviour of Wronskian $W$ as a function of wave number $k$ for different choices of the connecting point $r_0$. The left plot is for the background profiles at $T\approx 0.887\, T_N$ and the right one is for the profiles at $T\approx 0.484 \,T_N$. We consider the same theory as in figure~\protect\ref{fig:caseaf} with $\textbf{x}=0.5$ and $T_N\approx 0.15863$.}
\label{fig:afwronski}
\end{center}
\end{figure}

We plot $W(r_0)$ as a function of $k$ for several values of  the connecting point $r_0$ in figure~\ref{fig:afwronski}. In the left plot which is for the background profiles at $T\approx 0.887\, T_N$ with $T_N$ the critical temperature for the AF phase transition, $W$ dose not vanish, so there is no zero mode. In this case the AF order can not coexist with the striped order. They compete with each other and only the one which has a lower free energy can win. In contrast, for the background profiles at $T\approx 0.484\, T_N$ in the right plot, all curves intersect at $W=0$, which points out the value of $k$ at which the zero mode exists. It suggests that the homogeneous phase of antiferromagnetism might be unstable. There might be a mixed phase with spatially modulated condensate of antiferromagnetism and CDWs in temperatures much lower than $T_N$.

Similarly, we can also look for the possible spatially modulated static mode in the spectrum of fluctuations around the superconducting background~\eqref{scansatz}. A set of consistent perturbations with momentum $k$ in the $x$ direction up to linear order of $\lambda$ read
\begin{equation}
\begin{split}
&\delta g_{ty}= \lambda  h_{ty}(r)\sin(kx)+\mathcal{O}(\lambda^2),\quad \delta \alpha= \lambda w(r)\cos(kx)+\mathcal{O}(\lambda^2),\\
&\delta B_{y}= \lambda b_{y}(r)\sin(kx)+\mathcal{O}(\lambda^2), \quad \delta A_{y}= \lambda a_{y}(r)\sin(kx)+\mathcal{O}(\lambda^2),\\
&\delta \chi= 0+\mathcal{O}(\lambda^2),
\end{split}
\end{equation}
with all other perturbations set to zero at the linear order. Inspired by the above results around AF background, it seems most likely that spatially modulated static modes around the superconducting background could exist. As a consequence, there might be a striped superconductor in some doping range.


\end{document}